\documentclass[a4paper,12pt,preprint,amsmath,showpacs,nofootinbib,superscriptaddress]{revtex4-1}
\usepackage{amsmath}
\usepackage{graphicx}
\usepackage{bm,braket}
\usepackage{multirow}
\usepackage{color}
\usepackage{slashed}

\newcommand{\als}{\alpha_s}
\newcommand{\ep}{\epsilon}
\newcommand{\nn}{\nonumber}
\newcommand{\lp}{L_{\perp}}

\begin{document}

\preprint{SLAC-PUB-15443}

\title{ Top quark pair production at small transverse momentum in hadronic collisions}
\author{Hai Tao Li}
\affiliation{School of Physics and State Key Laboratory of Nuclear Physics and Technology, Peking University, Beijing 100871, China}
\author{Chong Sheng Li}
\email{csli@pku.edu.cn}
\affiliation{School of Physics and State Key Laboratory of Nuclear Physics and Technology, Peking University, Beijing 100871, China}
\affiliation{Center for High Energy Physics, Peking University, Beijing 100871, China}
\author{Ding Yu Shao}
\affiliation{School of Physics and State Key Laboratory of Nuclear Physics and Technology, Peking University, Beijing 100871, China}
\author{Li Lin Yang}
\email{yanglilin@pku.edu.cn}
\affiliation{School of Physics and State Key Laboratory of Nuclear Physics and Technology, Peking University, Beijing 100871, China}
\affiliation{Center for High Energy Physics, Peking University, Beijing 100871, China}
\author{Hua Xing Zhu}
\email{hxzhu@slac.stanford.edu}
\affiliation{SLAC National Accelerator Laboratory, Stanford University, Stanford, CA 94309, USA}

\begin{abstract}
  We investigate the transverse momentum resummation for top quark pair production at hadron colliders using the soft-collinear effective theory  and the heavy-quark effective theory. We derive the factorization formula for $t\bar{t}$ production at small pair transverse momentum, and show in detail the procedure for calculating the key ingredient of the factorization formula: the next-to-leading order soft functions. We compare our numerical results with experimental data and find that they are consistent within theoretical and experimental uncertainties. To verify the correctness of our resummation formula, we expand it to the next-to-leading order and the next-to-next-to-leading order, and compare those expressions with the exact fixed-order results numerically. Finally, using the results of transverse momentum resummation, we discuss the  transverse-momentum-dependent forward-backward asymmetry at the Tevatron.
 \end{abstract}

\maketitle

\section{Introduction}
\label{sec:introduction}

Top quark physics is one of the major research topics in current and future theoretical and experimental particle physics. The top quark is the most massive known particle, and it plays a special role in the Standard Model (SM) and in many possible extensions of the SM. Once produced, the top quark immediately decays to a $W$ boson and a bottom quark before hadronization. This gives us a great opportunity to study many properties of the top quark. Up to date, tens of thousands of top quark events have been produced and studied at the Tevatron. The experiments at the LHC are accumulating data and are expected to observe millions of top quark events with the increase of the integrated luminosity. The expected high precision of the experimental measurements poses high demand on equally precise theoretical predictions for many observables including total and differential cross sections. In the SM, the main source of top quark events at hadron colliders is the top quark pair production. Studying this process on
the one hand provides a precision test of the SM, and on the other hand, can control the background of many new physics (NP) signals. Therefore, it is worthwhile and important to make precise theoretical predictions for top quark pair production at hadron colliders.

Actually, the efforts on obtaining precision predictions for $t\bar{t}$ production at hadron colliders have a long history since the eighties of the last century, when the next-to-leading order (NLO) quantum chromodynamics (QCD) corrections to this process were first calculated \cite{Nason:1987xz, Beenakker:1988bq, Beenakker:1990maa}. The NLO electroweak corrections  \cite{Beenakker:1993yr, Bernreuther:2005is} were known shortly after. Using narrow-width approximation, fully differential top quark pair production and decay at NLO are also known \cite{Melnikov:2009dn, Bernreuther:2010ny, Campbell:2012uf}. Furthermore, off-shell effects in top quark pair production have also been investigated \cite{Denner:2010jp, Bevilacqua:2010qb, Falgari:2013gwa}. Beyond NLO, threshold soft gluon resummation has been calculated to the next-to-next-to-leading logarithmic (NNLL) accuracy \cite{Czakon:2009zw, Beneke:2009rj, Ahrens:2010zv, Kidonakis:2010dk, Cacciari:2011hy}. Recently, calculations of the full next-to-next-to-
leading order (NNLO) QCD corrections to $t\bar{t}$ production have been finished \cite{Baernreuther:2012ws, Czakon:2012zr, Czakon:2012pz, Czakon:2013goa}. The NNLO fully differential top quark decay was also obtained using two different methods \cite{Gao:2012ja, Brucherseifer:2013iv}.

Recently, the inclusive and differential cross sections of top quark pair production at the LHC have been measured by the ATLAS and CMS Collaborations \cite{Aad:2012hg, Aad:2012vip, Chatrchyan:2012bra, CMS:pas-top-12-027, CMS:pas-top-12-028}. One interesting differential observable is the transverse momentum of the top quark pair. This is particularly the case since the D0 and CDF Collaborations showed that the $t\bar{t}$ charge asymmetry at the Tevatron has strong dependence on the $t\bar{t}$ transverse momentum \cite{Aaltonen:2012it, Abazov:2011rq}. It has also been shown in \cite{Kuhn:2011ri, Alvarez:2012vq} that a kinematic cut on the top quark pair transverse momentum leads to a significant enhancement on the charge asymmetry. It's therefore important to have better theoretical understanding of this observable, in particular at low transverse momentum.

It was well-known that for the Drell-Yan process, at small transverse momentum, $q_T \ll M$ where $M$ is the mass of the Drell-Yan pair, the collinear factorization must be replaced by transverse-momentum-dependent (TMD) factorization (also called $k_T$-factorization). This has been clearly demonstrated in the pioneering works of Collins, Soper and Sterman (CSS) \cite{Collins:1981uk, Collins:1981uw, Collins:1984kg}. Using $k_T$-factorization, the large logarithms of the form $\ln^n(q_T/M)$ can be systematically resummed to all orders in the strong coupling constant $\alpha_s$ provided that $q_T$ is in the perturbative domain, i.e., $q_T \gg \Lambda_{\text{QCD}}$. This is the so-called CSS formalism and for a recent review, see e.g. \cite{Collins:2011zzd}. The CSS formalism has since then been extended and applied to many processes including Higgs production and diphoton production \cite{Balazs:1997hv, Balazs:1999yf, Balazs:2000sz, Balazs:2000wv, Nadolsky:2002gj, Bozzi:2003jy, Bozzi:2005wk, Balazs:2006cc, Cao:2009md,
 deFlorian:2011xf, Wang:2012xs}. Note that the CSS formula receives corrections suppressed by powers of $\Lambda_{\text{QCD}}/q_T$; therefore it breaks down for $q_T \lesssim \Lambda_{\text{QCD}}$, and one needs to introduce some non-perturbative form factors \cite{Collins:1984kg, Landry:1999an, Landry:2002ix}.

For processes involving strong-interacting particles in the final state, however, things are much more complicated. For back-to-back hadron production, it has been shown in \cite{Collins:2007nk, Rogers:2010dm} by a counterexample that TMD factorization is not valid due to the appearance of process-dependent Wilson lines in the transverse-momentum-dependent parton distribution functions and fragmentation functions. On the other hand, the top quark is different from light partons in the sense that its mass is much larger than $\Lambda_{\text{QCD}}$ and it also decays before forming hadrons. It is therefore hopeful that for top quark pair production, TMD factorization and hence transverse momentum resummation may work. In fact, by extending the CSS formalism, resummation of the initial state radiations and the final state radiations at a partially next-to-leading logarithmic (NLL) accuracy has been studied in Refs.~\cite{Berger:1993yp, Mrenna:1996cz}. However, they did not include the soft gluon exchanges
between the initial and final state partons and it was not known how to extend the analysis beyond the NLL accuracy.

In the past decade, soft-collinear effective theory (SCET) \cite{Bauer:2000yr, Bauer:2001yt, Beneke:2002ph} has been proven to be a very efficient tool to deal with soft and collinear radiations and solve factorization and resummation problems. In the case of transverse momentum resummation, frameworks equivalent to the CSS formalism have been developed for both the Drell-Yan process and Higgs production \cite{Gao:2005iu, Idilbi:2005er, Becher:2010tm, Becher:2011xn, GarciaEchevarria:2011rb, Chiu:2012ir, Becher:2012yn}. Based on these works, in our recent paper \cite{Zhu:2012ts}, we developed for the first time a systematic all-order framework for the TMD factorization in top quark pair production and performed the transverse momentum resummation at the NNLL accuracy. In this paper, we show the details of our framework and the calculations of the resummation. We also perform a non-trivial check of the $q_T$ spectrum at NNLO in fixed-order perturbation theory, and give more numerical results based on our
formula. Finally, we point out how to use our results to construct a subtraction framework for $t\bar{t}$ production at NNLO based on the method of Ref.~\cite{Catani:2007vq}.  Recently, it was pointed out in Ref.~\cite{Mitov:2012gt} by studying the interactions between top quarks and beam remnants that the power corrections to TMD factorization in $t\bar{t}$ production are of the order $\Lambda_{\text{QCD}}/p'_T$, where $p'_T$ is the transverse momentum of the hardest parton recoiling against the $t\bar{t}$ pair. In the sense of power corrections, $\Lambda_{\text{QCD}}/p'_T$ is not so different from our previous expectation $\Lambda_{\text{QCD}}/q_T$, and the actual size of the power corrections can only be estimated by comparing to experimental data or employing some non-perturbative methods, which is beyond the scope of our paper.

This paper is structured as follows. In the following section, we briefly review the basic ideas of the effective field theory method and apply it to $t\bar{t}$ production at hadron colliders to obtain a factorization formula at small pair transverse momentum. In Section~\ref{sec:functions}, we present the calculation of the soft functions at the NLO. We show in Section~\ref{sec:resum} the renormalization group (RG) equations for the hard and soft functions and the TMD PDFs. By solving these RG equations we arrive at the final resummation formula. We expand the resummation formula to NLO and NNLO in Section~\ref{sec:fixed} and compare them with fixed-order calculations at small transverse momentum. The phenomenological implications are discussed in Section~\ref{sec:numerics}. We draw our conclusions in Section~\ref{sec:conclusion}. Some expressions are collected in the Appendices for readers' convenience.

\section{$k_T$-factorization for $t\bar{t}$ production}
\label{sec:factorization}

In this section we present the derivation of $k_T$ factorization for $t\bar{t}$ production using the SCET and the heavy quark effective theory (HQET) \cite{Eichten:1989zv, Georgi:1990um, Grinstein:1990tj, Mannel:1991mc, Neubert:1993mb, Manohar:2000dt}. The HQET was originally developed to study decays of charmed and beauty hadrons. The first application of SCET and HQET in top quark physics at hadron colliders was performed in \cite{Yang:2006gs} in the context of threshold resummation for direct top quark production. The extension to top quark pair production was shown in \cite{Ahrens:2010zv, Ahrens:2011mw, Ahrens:2011px, Ahrens:2011uf, Ferroglia:2012ku, Ferroglia:2012uy}, where the threshold resummation for the total cross section and various differential cross sections are performed. The transverse momentum resummation discussed in this paper shares some similarity with threshold resummation but is also genuinely different from that. In particular, the treatment of hard fluctuations is exactly the same as
in threshold resummation, and we will therefore reiterate certain derivations in \cite{Ahrens:2010zv}. On the other hand, the treatment of soft and collinear radiations are completely different from threshold resummation.

We consider the process
\begin{align}
  N_1(P_1) + N_2(P_2) \rightarrow t(p_3) + \bar{t}(p_4) + X(p_X) \, ,
\end{align}
where $X$ is an inclusive hadronic final state. At the leading order (LO), there are two partonic processes, namely, the quark-antiquark annihilation process and gluon fusion process,
\begin{align}
  q(p_1)+\bar{q}(p_2) &\rightarrow t(p_3)+\bar{t}(p_4) \, , \nn
  \\
  g(p_1)+g(p_2) &\rightarrow t(p_3)+\bar{t}(p_4) \, ,
\end{align}
where $p_1=
\xi_1P_1$ and $p_2=\xi_2P_2$. For later convenience, we define the following kinematic variables
\begin{align}
  \label{eq:stu}
  s&=(P_1+P_2)^2 \, , \quad \hat{s}=(p_1+p_2)^2 \, , \quad M^2=(p_3+p_4)^2 \, ,\nn
  \\
  t_1&=(p_1-p_3)^2-m_t^2 \, , \quad u_1=(p_1-p_4)-m_t^2 \, , \quad \tau=\frac{M^2+q_T^2}{s} \, ,
\end{align}
where $q_T$ is the transverse momentum of the $t\bar{t}$ pair and $m_t$ is the top quark mass. The kinematic region which we are interested in is
\begin{align}
  \hat{s},M^2,|t_1|,|u_1|,m_t^2 \gg q_T^2 \gg \Lambda_{\text{QCD}}^2 \, .
\end{align}
In this limit, only soft or collinear emissions can contribute and to study them, it is convenient to introduce two light-like vectors $n$ and $\bar{n}$ along the directions of the colliding partons, which satisfy $n \cdot \bar{n} = 2$. In the lab frame, they can be written as
\begin{align}
  n=(1,0,0,1), \quad \bar{n}=(1,0,0,-1) \, .
\end{align}
With the help of these two vectors, any four vector can be decomposed as
\begin{align}
  k^{\mu} = n \cdot k \frac{\bar{n}^\mu}{2} + \bar{n} \cdot k \frac{n^\mu}{2} + k_{\perp}^\mu \equiv k^+ \frac{\bar{n}^\mu}{2} + k^- \frac{n^\mu}{2} + k_{\perp}^\mu \, .
\end{align}

In the small $q_T$ limit, we need to distinguish four different momentum regions
\begin{align*}
  \text{hard:} &\qquad k^\mu \sim M(1,1,1) \, ,
  \\
  \text{collinear:} &\qquad k^\mu \sim M(\lambda^2,1,\lambda) \, ,
  \\
  \text{anti-collinear:} &\qquad k^\mu \sim M(1,\lambda^2,\lambda) \, ,
  \\
  \text{soft:} &\qquad k^\mu \sim M(\lambda,\lambda,\lambda) \, ,
\end{align*}
where we denote momenta by their components $k^\mu=(k^+,k^-,k_\perp)$ and $\lambda=q_T/M$. The top quark momenta can be written as $p_i^\mu=m_tv_i^\mu+k_i^\mu \, (i=3,4)$ where $v_i^2=1$ and the residue momenta $k_i^\mu$ scale like the soft mode. Note that threshold resummation is different since it involves an ultrasoft region $k^\mu \sim M(\lambda^2,\lambda^2,\lambda^2)$ but no soft region. To deal with these momentum regions, the effective theory method is a very useful and generic framework, which can separate the different regions at the field theoretical level and convert multi-scale problems into single-scale problems.

To derive the factorization formula, we start with the effective Hamiltonian which contributes to $t\bar{t}$ production. It is the same as in threshold resummation and can be written as \cite{Ahrens:2010zv}
\begin{align}
  \label{eq:hamiltonian}
  \mathcal{H}_{\text{eff}}(x) = \sum_{I,m} \int dt_1 dt_2 \, e^{im_t(v_3+v_4) \cdot x} \left[ \tilde{C}_{Im}^{q\bar{q}}(t_1,t_2) O_{Im}^{q\bar{q}}(x,t_1,t_2) + \tilde{C}_{Im}^{gg}(t_1,t_2) O_{Im}^{gg}(x,t_1,t_2) \right] ,
\end{align}
where $I$ and $m$ label the color structure and Dirac structure, respectively.  The derivation will be very similar for the $q\bar{q}$ and the $gg$ channels, with subtleties arising in the $gg$ channel related to the Lorentz structure. We will therefore show the details in the $gg$ channel, with the factorization formula in the $q\bar{q}$ channel a natural extension. We will also suppress the $gg$ subscripts and superscripts unless necessary.

The operators in the $q\bar{q}$ channel can be written as \cite{Ahrens:2010zv}
\begin{align}
  O^{q\bar{q}}_{Im}(x,t_1,t_2) = \sum_{\{a\},\{b\}} (c^{q\bar{q}}_I)_{\{a\}} \, [O_m^h(x)]^{b_3b_4} \, [O_m^c(x,t_1,t_2)]^{b_1b_2} \, [O^s(x)]^{\{a\},\{b\}} \, ,
\end{align}
with
\begin{gather}
  [O_m^h(x)]^{b_3b_4} = \bar{h}_{v_3}^{b_3}(x) \, \Gamma''_m \, h_{v_4}^{b_4}(x) \, , \quad [O^c(x,t_1,t_2)]^{b_1b_2} = \bar{\chi}_{\bar{n}}^{b_2}(x+t_2n) \, \Gamma'_m \, \chi_n^{b_1}(x+t_1\bar{n}) \, , \nn
  \\
  [O^s(x)]^{\{a\},\{b\}} = [S^\dagger_{v_3}(x)]^{b_3a_3} \, [S_{v_4}(x)]^{a_4b_4} \, [S^{\dagger}_{\bar{n}}(x)]^{b_2a_2} \, [S_n(x)]^{a_1b_1} \, ,
  \label{eq:qqoperators}
\end{gather}
while the operators in the $gg$ channel can be written as
\begin{align}
  O^{gg}_{Im}(x,t_1,t_2) = \sum_{\{a\},\{b\}} (c^{gg}_I)_{\{a\}} \, [O_m^h(x)]_{b_3b_4}^{\mu\nu} \, [O^c(x,t_1,t_2)]_{\mu\nu}^{b_1b_2} \, [O^s(x)]^{\{a\},\{b\}} \, ,
\end{align}
where
\begin{gather}
  [O_m^h(x)]_{b_3b_4}^{\mu\nu} = \bar{h}_{v_3}^{b_3}(x) \, \Gamma_m^{\mu\nu} \, h_{v_4}^{b_4}(x) \, , \quad [O^c(x,t_1,t_2)]_{\mu\nu}^{b_1b_2} = \mathcal{A}_{n\mu\perp}^{b_1}(x+t_1\bar{n}) \, \mathcal{A}_{\bar{n}\nu\perp}^{b_2}(x+t_2n) \, , \nn
  \\
  [O^s(x)]^{\{a\},\{b\}} = [S^\dagger_{v_3}(x)]^{b_3a_3} \, [S_{v_4}(x)]^{a_4b_4} \, [S^{\text{adj}\dagger}_{\bar{n}}(x)]^{b_2a_2} \, [S^{\text{adj}}_n(x)]^{a_1b_1} \, .
  \label{eq:ggoperators}
\end{gather}
In the above, $h_v$, $\chi_n$ $\mathcal{A}_{n\mu\perp}$ are gauge-invariant fields for heavy quarks, collinear quarks and collinear gluons in HQET and SCET, respectively. The index $m$ labels different Dirac structures and $\Gamma_m^{\mu\nu}$, $\Gamma'_m$, $\Gamma''_m$ are combinations of Dirac matrices and the external vectors $n$, $\bar{n}$, $v_3$ and $v_4$. The indices $a_i$ and $b_i$ with $i=1,2,3,4$ are color indices which can be in either the fundamental or adjoint representation depending on the particle involved. The tensors $c_I$ define a basis in color space, which we choose the same way as in \cite{Ahrens:2010zv}. They are given by
\begin{gather}
  \left(c_1^{q\bar{q}}\right)_{\{a\}} = \delta_{a_1a_2} \, \delta_{a_3a_4} \, , \quad \left(c_2^{q\bar{q}}\right)_{\{a\}} = t_{a_1a_2}^c \, t_{a_3a_4}^c \, , \nn
  \\
  \left(c_1^{gg}\right)_{\{a\}} = \delta^{a_1a_2} \, \delta_{a_3a_4} \, , \quad \left(c_2^{gg}\right)_{\{a\}} = i f^{a_1a_2c} \, t^{c}_{a_3a_4} \, , \quad \left(c_3^{gg}\right)_{\{a\}} = d^{a_1 a_2 c} \, t^c_{a_3a_4} \, .
  \label{eq:colorbasis}
\end{gather}
The soft Wilson lines are defined by
\begin{align}
  [S_n(x)]^{ab} &= \mathcal{P} \exp \left( ig\int_{-\infty}^0 dt \, n \cdot A_s^c(x+tn) \, t^c_{ab} \right) , \nn
  \\
  [S^{\text{adj}}_n(x)]^{ab} &= \mathcal{P} \exp \left( ig\int_{-\infty}^0 dt \, n \cdot A_s^c(x+tn) \, (-if^{cab}) \right) ,
\end{align}
and similarly for the $\bar{n}$, $v_3$ and $v_4$ directions.

To deal with the color indices, it is useful to introduce the color-space formalism of \cite{Catani:1996jh, Catani:1996vz}, which was extensively discussed in \cite{Ahrens:2010zv}. In this formalism, scattering amplitudes and similar objects are treated as vectors in an abstract vector space, while color generators and any objects involving them are treated as matrices in this vector space. We will use boldface letters to denote color space matrices. For example, the soft operator in Eq.~(\ref{eq:ggoperators}) is a matrix $\bm{O}^s(x)$, while the color basis in Eq.~(\ref{eq:colorbasis}) are vectors $\ket{c_I}$. As a result, the scattering amplitude for the $gg$ channel can be written as
\begin{align}
  \ket{\mathcal{M}(x)} &= \sum_{m} \int dt_1dt_2 \, e^{im_t(v_3+v_4) \cdot x} \nn
  \\
  &\times \Braket{ t\bar{t}X | [O^h_m(x)]^{\mu\nu} \, [O^c(x,t_1,t_2)]_{\mu\nu} \, \bm{O}^s(x) | N_1N_2 } \Ket{\tilde{C}_m(t_1,t_2)} \, ,
\end{align}
where the vectors of Wilson coefficients are defined as
\begin{align}
  \Ket{\tilde{C}_m(t_1,t_2)} = \sum_{I} \tilde{C}_{Im}(t_1,t_2) \ket{c_I} \, .
\end{align}

We can now write down the differential cross section for $t\bar{t}$ production
\begin{align}
  \label{eq:dsigma}
  d\sigma &= \frac{1}{2s} \frac{d^3\vec{p}_3}{(2\pi)^3 2E_3} \frac{d^3\vec{p}_4}{(2\pi)^3 2E_4} \sum_{X} \int d^4x \braket{\mathcal{M}(x)|\mathcal{M}(0)} \, .
\end{align}
Using the fact that the fields in different sectors of the effective theory do not interact with each other after absorbing the interactions into Wilson lines, we can factorize the squared matrix element as
\begin{align}
  \label{eq:M2}
  \sum_X \braket{\mathcal{M}(x)|\mathcal{M}(0)} &= \sum_{m,m'} \int dt_1dt_2dt'_1dt'_2 \, e^{-i(p_3+p_4) \cdot x} \nn
  \\
  &\times \Braket{0 | [O^{h\dagger}_{m'}(0)]^{\rho\sigma} | t(p_3)\bar{t}(p_4)} \Braket{t(p_3)\bar{t}(p_4) | [O^h_m(0)]^{\mu\nu} | 0} \nn
  \\
  &\times \sum_{X_c} \Braket{N_1(P_1) | \mathcal{A}_{n\rho\perp}(x^++x_\perp+t'_1\bar{n}) | X_c} \Braket{X_c | \mathcal{A}_{n\mu\perp}(t_1\bar{n}) | N_1(P_1)} \nn
  \\
  &\times \sum_{X_{\bar{c}}} \Braket{N_2(P_2) | \mathcal{A}_{\bar{n}\sigma\perp}(x^-+x_\perp+t'_2n) | X_{\bar{c}}} \Braket{X_{\bar{c}} | \mathcal{A}_{\bar{n}\nu\perp}(t_2n) | N_2(P_2)} \nn
  \\
  &\times \sum_{X_s} \bra{\tilde{C}_{m'}(t'_1,t'_2)} \Braket{0 | \bm{O}^{s\dagger}(x_\perp) | X_s} \Braket{X_s | \bm{O}^s(0) | 0} \ket{\tilde{C}_m(t_1,t_2)} ,
\end{align}
where $X_c$, $X_{\bar{c}}$ and $X_s$ denote the collinear, anti-collinear and soft final states, respectively, and we have performed the multipole expansion for the collinear and soft fields.

In the above formula, the collinear matrix elements correspond to the PDFs. Since we have $x_\perp$ in the arguments of the collinear fields, we need the TMD PDFs \cite{Collins:1981uk, Collins:1981uw} which in the transverse position space are defined as \cite{Becher:2012yn}
\begin{align}
  \label{eq:pdf}
  \mathcal{B}^{n}_{q/N}(z,x_T^2,\mu) &= \frac{1}{2\pi} \int dt \, e^{-izt \bar{n} \cdot p} \, \braket{N(p) | \bar{\chi}^a(t\bar{n}+x_\perp) \, \frac{\slashed{\bar{n}}}{2} \, \chi^a(0) | N(p)} \, , \nn
  \\
  \mathcal{B}^{\bar{n}}_{q/N}(z,x_T^2,\mu) &= \frac{1}{2\pi} \int dt \, e^{-izt n \cdot p} \, \braket{N(p) | \bar{\chi}^a(tn+x_\perp) \, \frac{\slashed{n}}{2} \, \chi^a(0) | N(p)} \, , \nn
  \\
  \mathcal{B}_{g/N}^{\mu\nu,n}(z,x_\perp,\mu) &= -\frac{z \bar{n} \cdot p}{2 \pi} \int dt \, e^{-i z t \bar{n} \cdot p } \braket{ N(p) | \mathcal{A}^{\mu a}_{n\perp}(t\bar{n}+x_\perp) \mathcal{A}^{\nu a}_{n\perp}(0) | N(p)} , \nn
  \\
  \mathcal{B}_{g/N}^{\mu\nu,\bar{n}}(z,x_\perp,\mu) &= -\frac{z n \cdot p}{2 \pi} \int dt \, e^{-i z t n \cdot p } \braket{ N(p) | \mathcal{A}^{\mu a}_{\bar{n}\perp}(tn+x_\perp) \mathcal{A}^{\nu a}_{\bar{n}\perp}(0) | N(p)} ,
\end{align}
where we have used the superscripts $n$ and $\bar{n}$ to label the moving direction of the initial hadron. Inverting the definition above, we obtain the matrix elements that appeared in Eq.~(\ref{eq:M2}):
\begin{align}
  &\braket{N_1(P_1) | \mathcal{A}^{a}_{n\rho\perp}(x^++x_\perp+t'_1\bar{n}) \, \mathcal{A}^{b}_{n\mu\perp}(t_1\bar{n}) | N_1(P_1)} \nn
  \\
  &\hspace{10em} = - \frac{2\delta_{ab}}{d_g} \int_0^1 \frac{dz_1}{z_1} \, \mathcal{B}_{g/N}^{\mu\rho,n}(z_1,x_\perp,\mu) \, e^{i(x^++(t'_1-t_1)\bar{n}) \cdot p_1} \, , \nn
  \\
  &\braket{N_2(P_2) | \mathcal{A}^{a}_{\bar{n}\sigma\perp}(x^-+x_\perp+t'_2n) \, \mathcal{A}^{b}_{\bar{n}\nu\perp}(t_2n) | N_2(P_2)} \nn
  \\
  &\hspace{10em} = - \frac{2\delta_{ab}}{d_g} \int_0^1 \frac{dz_2}{z_2} \, \mathcal{B}_{g/N}^{\nu\sigma,\bar{n}}(z_2,x_\perp,\mu) \, e^{i(x^-+(t'_2-t_2)n) \cdot p_2} \, ,
\end{align}
where we have identified $p_1=z_1P_1$ and $p_2=z_2P_2$, and $d_g=N^2-1$ with $N=3$ for QCD. For the $q\bar{q}$ channel we have similar equations
\begin{align}
  &\braket{N_1(P_1) | \bar{\chi}^a(t\bar{n}+x_\perp) \, \frac{\slashed{\bar{n}}}{2} \, \chi^b(0) | N_1(P_1))} \nn
  \\
  &\hspace{10em} = \frac{\delta_{ab}}{d_q} \int_0^1 \frac{dz_1}{z_1} \left[ \mathcal{B}_{q/N_1}^{n}(z,x_T^2,\mu) + \mathcal{B}_{\bar{q}/N_1}^{n}(z,x_T^2,\mu) \right] e^{it \bar{n} \cdot p_1} \, \bar{n} \cdot p_1 \, , \nn
  \\
  &\braket{N_2(P_2) | \bar{\chi}^a(tn+x_\perp) \, \frac{\slashed{n}}{2} \, \chi^b(0) | N_2(P_2))} \nn
  \\
  &\hspace{10em} = \frac{\delta_{ab}}{d_q} \int_0^1 \frac{dz_2}{z_2} \left[ \mathcal{B}_{q/N_2}^{\bar{n}}(z,x_T^2,\mu) + \mathcal{B}_{\bar{q}/N_2}^{\bar{n}}(z,x_T^2,\mu) \right] e^{it n \cdot p_2}  \, n \cdot p_2 \, ,
\end{align}
where $d_q=N$.

The integrals over $t_1$, $t_2$, $t'_1$, $t'_2$ now give rise to the momentum-space Wilson coefficients
\begin{align}
  \ket{C_m} = \ket{C_m(M,m_t,\cos\theta,\mu)} = \int dt_1dt_2 \, e^{- i t_1 \bar{n} \cdot p_1 - i t_2 n \cdot p_2} \Ket{\tilde{C}_m(t_1,t_2)} .
\end{align}
We can then define the hard function in the $gg$ channel as
\begin{align}
  \bm{H}^{\mu\nu\rho\sigma}_{gg}(M,m_t,v_3,\mu) &= \frac{3}{8} \frac{1}{(4\pi)^2} \frac{1}{4d_g} \sum_{m,m'} \ket{C_m} \bra{C_{m'}} \nn
  \\
  &\times \Braket{0 | [O^{h\dagger}_{m'}(0)]^{\rho\sigma} | t(p_3)\bar{t}(p_4)} \Braket{t(p_3)\bar{t}(p_4) | [O^h_m(0)]^{\mu\nu} | 0}  .
\end{align}

We should now mention that the definitions of the TMD PDFs in Eq.~(\ref{eq:pdf}) involve light-cone singularities which are not regularized by dimensional regularization. These divergences can be regularized in various ways \cite{Becher:2010tm, Becher:2011dz, Chiu:2012ir, Collins:2011zzd, Echevarria:2012qe}, and the product of two such TMD PDFs are free from the light-cone singularities. However, anomalous dependence on the hard scale $M$ remains, which was called ``collinear anomaly'' in \cite{Becher:2010tm}. In our framework we adopt the analytic regularization of \cite{Becher:2011dz}, and the product of the two TMD PDFs can be refactorized as \cite{Becher:2012yn}
\begin{align}
  \mathcal{B}^{\mu\nu,n}_{g/N_1}(z_1,x_\perp,\mu) \, \mathcal{B}^{\rho\sigma,\bar{n}}_{g/N_2}(z_2,x_\perp,\mu) = \left( \frac{x_T^2M^2}{4e^{-2\gamma_E}} \right)^{-F_{gg}(x_T^2,\mu)} B^{\mu\nu}_{g/N_1}(z_1,x_\perp,\mu) \, B^{\rho\sigma}_{g/N_2}(z_2,x_\perp,\mu) \, ,
\end{align}
where $x_T^2=-x_\perp^2$. The anomalous dependence on $M$ is factorized out and is controlled by the function $F_{gg}$, while the $B^{\mu\nu}_{g/N}$ functions are independent of $M$. Note also that we don't need to distinguish the $n$ and $\bar{n}$ directions for the $B^{\mu\nu}_{g/N}$ functions. There are two possible Lorentz structures for the $B^{\mu\nu}_{g/N}$ functions, which we choose as
\begin{align}
    B^{\mu\nu}_{g/N}(z,x_\perp,\mu) = \frac{g_\perp^{\mu\nu}}{2} B_{g/N}(z,x_T^2,\mu) + \left(\frac{g_\perp^{\mu\nu}}{2} + \frac{x_\perp^\mu x_\perp^\nu}{x_T^2} \right) B'_{g/N}(z,x_T^2,\mu) \, .
\end{align}

We finally turn to the soft part of Eq.~(\ref{eq:M2}), and define position space soft functions as matrices in color space
\begin{align}
  \label{eq:softfunction}
  \bm{W}(x_\perp,\mu) &= \frac{1}{d_R} \Braket{0 | \bar{\mathbf{T}}[\bm{O}^{s\dagger}(x_\perp)] \, \mathbf{T}[\bm{O}^{s}(0)] | 0} ,
\end{align}
In the intermediate steps of calculating the soft functions, we also encounter light-cone singularities of the same nature as in the TMD PDFs. For those we also use the scheme of \cite{Becher:2011dz} for the regularization. In the final soft function, such singularities cancel when combining different contributions. Note that for the Drell-Yan process and Higgs production, the soft functions are trivial when adopting the regularization method of \cite{Becher:2011dz}. The appearance of the soft function matrices is a reflection of the color exchange among initial state and final state particles, and is a genuinely new feature of our framework.

Since we are interested in the momentum of the $t\bar{t}$ pair, it is helpful to define $q=p_3+p_4$ and insert the following identity into Eq.\ (\ref{eq:dsigma}):
\begin{align}
  1 = \int d^4q \, dM^2 \, \delta^{(4)}(q-p_3-p_4) \, \delta(M^2-q^2) \, .
\end{align}
Performing the $\vec{p}_4$ and $|\vec{p}_3|$ integrals using the $\delta$-functions, we arrive at a factorization formula for the differential cross sections in the limit of small pair transverse momentum
\begin{align}
  &d\sigma = \frac{\beta_t}{6\pi^2s M}  \, dq_T^2dy \, dM \, d\cos\theta \int x_Tdx_T \, d\phi_x \, d\phi_q \, d\phi_t \, e^{iq_Tx_T\cos(\phi_q-\phi_x)} \, \bigg\{ \nn
  \\
  &\left( \frac{x_T^2M^2}{4e^{-2\gamma_E}} \right)^{-F_{gg}(x_T^2,\mu)} 4 B^{\mu\rho}_{g/N_1}(\xi_1,x_\perp,\mu) \, B^{\nu\sigma}_{g/N_2}(\xi_2,x_\perp,\mu) \, \mathrm{Tr} \big[ \bm{H}^{\mu\nu\rho\sigma}_{gg}(M,m_t,v_3,\mu) \, \bm{W}_{gg}(x_\perp,\mu) \big] \nn
  \\
  &+ \left( \frac{x_T^2M^2}{4e^{-2\gamma_E}} \right)^{-F_{q\bar{q}}(x_T^2,\mu)} B_{q/N_1}(\xi_1,x_T^2,\mu) \, B_{\bar{q}/N_2}(\xi_2,x_T^2,\mu) \, \mathrm{Tr} \big[ \bm{H}_{q\bar{q}}(M,m_t,\cos\theta,\mu) \, \bm{W}_{q\bar{q}}(x_\perp,\mu) \big] \nn
  \\
  &+ (q \leftrightarrow \bar{q}) \bigg\} \, ,
\end{align}
where we have also included the $q\bar{q}$ channel, and $\beta_t=\sqrt{1-4m_t^2/M^2}$, $\xi_1 = \sqrt{\tau} e^y$, $\xi_2 = \sqrt{\tau} e^{-y}$. In the above formula, $\theta$ is the scattering angle of the top quark in the $t\bar{t}$ rest frame, $\phi_t$, $\phi_q$ and $\phi_x$ are the azimuthal angles of $v_3$, $q_\perp$ and $x_\perp$, respectively. The azimuthal integrals can be simplified by noting that the integrand only depends on the two differences $\phi_q-\phi_x$ and $\phi_t-\phi_x$. Therefore
\begin{align}
  \label{eq:factorization}
  &\frac{d^4\sigma}{dq_T^2 \, dy \, dM \, d\cos\theta} = \frac{8\pi\beta_t}{3s M} \frac{1}{2} \int x_Tdx_T \, \frac{d\phi}{2\pi} \, J_0(x_Tq_T) \, \bigg\{ \nn
  \\
  &\left( \frac{x_T^2M^2}{4e^{-2\gamma_E}} \right)^{-F_{gg}(x_T^2,\mu)} 4 B^{\mu\rho}_{g/N_1}(\xi_1,x_\perp,\mu) \, B^{\nu\sigma}_{g/N_2}(\xi_2,x_\perp,\mu) \, \mathrm{Tr} \big[ \bm{H}^{\mu\nu\rho\sigma}_{gg}(M,m_t,v_3,\mu) \, \bm{W}_{gg}(x_\perp,\mu) \big] \nn
  \\
  &+ \left( \frac{x_T^2M^2}{4e^{-2\gamma_E}} \right)^{-F_{q\bar{q}}(x_T^2,\mu)} B_{q/N_1}(\xi_1,x_T^2,\mu) \, B_{\bar{q}/N_2}(\xi_2,x_T^2,\mu) \, \mathrm{Tr} \big[ \bm{H}_{q\bar{q}}(M,m_t,\cos\theta,\mu) \, \bm{W}_{q\bar{q}}(x_\perp,\mu) \big] \nn
  \\
  &+ (q \leftrightarrow \bar{q}) \bigg\} \, ,
\end{align}
where $\phi$ is now the relative azimuthal angle between $x_\perp$ and $v_3$.

Eq.~(\ref{eq:factorization}) is the master factorization formula of our paper, which is valid to all orders in $\alpha_s$ and to any logarithmic accuracy, up to power corrections of the sizes $q_T^2/M^2$ and $\Lambda_{\text{QCD}}^2/q_T^2$. The appearance of the tensor structures in the $gg$ channel was noted before in the studies of the Higgs production \cite{Catani:2010pd, Becher:2012qa, Becher:2012yn}. The case for $t\bar{t}$ production, however, is even more complicated since the hard matching coefficient itself is a tensor. The situation can be simplified if we restrict ourselves up to the NNLL accuracy. At this order, the second Lorentz structure in the $B^{\mu\nu}_{g/N}$ functions does not contribute. This is guaranteed since $B'_{g/N}$ vanishes at the leading order, and
\begin{align}
  \int_0^{2\pi} d\phi \, g^{\mu\rho}_\perp \left( \frac{g^{\nu\sigma}}{2} + \frac{x^\nu_\perp x^\sigma_\perp}{x_T^2} \right) \bm{H}^{(0),\mu\nu\rho\sigma}_{gg}(M,m_t,v_3,\mu) = 0 \, ,
\end{align}
where $\bm{H}^{(0),\mu\nu\rho\sigma}_{gg}$ is the leading order coefficient of $\bm{H}^{\mu\nu\rho\sigma}_{gg}$ in the perturbative expansion in $\alpha_s$. Once this is true, the dependence on $\phi$ in the integrand of Eq.~(\ref{eq:factorization}) now resides only in the soft functions. This fact motivates us to define new soft functions as
\begin{align}
  \label{eq:softphiint}
  \bm{S}_{i\bar{i}}(L_\perp,M,m_t,\cos\theta,\mu) = \int \frac{d\phi}{2\pi} \, \bm{W}(x_\perp,\mu) \, ,
\end{align}
where
\begin{align}
  L_\perp = \ln\frac{x_T^2\mu^2}{4e^{-2\gamma_E}} \, .
\end{align}
Note that the soft function defined in this way doesn't obey non-abelian exponentiation theorem. The reason is that the extra phase space integration over $\phi$ doesn't factorize. This means that at NNLO, the scale independent terms proportional to $C^2_F$ cannot be obtained by simply exponentiating the NLO results, but have to be recalculated. Fortunately, for the logarithmic accuracy studied in this paper, those terms are not needed.
The simplified factorization formula, valid up to the NNLL accuracy, now reads
\begin{multline}
  \label{eq:factorizationNNLL}
  \frac{d^4\sigma}{dq_T^2 \, dy \, dM \, d\cos\theta} = \sum_{i=q,\bar{q},g} \frac{8\pi\beta_t}{3s M} \, \frac{1}{2} \int x_Tdx_T \, J_0(x_Tq_T) \left( \frac{x_T^2M^2}{4e^{-2\gamma_E}} \right)^{-F_{i\bar{i}}(x_T^2,\mu)}
  \\
  \times B_{i/N_1}(\xi_1,x_T^2,\mu) \, B_{\bar{i}/N_2}(\xi_2,x_T^2,\mu) \, \mathrm{Tr} \Big[ \bm{H}_{i\bar{i}}(M,m_t,\cos\theta,\mu) \, \bm{S}_{i\bar{i}}(L_\perp,M,m_t,\cos\theta,\mu) \Big] \, .
\end{multline}
This formula will be the starting point of our NNLL transverse momentum resummation in the following.

\section{NLO results for the hard and soft functions and the TMD PDFs}
\label{sec:functions}

In this section, we present the NLO calculations for the hard and soft functions and the TMD PDFs which are relevant for the NNLL transverse momentum resummation. While the hard functions and the TMD PDFs at NLO are already available in the literature, the transverse soft function is new in our framework and is a major difference from the Drell-Yan process or Higgs production. Therefore, we will first discuss the calculation of the soft function.

The soft functions are defined in Eq.~(\ref{eq:softfunction}) and Eq.~(\ref{eq:softphiint}). We define the perturbative expansions of them as
\begin{align}
  \bm{S}_{i\bar{i}}(\lp,M,m_t,\cos\theta,\mu) = \sum_{n=0}^\infty \bm{S}^{(n)}_{i\bar{i}} \left(\frac{\als}{4\pi}\right)^n .
\end{align}
Up to now we have been treating the soft functions as abstract matrices in color space. In practice, it is more convenient to cast them into a matrix form by defining the matrix elements
\begin{align}
  S_{IJ} = \braket{c_I | \bm{S} | c_J} \, .
\end{align}
In this form, the LO soft functions for the $q\bar{q}$ and $gg$ channels are given by
\begin{align}
  \bm{S}_{q\bar{q}}^{(0)} =
  \begin{pmatrix}
    N & 0
    \\
    0 & \frac{C_F}{2}
  \end{pmatrix}
  , \quad
  \bm{S}_{gg}^{(0)}=
  \begin{pmatrix}
    N & 0 & 0
    \\
    0 & \frac{N}{2} & 0
    \\
    0 & 0 & \frac{N^2-4}{2N}
  \end{pmatrix}
  .
\end{align}

\begin{figure}[t!]
  \includegraphics[width=0.9\textwidth]{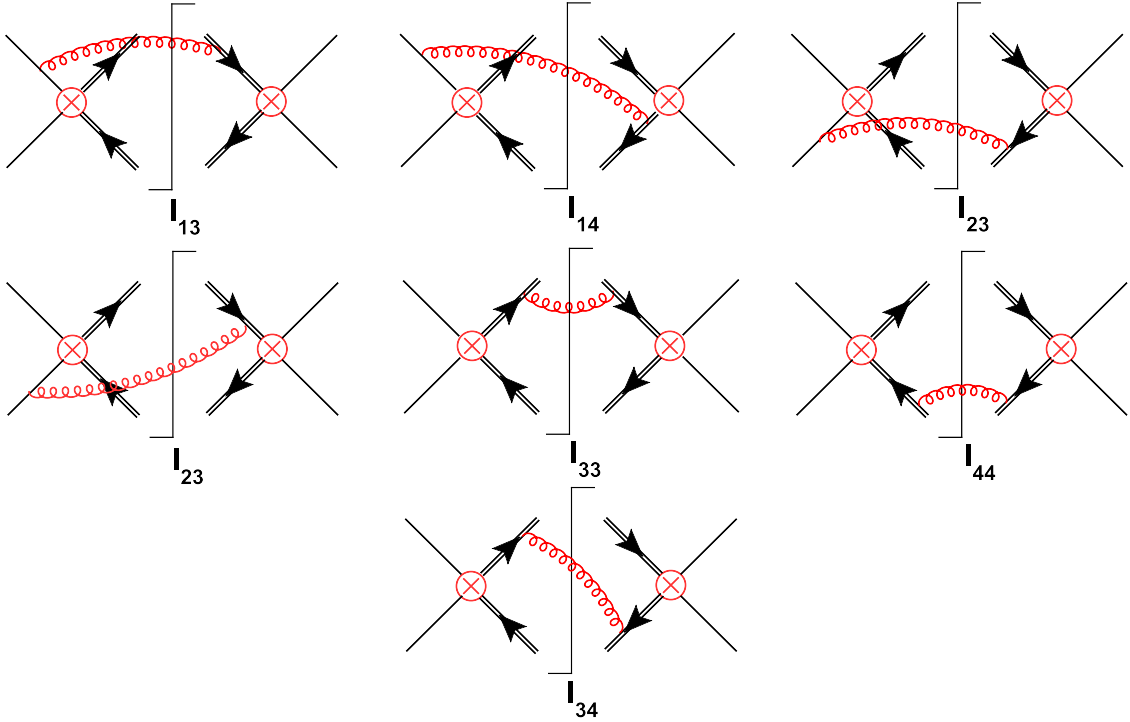}
  \vspace{-2ex}
  \caption{\label{fig:soft} Feynman diagrams contributing to the NLO soft functions. The double lines represent the Wilson lines in the directions along the top and anti-top quarks' movement. The single lines are the Wilson lines in the light-cone directions.}
\end{figure}

At the NLO, the soft functions receive contributions from the diagrams depicted in Fig.~\ref{fig:soft}. We can write the bare soft functions as
\begin{align}
  \label{eq:soft_NLO_bare}
  \bm{S}^{(1),\text{bare}}_{i\bar{i}} = \sum_{j,k} \bm{w}_{jk}^{i\bar{i}} \, I_{jk} \, ,
\end{align}
where $\bm{w}^{i\bar{i}}_{jk}$ is the NLO color matrices defined by
\begin{align}
  \left(\bm{w}^{i\bar{i}}_{jk}\right)_{IJ} = \frac{1}{d_R} \braket{c_I | \bm{T}_j \cdot \bm{T}_k | c_J } \, ,
\end{align}
with $\bm{T}_j$ the color generator associated with the parton $j$. These matrices can be found in Ref.~\cite{Ahrens:2010zv}.  $I_{jk}$ are integrals of the form
\begin{align}
  \label{eq:Ijk}
  I_{jk} &= -\frac{(4\pi\mu^2)^\epsilon}{\pi^{2-\epsilon}} \int^{2\pi}_0  d\phi \int d^dk \left( \frac{\nu}{n \cdot k} \right)^\alpha \delta(k^2) \, \theta(k^0) \, \frac{v_j \cdot v_k \, e^{-i x_\perp \cdot k_\perp}}{v_j \cdot k \; v_k \cdot k} \, ,
\end{align}
where the analytic regularization method of Ref.~\cite{Becher:2011dz} is used.
 We show an example for calculating $I_{13}$ in Appendix~\ref{sec:softcalc}. The results for the non-vanishing integrals  are
\begin{align}
  I_{13} &= \left( L_\perp + \frac{1}{\ep} \right) \left( -\frac{2}{\alpha} + \ln\frac{\mu^2}{\nu^2} + 2\ln\frac{-t_1}{m_tM} \right) + \frac{1}{\ep^2} - \frac{L^2_\perp}{2} - \frac{\pi^2}{12} - \mathrm{Li}_2 \left( 1 - \frac{t_1 u_1}{m_t^2 M^2} \right) , \nn
  \\
  I_{23} &= \left( L_\perp + \frac{1}{\ep} \right) \left( \frac{2}{\alpha} - \ln\frac{\mu^2}{\nu^2} + 2\ln\frac{-u_1}{m_tM} \right) - \frac{1}{\ep^2} + \frac{L^2_\perp}{2} + \frac{\pi^2}{12} - \mathrm{Li}_2 \left( 1 - \frac{t_1 u_1}{m_t^2 M^2} \right) , \nn
  \\
  I_{34} &= -\frac{(1+\beta_t^2)\ln x_s}{\beta_t} \left( L_\perp + \frac{1}{\ep} + f_{34} \right) , \nn
  \\
  I_{33} &= I_{44} = 2L_\perp + \frac{2}{\ep} - 2 \ln \left( \frac{t_1 u_1}{m_t^2 M^2} \right) , \nn
  \\
  I_{14} &= I_{13}(t_1 \leftrightarrow u_1) \, , \quad I_{24}=I_{23}(u_1 \leftrightarrow t_1) \, ,
  \label{eq:soft_Iij}
\end{align}
where $x_s = (1-\beta_t)/(1+\beta_t)$ and
\begin{align}
  f_{34} &=  - \mathrm{Li}_2 \left( -x_s \tan^2\frac{\theta}{2} \right) + \mathrm{Li}_2 \left( -\frac{1}{x_s} \tan^2\frac{\theta}{2} \right) + 4\ln x_s \ln\cos\frac{\theta}{2} \, .
\end{align}
The contribution from soft gluon exchange between initial states,  i.e. $I_{12}$, vanishes because the corresponding integral is scaleless in the regularization scheme we adopt.  If we had adopted other regularization methods such as that of Ref.~\cite{Chiu:2012ir}, $I_{12}$ could give a non-vanishing contribution. However, this contribution can always be absorbed into the two collinear sectors, so that our framework, in particular our soft function, is independent of the regularization scheme used.

Combining the integrals with the color matrices, and renormalizing in the $\overline{\text{MS}}$ scheme, we arrive at the final form of the NLO soft functions
\begin{align}
  \bm{S}_{i\bar{i}}^{(1)} &= 4 L_\perp \left( 2\bm{w}^{13}_{i\bar{i}}  \ln\frac{-t_1}{m_tM}  + 2\bm{w}^{23}_{i\bar{i}} \ln\frac{-u_1}{m_tM} + \bm{w}^{33}_{i\bar{i}} \right)  - 4 \left( \bm{w}^{13}_{i\bar{i}} + \bm{w}^{23}_{i\bar{i}} \right) \mathrm{Li}_2 \Biggl( 1 - \frac{t_1u_1}{m_t^2M^2} \Biggr) \nn
  \\
  &+ 4\bm{w}^{33}_{i\bar{i}} \ln\frac{t_1u_1}{m_t^2M^2} -  2\bm{w}^{34}_{i\bar{i}} \, \frac{1+\beta_t^2}{\beta_t} \, \bigl[ L_\perp \ln x_s + f_{34} \bigr] \, .
\end{align}
Note that while the individual integrals in Eq.~(\ref{eq:soft_Iij}) contain poles in the analytic regulator $\alpha$, these divergences cancel in the final soft functions, together with the dependence on the unphysical scale $\nu$.

We now turn to the hard functions. They are the absolute values squared of the Wilson coefficients of the operators, which can be obtained by matching the full theory onto SCET. The LO hard functions are simple to calculate,  which are just the tree-level amplitude squared, decomposed into the color basis in Eq.~(\ref{eq:colorbasis}). For the NLO hard functions, we work with on-shell external particles. As a result, the loop integrals in SCET are scaleless and vanish. Therefore, the NLO hard functions can be obtained by computing the one-loop virtual diagrams for top quark pair production. The NLO virtual corrections for $t\bar{t}$ production have been calculated long ago \cite{Nason:1987xz, Beenakker:1988bq, Beenakker:1990maa}.  However, they are not color-decomposed and are not suitable for extracting the hard functions. In Ref.~\cite{Ahrens:2010zv}, the one-loop amplitudes decomposed into the color basis were calculated,  and the NLO hard functions were extracted there, which we will take over. Up to NLO,
the hard functions can be expressed as
\begin{align}
  \bm{H}_{i\bar{i}} = \frac{3 \als^2}{8 d_i} \left( \bm{H}_{i\bar{i}}^{(0)} + \frac{\als}{4\pi} \, \bm{H}_{i\bar{i}}^{(1)} \right) .
\end{align}
Similar to the soft functions, the matrix form of the hard functions is defined by
\begin{align}
  H_{IJ} = \frac{1}{\braket{c_I|c_I} \, \braket{c_J|c_J}} \, \braket{c_I | \bm{H} | c_J} \, .
\end{align}
The leading order hard function matrices are
\begin{align}
  \bm{H}_{q\bar{q}}^{(0)} &=
  \begin{pmatrix}
    0 & 0
    \\
    0 & 2
  \end{pmatrix}
  \Bigg[ \frac{t_1^2 + u_1^2}{M^4} + \frac{2m_t^2}{M^2} \Bigg] \, , \nn
  \\
  \bm{H}_{gg}^{(0)} &=
  \begin{pmatrix}
    \frac{1}{N^2} & \frac{1}{N}\,\frac{t_1-u_1}{M^2} & \frac{1}{N}
    \\
    \frac{1}{N}\,\frac{t_1-u_1}{M^2} & \frac{(t_1-u_1)^2}{M^4} & \frac{t_1-u_1}{M^2}
    \\
    \frac{1}{N} & \frac{t_1-u_1}{M^2} & 1
  \end{pmatrix}
  \frac{M^4}{2t_1u_1} \Bigg[ \frac{t_1^2+u_1^2}{M^4} + \frac{4m_t^2}{M^2} -
  \frac{4m_t^4}{t_1u_1} \Bigg] \, .
\end{align}
The expressions for the NLO hard functions are rather lengthy, and we have obtained them from the electronic file associated with the arXiv submission of Ref.~\cite{Ahrens:2011px}.

We finally discuss the TMD PDFs. The $B_{q/N}$ and $B_{g/N}$ functions introduced in Section~\ref{sec:factorization} are intrinsically non-perturbative objects. For $x_T \ll 1/\Lambda_{\text{QCD}}$, they can be matched onto the normal PDFs \cite{Becher:2010tm} via
\begin{align}
  \label{eq:Imatch}
  B_{q/N}(z,x_T^2,\mu) &= \sum_{i} \int \frac{d\xi}{\xi} \, I_{q \leftarrow i}(\xi,L_\perp,\mu) \, \phi_{i/N}(z/\xi,\mu) \, , \nn
  \\
  B_{g/N}(z,x_T^2,\mu) &= \sum_{i} \int \frac{d\xi}{\xi} \, I_{g \leftarrow i}(\xi,L_\perp,\mu) \, \phi_{i/N}(z/\xi,\mu) \, ,
\end{align}
with perturbatively calculable matching coefficient functions $I_{i \leftarrow j}$. At leading order the $I_{i \leftarrow j}$ functions are given by $I^{(0)}_{i \leftarrow j}(z,L_\perp,\mu) = \delta_{ij} \, \delta(1-z)$. The NLO results for them have been calculated in Refs.~\cite{Becher:2010tm, Becher:2012yn}. We collect those results in Appendix~\ref{sec:expressions} for the readers' convenience.

\section{RG evolution and transverse momentum resummation}
\label{sec:resum}

In our formalism, the factorization scale $\mu$ is chosen at the typical soft/collinear scale where the soft functions and the TMD PDFs can be expanded in the perturbation theory. The resummation of large logarithms is achieved by evolving the hard functions from the hard scale $\mu_h$ down to the factorization scale. In this section, we present the RG equations for the various functions and the final form of the resummation formula. The explicit expressions for the relevant anomalous dimensions are collected in Appendix~\ref{sec:expressions}.

The treatment for the $I_{i \leftarrow j}$ functions are the same as in Ref.~\cite{Becher:2011xn}. For completeness, we briefly summarize it here. The $I_{i \leftarrow j}$ functions, as defined in Eq.~(\ref{eq:Imatch}), satisfy the following RG equation
\begin{align}
  \label{eq:RGI}
  \frac{d}{d\ln\mu} I_{i \leftarrow j} (z, L_\perp, \als) &= \left[ \Gamma_{\text{cusp}}^{i}(\als) \, L_\perp - 2\gamma^i(\als) \right] I_{i \leftarrow j}(z,L_\perp,\als) \nn
  \\
  &\hspace{5em} - \sum_k \int_z^1 \frac{d\xi}{\xi} \, I_{i \leftarrow k}(\xi,L_\perp,\als) \, \mathcal{P}_{kj}(z/\xi,\als) \, ,
\end{align}
where $\Gamma_{\text{cusp}}^i$ are the cusp anomalous dimensions, $\gamma^i$ are the single parton anomalous dimensions and $\mathcal{P}_{kj}$ are the DGLAP splitting functions. While we are going to choose the scale $\mu$ such that the $I_{i \leftarrow j}$ functions contain no large logarithms, it has been shown in Ref.~\cite{Becher:2011xn} that it is essential to exponentiate the double logarithmic terms in the $I_{i \leftarrow j}$ functions. This can be achieved by defining
\begin{align}
  \label{eq:Iexp}
  I_{q \leftarrow i}(z,L_\perp,\als) &\equiv e^{h_q(L_\perp,\als)} \, \bar{I}_{q \leftarrow i}(z,L_\perp,\als) \, , \nn
  \\
  I_{g \leftarrow i}(z,L_\perp,\als) &\equiv e^{h_g(L_\perp,\als)} \, \bar{I}_{g \leftarrow i}(z,L_\perp,\als) \, ,
\end{align}
where $h_i(L_\perp,\als)$ satisfies the evolution equation
\begin{align}
  \frac{d}{d\ln\mu} h_i(L_\perp,\als) = \Gamma_{\text{cusp}}^{i}(\als) \, L_\perp - 2\gamma^i(\als) \, .
\end{align}
The expressions for $h_i(L_\perp,\als)$ have been given in Ref. \cite{Becher:2011xn}. The new functions $\bar{I}_{g\leftarrow i}(z,L_\perp,\als)$ evolve exactly following the DGLAP equations with an opposite sign.  With Eq.~(\ref{eq:Imatch}) and Eq.~(\ref{eq:Iexp}), the NNLL factorization formula Eq.~(\ref{eq:factorizationNNLL}) now becomes
\begin{multline}
  \label{eq:master}
  \frac{d^4\sigma}{dq_T^2 \, dy \, dM \, d\cos\theta} = \sum_{i=q,\bar{q},g} \sum_{a,b} \frac{8\pi\beta_tM}{3s(M^2+q_T^2)} \int_{\xi_1}^1 \frac{dz_1}{z_1} \int_{\xi_2}^1 \frac{dz_2}{z_2} \, f_{a/N_1}(\xi_1/z_1,\mu) \, f_{b/N_2}(\xi_2/z_2,\mu)
  \\
  \times C_{i\bar{i} \leftarrow ab}(z_1,z_2,q_T,M,m_t,\cos\theta,\mu) \, ,
\end{multline}
where
\begin{multline}
  \label{eq:cii}
  C_{i\bar{i} \leftarrow ab} (z_1,z_2,q_T,M,\cos\theta,m_t,\mu) = \frac{1}{2} \int^\infty_0 x_Tdx_T \, J_0(x_Tq_T)  \, \exp \big[ g_i(\eta_i,L_\perp,\als) \big]
  \\
  \times \bar{I}_{i \leftarrow a}(z_1,L_\perp,\als) \, \bar{I}_{\bar{i} \leftarrow b}(z_2,L_\perp,\als) \, \mathrm{Tr} \Big[ \bm{H}_{i\bar{i}}(M,m_t,\cos\theta,\mu) \, \bm{S}_{i\bar{i}}(\lp,M,m_t,\cos\theta,\mu) \Big] .
\end{multline}
Here $\eta_i = (C_i\als/\pi)\ln(M^2/\mu^2)$ with $C_q = C_F$ and $C_g = C_A$. The $g_i$ function  is given by
\begin{align}
  g_i(\eta_i,L_\perp,\als) = - \left( \ln \frac{M^2}{\mu^2} + L_\perp \right) F_{i\bar{i}}(L_\perp,\als) + 2h_i(L_\perp,\als) \, .
\end{align}
As described in Ref.~\cite{Becher:2011xn}, for very small $q_T$ we must reorganize the resummation procedure by using the modified power counting scheme, where $(\alpha_sL_\perp)^n$ are counted as order $\epsilon^{n/2}$, with $\epsilon$ an auxiliary expansion parameter. The NLL accuracy should contain terms up to order $\epsilon^0$ and the NNLL accuracy means up to order $\epsilon^1$. Within this power counting scheme, the $g_i$ functions can be written as \cite{Becher:2011xn, Becher:2012yn}
\begin{align}
  \label{eq:gi}
  g_i(\eta_i,L_\perp,\als) = &- \left[ \eta_i L_\perp \right]_{\epsilon^{-1/2}} - \left[ a_s (\Gamma_0^i + \eta_i \beta_0 ) \, \frac{L_\perp^2}{2} \right]_{\epsilon^0} \nn
  \\
  &- \left[ a_s (2\gamma_0^i+\eta_iK) \, L_\perp + a_s^2 ( \Gamma_0^i + \eta_i \beta_0 ) \, \beta_0 \, \frac{L_\perp^3}{3}  \right]_{\epsilon^{1/2}}
  \\
  &- \left[ a_s \eta_i d_2 + a_s^2 (K\Gamma_0^i + 2\gamma_0^i \beta_0 + \eta_i (\beta_1 +2K\beta_0)) \, \frac{L_\perp^2}{2} +  a_s^3 (\Gamma_0^i + \eta_i \beta_0 ) \beta_0^2 \, \frac{L_\perp^4}{4} \right]_{\epsilon} \nn
  \\
  &- \mathcal{O}(\epsilon^{3/2}) \, , \nn
\end{align}
where
\begin{align}
    a_s=\frac{\alpha_s}{4\pi}, \ \
    K = \left( \frac{67}{9}-\frac{\pi^2}{3}\right) C_A - \frac{20}{9} T_F n_f, \ \
    d_2 = \left( \frac{202}{27}-7\zeta_3 \right) C_A -\frac{56}{27} T_F n_f.
\end{align}
And with NNLL accuracy, the $\bar{I}_{i \leftarrow j }(z,L_\perp,\alpha_s)$ functions are given by
\begin{align}
  \bar{I}_{i \leftarrow j}(z,L_\perp,\alpha_s) &= \delta_{ij} \, \delta(1-z) -\left[ a_s \mathcal{P}^{(1)}_{ij}(z) \frac{L_\perp}{2}\right]_{\epsilon^{1/2}} \nn
  \\
  &+ \left[ a_s \mathcal{R}^{(1)}_{ij}(z) + a_s^2 \left( \mathcal{D}_{ij}(z) - 2\beta_0 \mathcal{P}^{(1)}_{ij}(z) \right) \frac{L_\perp^2}{8} \right]_{\epsilon} + \mathcal{O}(\epsilon^{3/2}) \, ,
\end{align}
where $\mathcal{D}_{ij}(z)$ is defined as
\begin{align}
  \label{eq:Dij}
  \mathcal{D}_{ij}(z) = \sum_{k} \int_z^1 \frac{d\xi}{\xi} \, \mathcal{P}^{(1)}_{ik}(\xi) \, \mathcal{P}^{(1)}_{kj}(z/\xi) \, .
\end{align}

The RG evolution of the hard functions is the same as in the threshold resummation for top quark pair production studied in \cite{Ahrens:2010zv}, which can be written as
\begin{align}
  \label{eq:RG_H}
  \frac{d}{d\ln\mu} \bm{H}_{i\bar{i}}(M,m_t,\cos\theta,\mu) &= \bm{\Gamma}^H_{i\bar{i}}(M,m_t,\cos\theta,\mu) \, \bm{H}_{i\bar{i}}(M,m_t,\cos\theta,\mu) \nn
  \\
  &+ \bm{H}_{i\bar{i}}(M,m_t,\cos\theta,\mu) \, \bm{\Gamma}_{i\bar{i}}^{H\dagger}(M,m_t,\cos\theta,\mu) \, ,
\end{align}
where $\bm{\Gamma}^H_{i\bar{i}}$ are the anomalous dimensions of the hard functions and can be found in Ref.~\cite{Ferroglia:2009ii} for both the $q\bar{q}$ and $gg$ initial states. It will be convenient to split $\bm{\Gamma}^H_{i\bar{i}}$ into two parts
\begin{align}
  \bm{\Gamma}^{H}_{i\bar{i}} = \left[ \Gamma^i_{\text{cusp}}(\als) \left( \ln\frac{M^2}{\mu^2} - i\pi \right) + 2\gamma^i(\als) \right] \bm{1} + \gamma_{i\bar{i}}^{h}(M,m_t,\cos\theta,\als) \, .
\end{align}
 The solution to Eq.~(\ref{eq:RG_H}) is
\begin{align}
  \bm{H}_{i\bar{i}}(M,m_t,\cos\theta,\mu) = \bm{U}_{i\bar{i}}^{H}(M,m_t,\cos\theta,\mu_h,\mu) \, \bm{H}_{i\bar{i}}(M,m_t,\cos\theta,\mu_h) \, \bm{U}^{H\dagger}_{i\bar{i}}(M,m_t,\cos\theta,\mu_h,\mu) \, ,
\end{align}
where $\bm{U}^{H}_{i\bar{i}}$ is given by
\begin{align}
  \bm{U}^{H}_{i\bar{i}}(M,m_t,\cos\theta,\mu_h,\mu) &= \exp \big[ 2S_i(\mu_h,\mu) - 2a_{\gamma^i}(\mu_h,\mu) \big] \left( \frac{M^2}{\mu_h^2}\right)^{-a_{\Gamma_i}} \nn
  \\
  &\hspace{10em} \times \bm{u}^{h}_{i\bar{i}}(M,m_t, cos\theta,\mu_h,\mu) \, .
\end{align}
The functions $S_i(\mu_h,\mu)$ and $a_{\Gamma^i}$ are defined as
\begin{align}
  S_i(\mu_h,\mu) = - \int_{\als(\mu_h)}^{\als(\mu)} d\alpha \, \frac{\Gamma_{\text{cusp}}^{i}(\alpha)}{\beta(\alpha)} \int_{\als(\mu_h)}^\alpha \frac{d\alpha'}{\beta(\alpha')} \, , \quad a_{\Gamma^i}(\mu_h,\mu) = - \int_{\als(\mu_h)}^{\als(\mu)} d\alpha \, \frac{\Gamma_{\text{cusp}}^{i}(\alpha)}{\beta(\alpha)} \, ,
\end{align}
and similarly for $a_{\gamma^i}$. The matrix $\bm{u}^{h}_{i\bar{i}}$ is
\begin{align}
  \label{eq:uh}
  \bm{u}^{h}_{i\bar{i}}(M,m_t,\cos\theta,\mu_h,\mu) = \mathcal{P} \exp \int_{\als(\mu_h)}^{\als(\mu)} \frac{d\alpha}{\beta(\alpha)} \, \bm{\gamma}^{h}_{i\bar{i}}(M,m_t,\cos\theta,\mu_h,\mu) \, .
\end{align}
Using the method shown in Refs.~\cite{Buras:1991jm, Buchalla:1995vs}, the matrix function $\bm{u}^h_{i\bar{i}}$ can be obtained as follows:
\begin{align}
  \bm{u}_{i\bar{i}}^{h}(M,m_t,\cos\theta,\mu_h,\mu)  = \bm{V} \left( 1 + \frac{\als(\mu)}{4\pi} \bm{K} \right) \left[ \frac{\als(\mu_h)}{\als(\mu)} \right]^{\frac{\bm{\gamma}^{h(0)}_{i\bar{i}}}{2\beta_0}}_{D}  \left( 1 - \frac{\als(\mu_h)}{4\pi} \bm{K} \right) \bm{V}^{-1} \, ,
\end{align}
with
\begin{align}
  \bm{\gamma}^{h(0)}_{i\bar{i},D} &= \bm{V}^{-1} \, \bm{\gamma}^{h(0)}_{i\bar{i}} \, \bm{V} \nn
  \\
  \bm{K}_{IJ} &= \frac{\beta_1}{2\beta_0^2} \, \delta_{IJ}  \left( \bm{\gamma}^{h(0)}_{i\bar{i},D} \right)_{II} - \frac{ \left[ \bm{V}^{-1} \, \bm{\gamma}^{h(1)}_{i\bar{i}} \, \bm{V} \right]_{IJ}} {2\beta_0 + \left( \bm{\gamma}^{h(0)}_{i\bar{i},D} \right)_{II} - \left( \bm{\gamma}^{h(0)}_{i\bar{i},D} \right)_{JJ} } \, ,
\end{align}
where the subscript $D$ is used to label the diagonalized matrix. Finally, the solution of Eq.~(\ref{eq:RG_H}) is
\begin{align}
  \bm{H}_{i\bar{i}}(M,m_t,\cos\theta,\mu_h,\mu) &= \exp \big[ 4S_i(\mu_h,\mu) - 4a_{\gamma^i}(\mu_h,\mu) \big] \left( \frac{M^2}{\mu_h^2} \right)^{-2a_{\Gamma_i}} \nn
  \\
  &\hspace{-6em} \times \bm{u}^{h}_{i\bar{i}}(M,m_t,\cos\theta,\mu_h,\mu) \, \bm{H}_{i\bar{i}}(M,m_t,\cos\theta,\mu_h) \, \bm{u}^{h\dagger}_{i\bar{i}}(M,m_t,\cos\theta,\mu_h,\mu) \, .
\end{align}

From the evolution equations of the hard functions and the TMD PDFs, we can derive the RGEs of the soft functions as follows
\begin{align}
  \frac{d}{d\ln\mu} \bm{S}_{i\bar{i}}(\lp,M,m_t,\cos\theta,\mu) = &- \bm{\gamma}^{s\dagger}_{i\bar{i}}(M,m_t,\cos\theta,\mu) \, \bm{S}_{i\bar{i}}(\lp,M,m_t,\cos\theta,\mu) \nn
  \\
  &- \bm{S}_{i\bar{i}}(\lp,M,m_t,\cos\theta,\mu) \, \bm{\gamma}^{s}_{i\bar{i}}(M,m_t,\cos\theta,\mu) \, ,
\end{align}
with $\bm{\gamma}_{i\bar{i}}^s = \bm{\gamma}_{i\bar{i}}^{h}$. Similar to the treatment of the $I$ functions, we can factor out certain logarithmic terms from the soft functions to the exponent by
\begin{align}
  \bm{S}_{i\bar{i}}(\lp,M,m_t,\cos\theta,\mu) = \bm{u}^{s\dagger}_{i\bar{i}}(M,m_t,\cos\theta,\mu) \, \bar{\bm{S}}_{i\bar{i}}(\lp,M,m_t,\cos\theta,\alpha_s(\mu)) \, \bm{u}^{s}_{i\bar{i}}(M,m_t,\cos\theta,\mu) \, ,
\end{align}
where
\begin{align}
  &\frac{d}{d\ln\mu} \bm{u}^{s}_{i\bar{i}}(M,m_t,\cos\theta,\mu) = - \bm{u}^{s}_{i\bar{i}}(M,m_t,\cos\theta,\mu) \, \bm{\gamma}^{s}_{i\bar{i}}(M,m_t,\cos\theta,\mu) \, , \nn
  \\
  &\frac{d}{d\ln\mu} \bar{\bm{S}}_{i\bar{i}}(\lp,M,m_t,\cos\theta,\alpha_s(\mu)) = 0 \, .
\end{align}
Following the same procedure as for $\bm{u}_{h}^{i\bar{i}}$, we can obtain $\bm{u}_{s}^{i\bar{i}}$. And similar to the case of the $h_i$ functions, we choose the boundary conditions as $\bm{u}^{s}_{i\bar{i}}(M,m_t,\cos\theta,b_0/x_T)=1$, where $b_0=2e^{-2\gamma_E}$. In the $\bar{\bm{S}}$ functions, the $\mu$-dependence through $\alpha_s(\mu)$ and $\lp$ cancels each other, and up to the NLO, they are given by
\begin{align}
  \bar{\bm{S}}_{i\bar{i}}(\lp,M,m_t,\cos\theta,\alpha_s(\mu)) = \bm{S}_{i\bar{i}}(0,M,m_t,\cos\theta,\mu) \, .
\end{align}

With these RG improved hard and soft functions, we can now perform the $q_T$ resummation according to Eq.~(\ref{eq:master}) and (\ref{eq:cii}).
 To give precise prediction, we resum the singular terms to all orders and include the non-singular terms up to NLO, which can be written as
\begin{align}
  \label{eq:match}
  \frac{d\sigma^{\text{NNLL+NLO}}}{dq_T} = \frac{d\sigma^{\text{NNLL}}}{dq_T} + \left( \frac{d\sigma^{\text{NLO}}}{dq_T} - \frac{d\sigma^{\text{NNLL}}}{dq_T}\bigg|_{\text{expanded to NLO}} \right) .
\end{align}
In Eq.~(\ref{eq:match}), the exact NLO QCD corrections can be calculated by public codes such as MCFM \cite{Campbell:2000bg} and the expansion of the resummed formula will be shown in the next section.

\section{The $q_T$ spectrum of $t\bar{t}$ at fixed order}
\label{sec:fixed}

To verify the correctness of our factorization formula and our soft functions, we can expand our $q_T$ spectrum to the NLO and the NNLO, and compare with existing results in the literature. We can also reproduce the NLO total cross sections using a variation of the $q_T$ subtraction method of \cite{Catani:2007vq}, which can be compared to the known results.

We start from Eq.~(\ref{eq:master}) and Eq.~(\ref{eq:cii}).  We will expand the $C_{i\bar{i} \leftarrow ab}$ functions to the NNLO, which will contain $L_\perp$ up to power 4. We will then need to carry out the Fourier transform to the $q_T$ space.
 The relevant transformation can be performed as follows \cite{Becher:2010tm}
\begin{align}
  \frac{1}{2} \int x_Tdx_T \, J_0(x_Tq_T) \, L_\perp^n \left( \frac{x_T^2\mu^2}{4e^{-2\gamma_E}} \right)^{-\eta} = \left( -\partial_\eta \right)^n \frac{1}{q_T^2} \left( \frac{\mu^2}{q_T^2} \right)^{-\eta} \frac{\Gamma(1-\eta)}{e^{2\eta\gamma_E}\Gamma(\eta)} \, .
\end{align}
This relation works only when $1/4<\eta<1$ and the range can be  analytically continued to $0<\eta<1$.  We can then take the limit $\eta \to 0$ after carry out the derivatives. The Fourier transformation of $L_\perp^n$ then corresponds to the following replacements
\begin{align}
  1  &\to \delta(q_T^2), \quad L_\perp \to - \left[ \frac{1}{q_T^2} \right]^{[\mu^2]}_* \, , \quad L_\perp^2 \to - \left[ \frac{2}{q_T^2} \ln\frac{\mu^2}{q_T^2} \right]^{[\mu^2]}_* \, , \nn
  \\
  L_\perp^3 &\to - \left[ \frac{3}{q_T^2} \ln^2\frac{\mu^2}{q_T^2} \right]^{[\mu^2]}_* - 4 \zeta_3 \, \delta(q_T^2) \, , \quad L_\perp^4 \to - \left[ \frac{4}{q_T^2} \ln^3\frac{\mu^2}{q_T^2} \right]^{[\mu^2]}_* + 16 \zeta_3 \left[ \frac{1}{q_T^2} \right]^{[\mu^2]}_*  \, ,
\end{align}
where the definition for the star-distribution can be found in, e.g., Ref.~\cite{Schwartz:2007ib}. We briefly show the properties of the star-distribution as follows
\begin{align}
  \left[ f(q_T^2) \right]^{[m^2]}_* &= f(q_T^2) \quad \text{for } q_T>0 \, , \nn
  \\
  \int_0^{m^2} dq_T^2 \left[ f(q_T^2) \right]^{[m^2]}_*  \, g(q_T^2) &= \int_0^{m^2} dq_T^2 \, f(q_T^2) \, \left[ g(q_T^2) - g(0) \right] , \nn
  \\
  \left[ f(q_T^2) \right]^{[m^2]}_* &= \left[ f(q_T)^2 \right]^{[m'^2]} + \delta(q_T^2) \int_{m^2}^{m'^2} dp_T^2 \, f(p_T^2) \, .
\end{align}
With the one-loop $I_{i \leftarrow j}$ and soft functions, we can get the $q_T$ distribution for top quark pair production at NLO in the small $q_T$ region. The results  are given by
\begin{align}
  \label{eq:nlo}
  \frac{d^4\sigma}{dq_T^2 \, dy \, dM \, d\cos\theta} &= \frac{\beta_t\alpha_s^3}{4sM} \sum_{i=q,\bar{q},g} \frac{1}{d_i} \nn
  \\
  &\hspace{-6em} \times
  \Bigg\{ f_{i/N_1}(\xi_1) \, f_{\bar{i}/N_2}(\xi_2) \, \mathrm{Tr} \left[ \bm{H}^{(0)}_{i\bar{i}} \left( \bm{A}_{i\bar{i}} \left[ \frac{1}{q_T^2} \ln\frac{M^2}{q_T^2} \right]^{[\mu^2]}_* + \bm{B}_{i\bar{i}} \left[ \frac{1}{q_T^2} \right]^{[\mu^2]}_* \right) + \bm{C}_{i\bar{i}} \, \delta(q_T^2) \right] \nn
  \\
  &\hspace{-5em} + \mathrm{Tr} \left[ \bm{H}^{(0)}_{i\bar{i}} \bm{S}^{(0)}_{i\bar{i}} \right] \bigg[ \sum_{a} \left[ \left( \frac{\mathcal{P}_{ia}^{(1)}}{2} \left[ \frac{1}{q_T^2} \right]^{[\mu^2]}_* + \mathcal{R}_{ia}^{(1)} \, \delta(q_T^2) \right) \otimes f_{a/N_1} \right](\xi_1) \, f_{\bar{i}/N_2}(\xi_2) \nn
  \\
  &\hspace{1em} + \sum_{b} f_{i/N_1}(\xi_1) \, \left[ \left( \frac{\mathcal{P}_{\bar{i}b}^{(1)}}{2} \left[ \frac{1}{q_T^2} \right]^{[\mu^2]}_* + \mathcal{R}_{\bar{i}b}^{(1)} \, \delta(q_T^2) \right) \otimes f_{b/N_2} \right](\xi_2) \bigg] \Bigg\}+\mathcal{O}(q_T^2) \, ,
\end{align}
where
\begin{align}
  \label{eq:nloab}
  \bm{A}_{i\bar{i}} &= \Gamma^i_0 \, \bm{S}^{(0)}_{i\bar{i}} \, , \nn
  \\
  \bm{B}_{i\bar{i}} &= 2\gamma^{i}_0 \, \bm{S}^{(0)}_{i\bar{i}} - 4 \bm{w}^{33}_{i\bar{i}} + \frac{2(1+\beta_t^2)\ln x_s}{\beta_t} \bm{w}^{34}_{i\bar{i}}- 8 \ln\frac{-t_1}{m_t M} \, \bm{w}^{13}_{i\bar{i}} - 8 \ln\frac{-u_1}{m_t M} \, \bm{w}^{23}_{i\bar{i}} \, , \nn
  \\
  \bm{C}_{i\bar{i}} &= \bm{H}^{(1)}_{i\bar{i}} \bm{S}^{(0)}_{i\bar{i}} + \left(\bm{H}^{(0)}_{i\bar{i}} \bm{S}^{(1)}_{i\bar{i}}\big|_{L_\perp \rightarrow 0} \right) .
\end{align}

\begin{figure}[t!]
  \centering
  \includegraphics[scale=0.6]{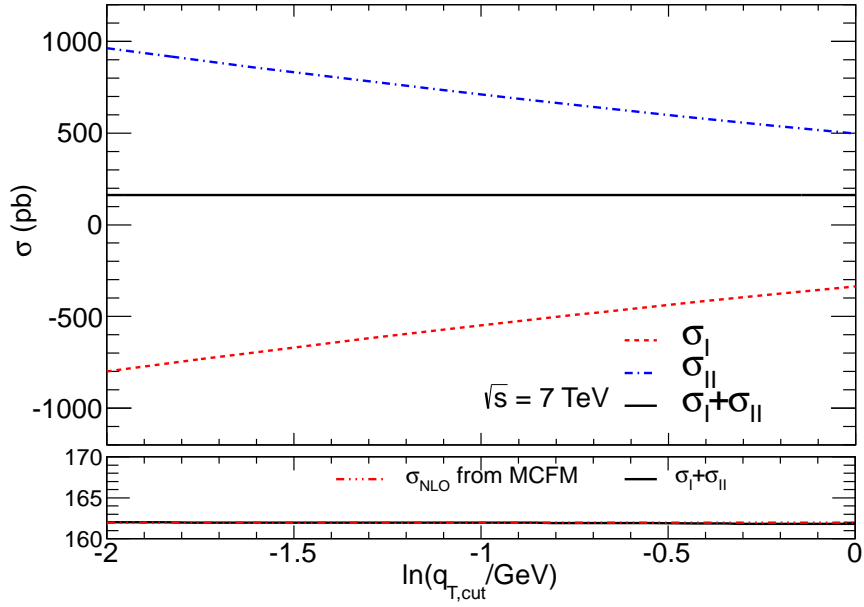}
  \vspace{-2ex}
  \caption{\label{fig:nlo_tot} The NLO total cross section for $t\bar{t}$ production at the LHC with $\sqrt{s}=7$ TeV. The red dotted line represents $\sigma_{\text{I}}$ and the blue dot-dashed line stands for $\sigma_{\text{II}}$. The total cross section is shown as the black solid line. In the lower plot, the red dash-dotted line represents the result calculated by MCFM, to which we found perfect agreement.}
\end{figure}

In the previous work \cite{Zhu:2012ts}, we have shown that the leading singular terms in Eq.~(\ref{eq:nlo}) agree well with the exact results  at the QCD NLO level in the small transverse momentum region.  However, the NLO $q_T$ spectrum receives no contribution from the $\delta(q_T^2)$ terms in  Eq.~(\ref{eq:nlo}).  To check the soft functions further, we now reproduce the total cross section at NLO for top quark pair production. Using the phase space slicing  method, the NLO total cross section can be divided into two parts: small $q_T$ region denoted by $\sigma_{\text{I}}$,
 which can be obtained by integrating the distribution in Eq.\ (\ref{eq:nlo}) in the approximation of neglecting $\mathcal{O}(q^2_T/M^2)$ terms, and the large $q_T$ part labeled by $\sigma_{\text{II}}$, which is infrared safe and can be numerically computed directly. Thus the total cross section  is given by
\begin{align}
  \label{eq:nlo_tot}
    \sigma_{\text{NLO}} = \int_0^{q^2_{T,\text{cut}}} dq^2_T \, \frac{d\sigma_{\text{NLO}}}{dq^2_{T}} + \int_{q^2_{T,\text{cut}}}^{\infty} dq^2_T \, \frac{d\sigma_{\text{NLO}}}{dq^2_{T}} = \sigma_{\text{I}}+\sigma_{\text{II}} \  .
\end{align}
We can make numerical calculation to clarify the correctness of  Eq.~(\ref{eq:nlo_tot}). In our numerical calculation, we use the MSTW2008NLO PDFs \cite{Martin:2009iq} and set $m_t=172.5$~GeV. We show the numerical results at the 7 TeV LHC in Fig.~\ref{fig:nlo_tot}. It can be seen that the dependence on $q_{T,{\rm{cut}}}$  is canceled when we sum $\sigma_{\text{I}}$ and $\sigma_{\text{II}}$.
 Fig.~\ref{fig:nlo_tot} shows that the total cross section  computed from our formula is in  perfect agreement with the one calculated by MCFM, and is independent of $q_{T,{\text{cut}}}$ when $q_{T,{\text{cut}}}$ is sufficiently small.
\begin{figure}[t!]
  \centering
  \includegraphics[scale=0.4]{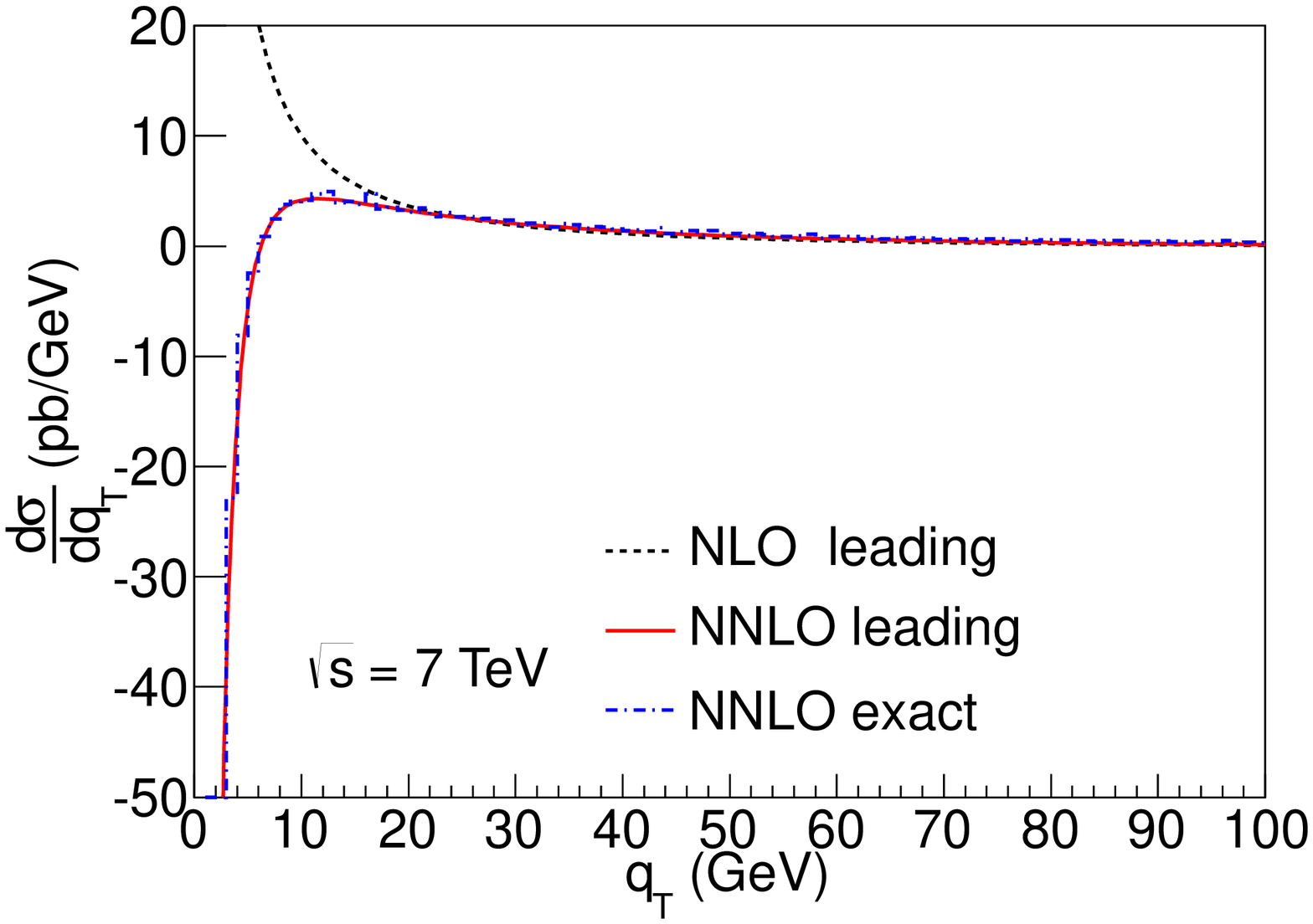} \hspace{0.5em}
  \includegraphics[scale=0.4]{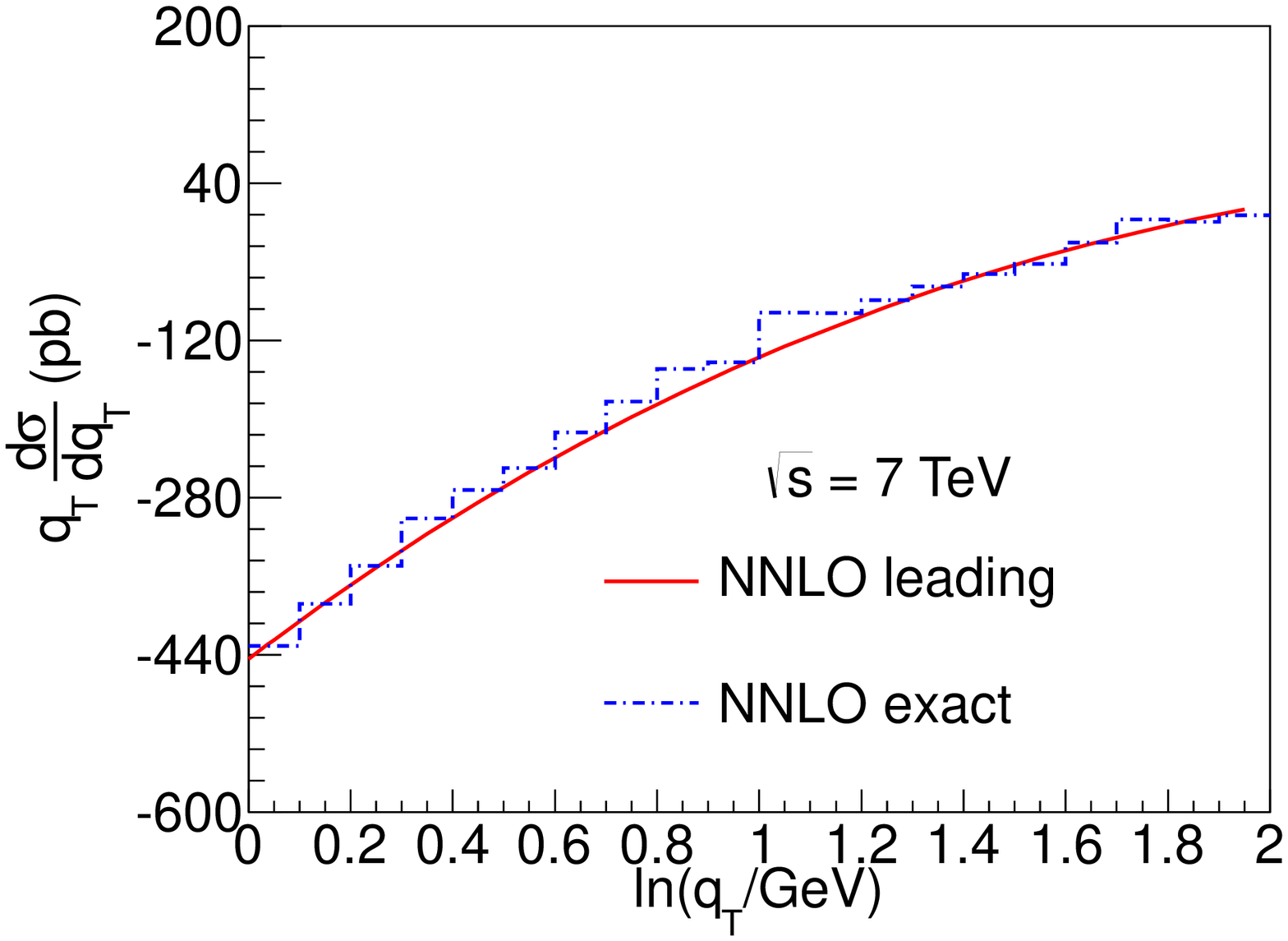}
  \vspace{-2ex}
  \caption{\label{fig:nnlo}  The hadronic $q_T$ distribution for top quark pair production.
  The left plot shows the leading singular $q_T$-distribution (red solid line) at NNLO compared with the exact result (blue dash-dotted line) computed using POWHEG BOX. The right plot shows the small $q_T$ region.}
\end{figure}

We can now proceed to calculate the leading singular $q_T$ distribution at the NNLO. For that we will need all the $L_\perp$-dependent terms in the NNLO functions $F^{(2)}$, $I^{(2)}_{i \leftarrow j}$ and $\bm{S}^{(2)}_{i\bar{i}}$, which can be obtained from the RG equations and the known results for the two-loop splitting functions and anomalous dimensions.
 We write the cross section as
\begin{multline}
     \label{eq:FOsigma}
     \frac{d^3 \sigma} {dq_T^2 dM d \cos\theta} =
         \sum_{i=q,\bar{q},g} \sum_{a,b} \frac{8 \pi \beta_t} {3 s M} \int_{z_1 z_2 > \tau} \frac{dz_1dz_2}{z_1 z_2}
         C_{i\bar{i}\leftarrow ab}(z_1,z_2,q_T,M,\cos\theta, m_t,\mu) \\ \times
         \int \frac{dx}{x} f_{a/N_1}(x,\mu) f_{b/N_2}(\frac{\tau}{xz_1z_2},\mu)\ ,
\end{multline}
where the perturbative expansion of the partonic cross section is
\begin{equation}
    C_{i\bar{i} \leftarrow ab}(z_1,z_2,q_T,M,m_t,\cos\theta,\mu) = \sum_{n}\left(\frac{\alpha_s}{4\pi}\right)^{n}
      C^{(n)}_{i\bar{i} \leftarrow ab}(z_1,z_2,q_T,M,m_t,\cos\theta,\mu)\ .
\end{equation}
 At the NNLO level, the partonic cross section can be  expanded as
\begin{multline}
  \label{eq:nnlo_qt}
  \int \frac{dz_1}{z_1} \frac{dz_2}{z_2} \, \delta(z-z_1 z_2) C^{(2)}_{i\bar{i} \leftarrow ab}(z_1,z_2,q_T,M,m_t,\cos\theta,\mu) = \left( \Sigma^{(2,0)}_{i\bar{i} \leftarrow ab} +4/3 \zeta_3 \Sigma^{(2,3)}_{i\bar{i}\leftarrow ab} \right) \delta(q_T^2)
  \\
  + \left( \Sigma^{(2,1)}_{i\bar{i} \leftarrow ab} + 4 \zeta_3 \Sigma^{(2,4)}_{i\bar{i} \leftarrow ab} \right) \frac{1}{q_T^2} + \Sigma^{(2,2)}_{i\bar{i}\leftarrow ab} \frac{1}{q_T^2} \ln\frac{q_T^2}{\mu^2} + \Sigma^{(2,3)}_{i\bar{i} \leftarrow ab} \frac{1}{q_T^2} \ln^2\frac{q_T^2}{\mu^2} + \Sigma^{(2,4)}_{i\bar{i} \leftarrow ab} \frac{1}{q_T^2} \ln^3\frac{q_T^2}{\mu^2}+\mathcal{O}(q_T^2) \, ,
\end{multline}
where the functions $C^{(2)}_{i\bar{i}\leftarrow ab}$ are the NNLO coefficients of the functions $C_{i\bar{i}\leftarrow ab}$ and we have suppressed the arguments of the functions $\Sigma^{(2,m)}_{i\bar{i} \leftarrow ab}(z,M,m_t,\cos\theta,\mu)$. From our current knowledge, we can calculate the coefficients $\Sigma^{(2,m)}_{i\bar{i} \leftarrow ab}$ for $m=1,2,3,4$, which are sufficient for the $q_T$ spectrum. The explicit expressions for them are too lengthy to be presented in the text, and we give them in a Mathematica notebook file associated with the arXiv submission of this paper.  The coefficient $\Sigma^{(2,0)}_{i\bar{i} \leftarrow ab}$ is crucial for computing the NNLO total cross section using the $q_T$ subtraction method. It receives contributions from the $L_\perp$-independent terms in $I^{(2)}_{i \leftarrow j}$ and $S^{(2)}_{i\bar{i}}$, as well as the NNLO hard functions $H_{i\bar{i}}$. These are still not available except the matching coefficient $I_{q \leftarrow q}$, whose NNLO result has been
calculated in Ref.~\cite{Gehrmann:2012ze}.  The left plot in Fig.~\ref{fig:nnlo} shows the numerical results of the NLO and NNLO $q_T$ distributions, calculated using Eq.~(\ref{eq:nlo}) and (\ref{eq:nnlo_qt}).  Shown together is the exact NNLO distribution calculated using POWHEG BOX \cite{Nason:2004rx, Frixione:2007vw, Alioli:2010xd, Alioli:2011as}, which implements the results of \cite{Dittmaier:2007wz}. It can be seen that the leading singular distribution at the NNLO agrees well with the exact result, which is a very strong support of our framework. The right plot shows the small $q_T$ region with a logarithmic scale for the $q_T$ axis. As expected from Eq.~({\ref{eq:nnlo_qt}}), the distribution is a cubic function of $\ln(q_T)$.

\section{Numerical results}
\label{sec:numerics}

\begin{figure}[t!]
  \centering
  \includegraphics[scale=0.4]{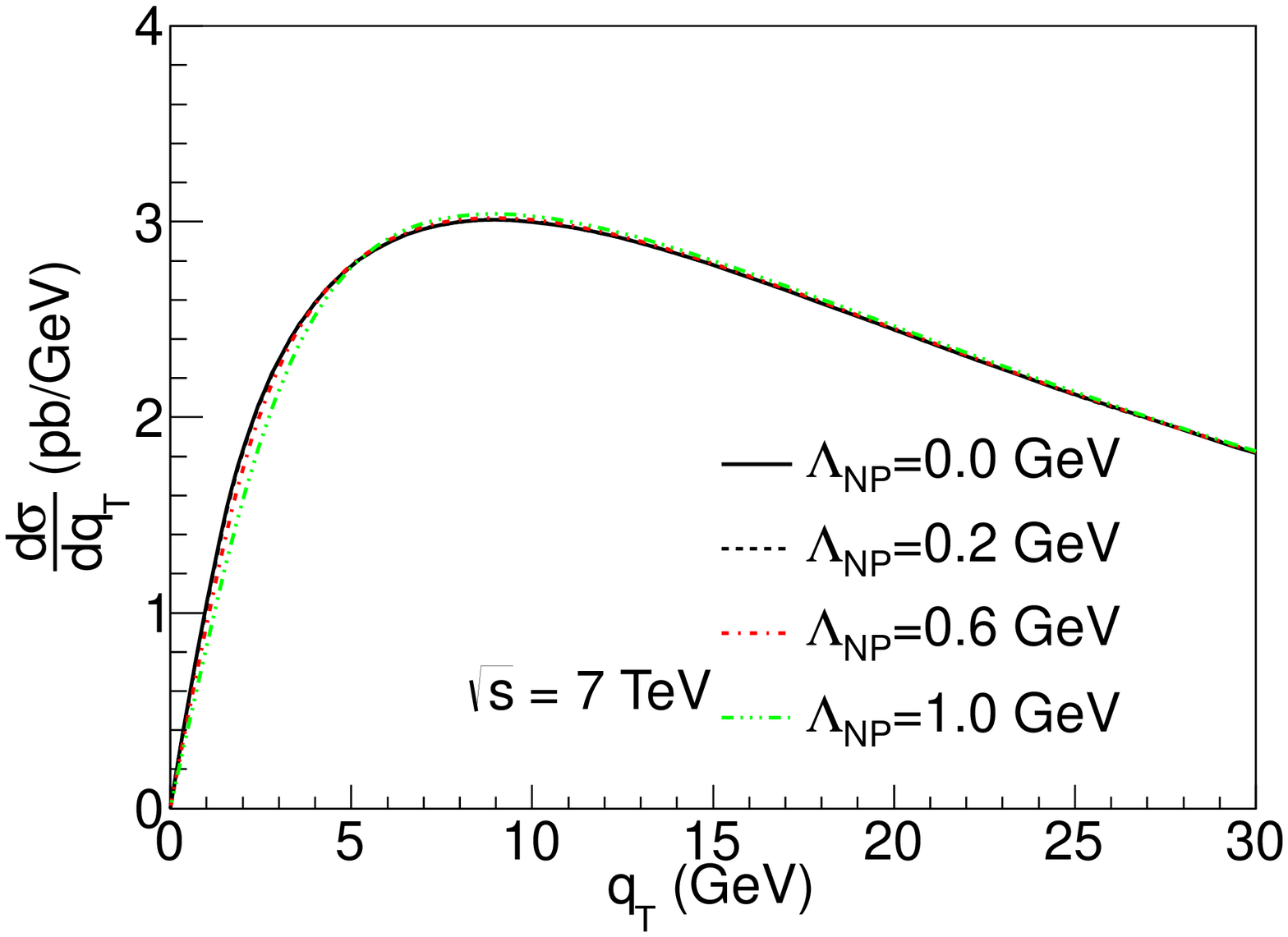} \hspace{0.5em}
  \includegraphics[scale=0.4]{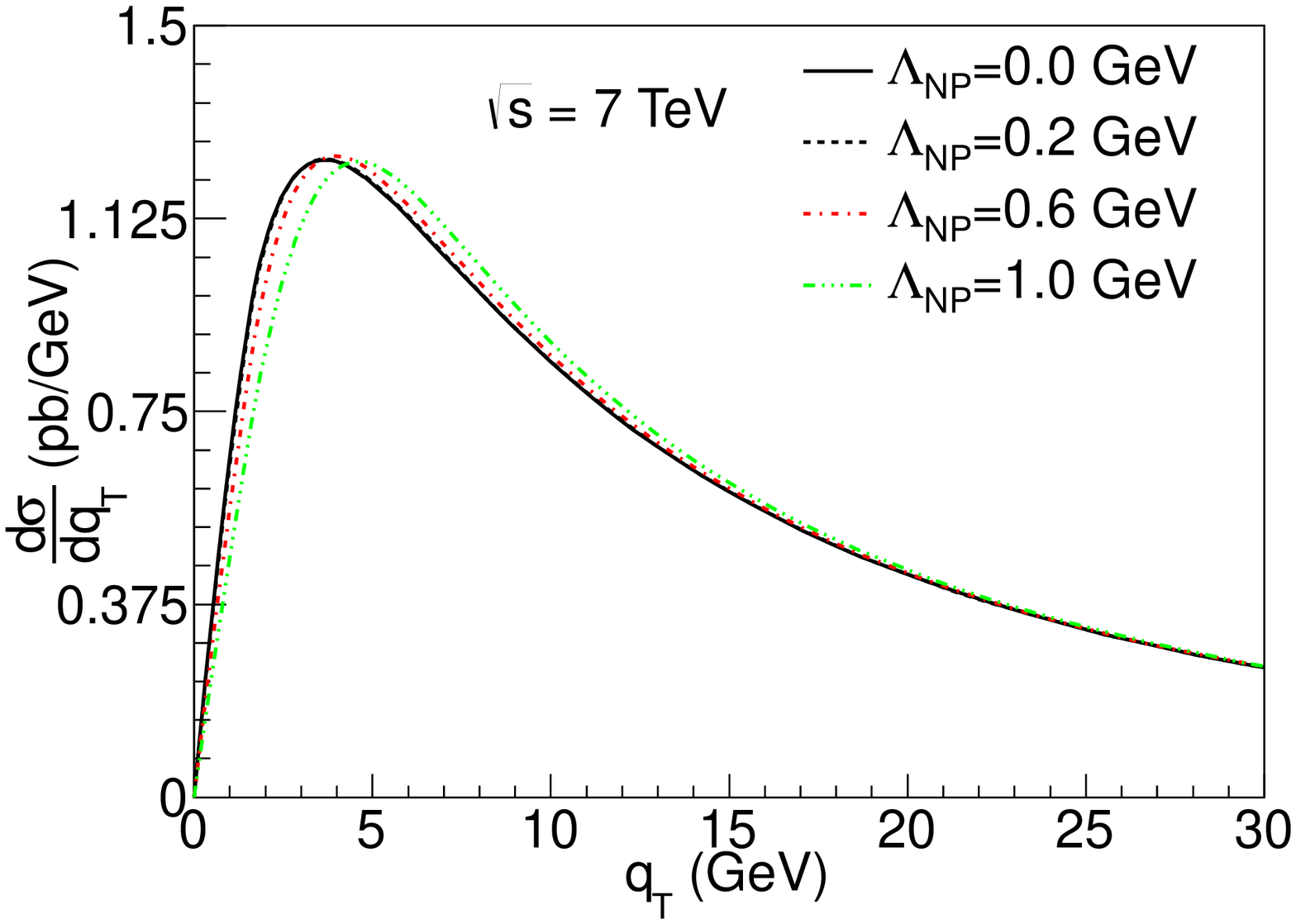}
  \vspace{-2ex}
  \caption{\label{fig:lam_NP} The transverse momentum distribution of top quark pair at the LHC with $\sqrt{s}=7$~TeV, using different values for $\Lambda_{\text{NP}}$. The left plot shows the total differential cross section at the NNLL accuracy. The right plot shows the contributions from $q\bar{q}$ initial states.}
\end{figure}

In this section, we present the numerical results of the NNLL transverse momentum resummation.  We set the top quark mass to be 172.5~GeV, and use MSTW2008NNLO PDFs.  Following Ref.~\cite{Becher:2011xn} for the Drell-Yan process and Ref.~\cite{Becher:2012yn} for the Higgs production, we choose the factorization scale to be $\mu^i=q^*_i + q_T$, where $q^*_i$ is determined from
\begin{align}
  q^*_i = M \exp \left( - \frac{2\pi}{\Gamma_0^i \, \als(q^*_i)} \right) .
\end{align}

Our factorization formula is formally valid in the region $\Lambda_{\text{QCD}} \ll q_T \ll M$. When $q_T \sim \Lambda_{\text{QCD}}$,  there are corrections in powers of $x_T\Lambda_{\text{QCD}}$, which  comes form the operator-product expansion of the transverse PDFs \cite{Becher:2011xn}. These power corrections are of non-perturbative origin and one must model them using some ansatz.  Following Ref.~\cite{Becher:2011xn},  we choose a simple model such that the TMD PDFs are replaced by
\begin{align}
  B_{i/N}(z,x_T^2,\mu) = B^{\text{pert}}_{i/N}(z,x_T^2,\mu) \, f_{\text{hadr}}(x_T \Lambda^i_{\text{NP}}) \, ,
\end{align}
where $\Lambda^i_{\text{NP}}$ is a hadronic scale with $i=q,g$, and $f_{\text{hadr}}(x_T \Lambda_{\text{NP}})$ is
\begin{align}
    f_{\text{hadr}}(x_T \Lambda_{\text{NP}}) = \exp(-\Lambda_{\text{NP}}^2 x_T^2) \, .
\end{align}
Fig.~\ref{fig:lam_NP} shows the $\Lambda_{\text{NP}}$ dependence of the NNLL resummed transverse momentum distribution of the top quark pair at the LHC with $\sqrt{s}=7$~TeV.  The left plot shows the total differential cross section while the right plot shows the contributions from the $q\bar{q}$ channel. Obviously, the non-perturbative form factor have very tiny effect on the total distribution. This can be understood since $q^*_g \gtrsim 14.0$~GeV is far away form the non-perturbative region, and the dominant contribution to $t\bar{t}$ production at the LHC comes from the $gg$ channel. For the $q\bar{q}$ channel, $q^*_q \gtrsim 3.0$~GeV, and the long-distance effects are expected to have a stronger influence, which can be seen from the right plot in Fig.~\ref{fig:lam_NP}. In our numerics in the following, we set $\Lambda^i_{\text{NP}}=0.6$~GeV to  simulate the nonperturbative effects for $t\bar{t}$ production.

\begin{figure}[t!]
  \centering
  \includegraphics[scale=0.6]{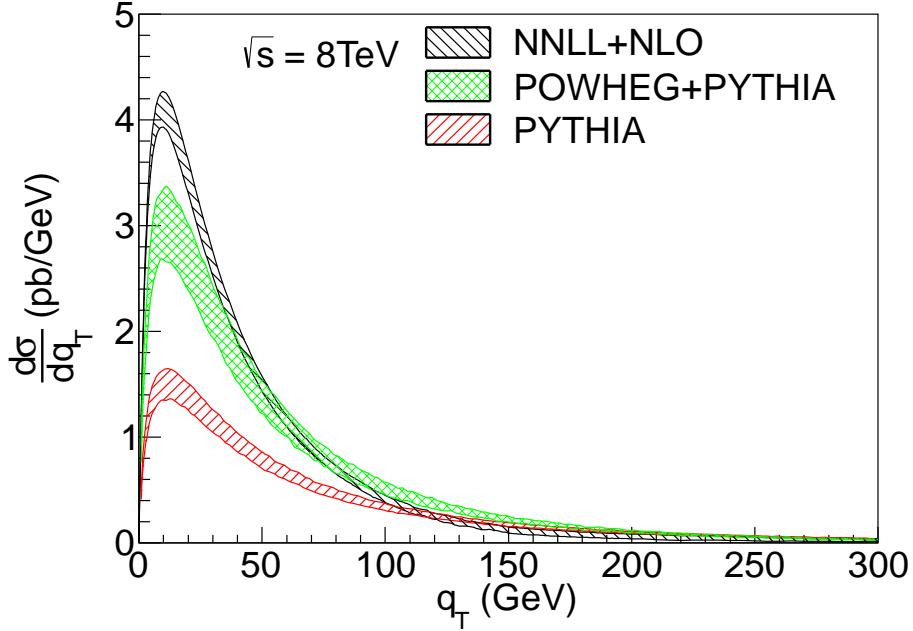} \hspace{0.5em}
  \vspace{-2ex}
  \caption{\label{fig:res_PS} The $q_T$ distributions with scale uncertainties. The black, green and red bands represent the predictions from our NNLL+NLO resummed formula, POWHEG+PYTHIA and PYTHIA, respectively.}
\end{figure}

We now make predictions for the transverse momentum distribution of the top quark pair, and compare them with results from parton shower (PS) methods and with experimental data.  According to Eq.~(\ref{eq:match}), we obtain the theoretical predictions at the NNLL+NLO accuracy.
In Fig.~\ref{fig:res_PS}, we compare the NNLL+NLO resummed distribution with the results from the LO+PS program PYTHIA 8.1 \cite{Sjostrand:2007gs} and from NLO+PS program POWHEG+PYTHIA \cite{Frixione:2007nw},  where the top quark pair production process is used. Here and in the following, we choose the central value for $\mu_h$ to be $M$, and the perturbative uncertainties are estimated by varying $\mu_h$, $\mu^q$ and $\mu^g$ by a factor of 2 around their central values. When  using PYTHIA, we turn off the hadronization and the decay of top quark. It can be observed that when going from LO+PS to NLO+PS, the differential cross sections are significantly increased, especially in the peak region. Our NNLL+NLO resummed result add another $~$30\% for the peak, while at large $q_T$ it's slightly smaller than the NLO+PS result. We also note that the uncertainties of the resummed predictions are much smaller than those obtained by POWHEG+PYTHIA and PYTHIA.

\begin{figure}[t!]
  \centering
  \includegraphics[scale=0.6]{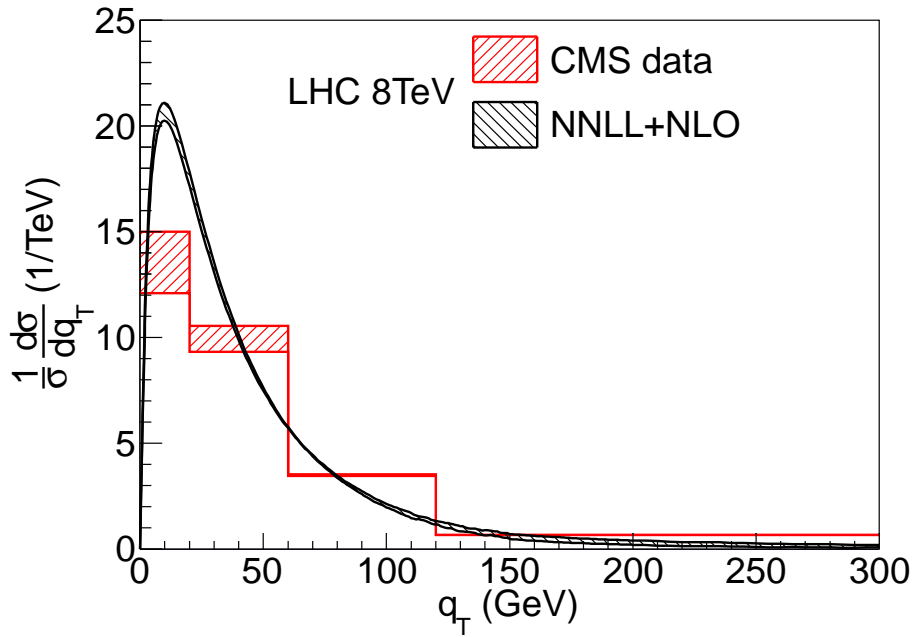}
  \vspace{-2ex}
  \caption{\label{fig:exp} Comparison of normalized $q_T$ distribution between the theoretical prediction and the experimental data from the CMS Collaboration at the 8~TeV LHC. The red band is the data measured by CMS. The black band is the resummed prediction at NNLL+NLO accuracy.}
\end{figure}

In Table~\ref{tab:resum_exp} and Fig.~\ref{fig:exp}, we compare the resummed prediction for the normalized differential cross section $1/\sigma d\sigma/dq_T$ with experimental data from the ATLAS and CMS Collaborations. In Table~\ref{tab:resum_exp}, the data was measured by the ATLAS Collaboration in the lepton+jets channel \cite{Aad:2012hg} with $\sqrt{s}=7$~TeV and is grouped in 3 bins up to 1.1~TeV. Our NNLL+NLO resummed predictions are consistent with the data in all 3 bins within theoretical and experimental uncertainties. We note that in the small $q_T$ region the theoretical uncertainties of our predictions are well under-control. At large transverse momentum, however, the scale dependence is rather large. This can be understood since for large $q_T$ the differential cross sections are dominated by hard gluon emissions, which are not captured by the resummation of soft and collinear emissions. The behavior at large $q_T$ can be improved by matching our resummed results to the NLO results for $t\bar{t}
$+jet production calculated in \cite{Dittmaier:2007wz,Dittmaier:2008uj,Melnikov:2010iu,Melnikov:2011qx}, which we leave for future update of our work. In Fig.~\ref{fig:exp}, we present the normalized $q_T$ distribution at the 8~TeV LHC and compare the theoretical predictions with the data measured by the CMS Collaboration \cite{CMS:pas-top-12-027}.
It can be seen that the resummed results are consistent with the experimental data.
 The only small difference is in the first bin, where our prediction is higher than the measured result. We note that in the first bin in Table~\ref{tab:resum_exp}, our prediction is also slightly higher than the ATLAS data, albeit with a better agreement. A possible effect which will decrease the differential cross section at small $q_T$ is again coming from matching to the NLO results for $t\bar{t}$+jet. This will increase the differential cross section at large $q_T$, and hence will affect the shape globally after normalizing.

\begin{table}[t!]
  \centering
  \begin{tabular}{c|c|c}
    \hline \hline
    \multirow{2}{*}{$q_T$ [GeV]} & \multicolumn{2}{c}{$1/\sigma d\sigma/dq_T$ [1/TeV]}
    \\
    \cline{2-3} & NNLL+NLO  & ATLAS \cite{Aad:2012hg}
    \\
    \hline \hline
    $0 \sim 40$ &  $15.5^{+0.7(+5\%)}_{-0.6(-4\%)}$ & $14 \pm 2 (\pm 14\%)$
    \\
    \hline
    $40 \sim 170$ & $2.4^{+0.3(+13\%)}_{-0.2(-8\%)}$ & $3.0 \pm 0.3 (\pm 10\%)$
    \\
    \hline
    $170 \sim 1100$ & $0.078^{+0.031(+40\%)}_{-0.026(-33\%)}$ & $0.051 \pm 0.008 (\pm 16\%)$
    \\
    \hline \hline
  \end{tabular}
  \caption{\label{tab:resum_exp} Normalized differential cross section $1/\sigma d\sigma/dq_T$ at the 7~TeV LHC. The experiment results are measured by the ATLAS Collaboration in the lepton+jets channel.}
\end{table}

\begin{figure}[t!]
  \centering
  \includegraphics[scale=0.6]{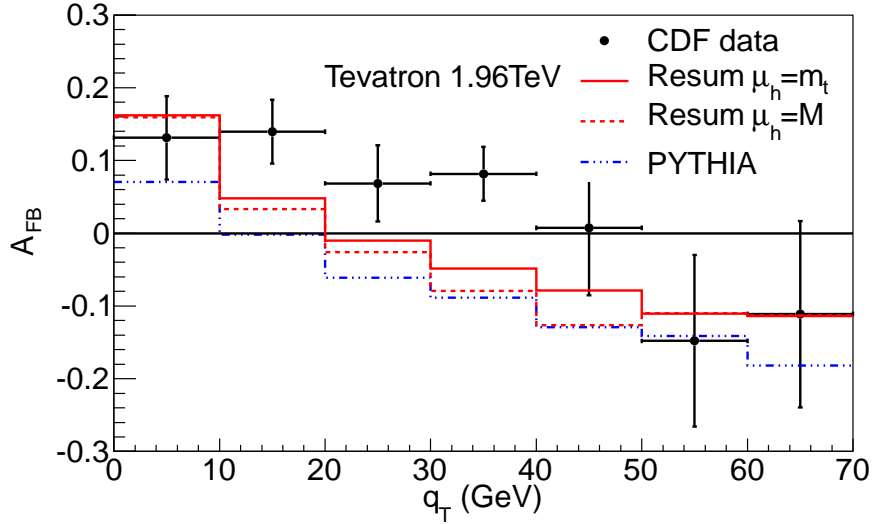} \hspace{0.5em}
  \vspace{-2ex}
  \caption{\label{fig:afb} The $q_T$-dependent forward-backward asymmetry. The black points represent the experimental data. The red solid line and the red dashed line are the resummed results at NNLL+NLO with different hard scales. The blue dotted line shows the prediction from PYTHIA.}
\end{figure}

Now we turn to the forward-backward asymmetry in top quark pair production at the Tevatron. The $q_T$-dependent asymmetry is defined as
\begin{align}
  A_{FB}(q_T) = \frac{\sigma_{F}(q_T)-\sigma_B(q_T)}{\sigma_{F}(q_T)+\sigma_B(q_T)} \, ,
\end{align}
where
\begin{align}
  \sigma_{F}(q_T) = \int^1_0 d\cos\theta \, \frac{d^2\sigma}{d\cos\theta dq_T} \, ,
  \quad
  \sigma_{B}(q_T) = \int^0_{-1} d\cos\theta \, \frac{d^2\sigma}{d\cos\theta dq_T} \, .
\end{align}
The forward-backward asymmetry has been measured by the CDF \cite{Aaltonen:2011kc, Aaltonen:2012it} and D0 \cite{Abazov:2011rq} Collaborations and was found to be larger than the SM prediction. To find out the possible origin of the asymmetry, it is instructive to study its kinematic dependence. For example, the invariant-mass-dependent asymmetry has been measured by the CDF Collaboration \cite{Aaltonen:2011kc} and it was found that the discrepancy is mainly in the high invariant mass region. In \cite{Kuhn:2011ri, Alvarez:2012vq}, it has been shown that the asymmetry also has intriguing dependence on the pair transverse momentum. In particular, by concentrating in the small $q_T$ region, one can increase the asymmetry which leads to a better signal. Apparently, in the small $q_T$ region, it is essential to use our framework to resum the soft and collinear gluon effects. Fig.~\ref{fig:afb} shows the theoretical predictions and the experimental results at the Tevatron.  Here the CDF data is extracted from Ref.
~\cite{Aaltonen:2012it}. We observe that our resummed results are consistent with the experimental data, and have better agreements than the results of PYTHIA.  We, however, note that the central values of our predictions in the intermediate $q_T$ region is slightly lower than the central values of the data. It would be interesting to see if matching to the NLO results for $t\bar{t}$+jet can reduce the difference.

\section{Summary and outlook}
\label{sec:conclusion}

Based on SCET and HQET, we have developed a framework for transverse momentum resummation  for top quark  pair production at hadron colliders.  We have shown the details of the derivation of the factorization formula in the limit of small pair transverse momentum. In the procedure of factorization, we first of all integrate out the hard fluctuations and obtain the hard functions. We then describe the collinear and soft emissions in terms of the TMD PDFs and the transverse soft functions, respectively. The resummation of large logarithms is achieved by renormalization group evolution.

In order to validate our resummation formula, we expand it to the NLO and the NNLO to obtain the leading singular terms at these orders. We then compare the transverse momentum distributions at the NNLO with the exact results and find perfect agreement in the small $q_T$ region. We also reproduced the NLO total cross section using a variation of the $q_T$ subtraction method. These consistency checks provide a strong support for our framework.

We then perform  the calculation of the transverse momentum  distributions at NNLL accuracy, and compare  them with experimental data and predictions from parton shower Monte Carlo programs.   Our results show that our resummed predictions agree well with measurements within theoretical and experimental uncertainties. Furthermore, we discuss the $q_T$-dependent forward-backward asymmetry at the Tevatron, which is sensitive to new physics effects. Our predictions are consistent with the experimental data, despite the fact that the measurements have very large error bars.

In our numerical results, we have matched our NNLL resummed formula to the NLO results for $t\bar{t}$ production, which is the leading order for $t\bar{t}$+jet production. Matching to the NLO results for $t\bar{t}$+jet production will definitely provide a big improvement in the large $q_T$ region, and hence also for the shape of the distribution. On another issue, it is interesting to study the non-perturbative contributions in the small $q_T$ region coming from the effects discussed in \cite{Mitov:2012gt}. In addition, it will be useful to calculate the $q_T$-dependent charge asymmetries at the LHC once such measurements are possible. These we leave for future updates of our work.

Our formalism can be easily generalized to other processes for massive colored particle production at hadron colliders, such as $b\bar{b}$, $c\bar{c}$ and colored supersymmetric partners. Besides,  our resummed formula can provide another approach to construct the subtraction terms for the NNLO calculations of top quark pair production, based on the $q_T$ subtraction method.   This will require the calculation of the NNLO hard and soft functions and the TMD PDFs. The NNLO hard functions may be extracted from the calculations in \cite{Baernreuther:2012ws, Czakon:2012zr, Czakon:2012pz, Czakon:2013goa}. The NNLO TMD PDFs for the quark to quark case has been calculated in \cite{Gehrmann:2012ze}, and the results for all channels are likely to be available soon. The only remaining issue is the $L_\perp$-independent terms in the NNLO soft functions.

\acknowledgments
We would like to thank Stefan Dittmaier for providing us with the NLO virtual corrections for $t\bar{t}$+jet, and Simone Alioli for help with the $t\bar{t}$+jet code in POWHEG BOX. We would also like to thank Jun Gao, Stefan Hoeche and George Sterman for useful discussions. This work is partially supported by the National Natural Science Foundation of China, under Grants No. 11021092, No. 11135003, and No. 11345001, and by the Department of Energy of the United States, under Contract No. DE-AC02-76SF00515.

\appendix

\section{Explicit expressions of the TMD PDFs and the anomalous dimensions}
\label{sec:expressions}

For the convenience of the readers, we collect in this Appendix explicit expressions of the matching coefficient functions for the TMD PDFs as well as the anomalous dimensions relevant for the NNLL resummation.

The cusp anomalous dimensions are given by
\begin{align}
  \Gamma^{i}_{\text{cusp}}(\alpha_s) = \sum_{n=0}^{\infty} C_i \, \gamma^{\text{cusp}}_{n} \left(\frac{\als}{4\pi}\right)^{n+1} ,
\end{align}
with $C_q=C_F$, $C_g=C_A$, and the first three coefficients are \cite{Korchemskaya:1992je, Moch:2004pa}
\begin{align}
  \gamma^{\text{cusp}}_{0} &= 4 \, , \nn
  \\
  \gamma^{\text{cusp}}_{1} &= C_A \left( \frac{268}{9} - \frac{4\pi^2}{3}\right) - \frac{80}{9} T_F n_f \, , \nn
  \\
  \gamma^{\text{cusp}}_{2} &= C_A^2 \left( \frac{490}{3} - \frac{536\pi^2}{27} + \frac{44\pi^4}{45} + \frac{88}{3} \zeta_3 \right) + C_A T_F n_f \left( - \frac{1672}{27} + \frac{160\pi^2}{27} - \frac{224}{3}\zeta_3 \right) \nn
  \\
  &+ C_F T_F n_f \left( - \frac{220}{3} + 64 \zeta_3 \right) - \frac{64}{27} T_F^2 n_f^2 \, .
\end{align}
The anomalous dimensions for massless partons are given by \cite{Becher:2009qa}
\begin{align}
  \gamma_0^q &= -3 C_F \, , \nn
  \\
  \gamma_1^q &= C_F^2 \left( - \frac{3}{2} + 2\pi^2 - 24 \zeta_3 \right) + C_A C_F \left( - \frac{961}{54} - \frac{11 \pi^2}{6} + 26 \zeta_3 \right) + C_F T_F n_f \left( \frac{130}{27} + \frac{2\pi^2}{3} \right) , \nn
  \\
  \gamma^g_0 &= - \frac{11}{3}C_A + \frac{4}{3}T_F n_f \, , \nn
  \\
  \gamma^g_1 &= C_A^2 \left( - \frac{692}{27} + \frac{11 \pi^2}{18} + 2\zeta_3 \right) + C_A T_F n_f \left( \frac{256}{27} - \frac{2\pi^2}{9} \right) + 4C_F T_F n_f \, .
\end{align}
The anomalous dimensions for massive quarks are given by \cite{Becher:2009kw}
\begin{align}
  \gamma_0^Q &= -2 C_F \, , \nn
  \\
  \gamma_1^Q &= C_F C_A \left( -\frac{98}{9} + \frac{2\pi^2}{3} - 4\zeta_3 \right) + \frac{40}{9}C_F T_F n_f \, .
\end{align}
The $\beta$ function is expanded perturbatively as
\begin{align}
  \beta(\als) =& -2\als \sum_{n=0}^\infty \beta_n \left(\frac{\als}{4\pi}\right)^{n+1} \, ,
\end{align}
with the expansion coefficients up to three loops being
\begin{align}
  \beta_0 &= \frac{11}{3}C_A - \frac{4}{3}T_F n_f \, , \nn
  \\
  \beta_1 &= \frac{34}{3}C_A^2 - \frac{20}{3} C_A T_F n_f - 4C_F T_F n_f \, , \nn
  \\
  \beta_2 &= \frac{2857}{54}C_A^3 + T_F n_f \left( 2C_F^2 - \frac{205}{9}C_F C_A - \frac{1415}{27}C_A^2 \right) + T_F^2 n_f^2 \left( \frac{44}{9}C_F + \frac{158}{27} C_A \right) .
\end{align}

The matching functions for the TMD PDFs up to the NLO can be
generically written as \cite{Becher:2010tm, Becher:2012yn}
\begin{align}
  \label{eq:Iij}
  I_{i \leftarrow j}(z,L_\perp,\mu) &= \delta(1-z) \, \delta_{ij} \left[1+ \frac{\alpha_s}{4\pi} \left( \Gamma^{i}_0 \frac{L_\perp^2}{4}-\gamma^i_0 L_\perp \right) \right] + \frac{\alpha_s}{4\pi}  \left( -\mathcal{P}^{(1)}_{ij}(z) \frac{L_\perp}{2}+\mathcal{R}^{(1)}_{ij}(z)\right) ,
\end{align}
with the one-loop splitting functions given by
\begin{align}
  \mathcal{P}^{(1)}_{qq}(z) &= 4C_F \left(\frac{1+z^2}{1-z}\right)_+ , \nn
  \\
  \mathcal{P}^{(1)}_{qg}(z) &= 4T_F (z^2+(1-z)^2) \, , \nn
  \\
  \mathcal{P}^{(1)}_{gg}(z) &= 8C_A \left( \frac{z}{(1-z)_+} + \frac{1-z}{z} + z(1-z) \right) + 2\beta_0 \, \delta(1-z) \, , \nn
  \\
  \mathcal{P}^{(1)}_{gq}(z) &= 4C_F \frac{1+(1-z)^2}{z} \, .
\end{align}
and the one-loop remainder functions given by
\begin{align}
  \mathcal{R}^{(1)}_{qq}(z) &= C_F \left( 2(1-z) - \frac{\pi^2}{6}\delta(1-z) \right) \, , \quad \mathcal{R}^{(1)}_{qg}(z) = 4 T_F z (1-z) \, , \nn
  \\
  \mathcal{R}^{(1)}_{gg}(z) &= -C_A \frac{\pi^2}{6} \, \delta(1-z) \, , \quad \mathcal{R}^{(1)}_{gq}(z) = 2C_F z \, .
\end{align}

\section{Calculation of soft functions}
\label{sec:softcalc}

In this appendix we show the details of calculating the soft functions. We will work in the momentum space, where the integrals in Eq.~(\ref{eq:Ijk}) are represented as
\begin{align}
  \tilde{I}_{jk} &= -\frac{\mu^{2\epsilon}e^{\epsilon\gamma_E}}{\pi^{1-\epsilon}} \int^{2\pi}_0  d\phi \int d^dk \left( \frac{\nu}{n \cdot k} \right)^\alpha \delta(k^2) \, \theta(k^0) \, \delta^{(2)}(k_\perp-q_\perp) \, \frac{v_j \cdot v_k}{v_j \cdot k \; v_k \cdot k} \, ,
\end{align}
where $\phi$ is now the azimuthal angle of $q_\perp$, and we have suppressed the $\overline{\text{MS}}$ factor $(4\pi)^\epsilon e^{-\epsilon\gamma_E}$. The $\phi$ integral can be performed with the help of the $\delta$-function
\begin{align}
  \delta^{(2)}(k_\perp-q_\perp) = 2 \delta(k_T^2-q_T^2) \, \delta(\phi) \, .
\end{align}
We then have
\begin{align}
  \tilde{I}_{jk} &= -\frac{2\mu^{2\epsilon}e^{\epsilon\gamma_E}}{\pi^{1-\epsilon}} \int d^dk \left( \frac{\nu}{n \cdot k} \right)^\alpha \delta(k^2) \, \theta(k^0) \, \delta(k_T^2-q_T^2) \, \frac{v_j \cdot v_k}{v_j \cdot k \; v_k \cdot k} \, .
\end{align}
We parametrize the vectors as
\begin{align}
  \label{eq:soft_para}
  k &= k_0 \, (1,\ldots, \sin\theta_1\sin\theta_2, \sin\theta_1\cos\theta_2, \cos\theta_1) \, , \nonumber
  \\
  v_3 &= \frac{1}{\sqrt{1-\beta_t^2}} \, (1, \ldots, 0, \beta_t \sin\theta, \beta_t \cos\theta) \, , \nonumber
  \\
  v_4 &= \frac{1}{\sqrt{1-\beta_t^2}} \, (1, \ldots, 0, -\beta_t \sin\theta, -\beta_t \cos\theta) \, .
\end{align}
The scalar products appearing in $\tilde{I}_{jk}$ are
\begin{align}
  v_3 \cdot v_4 &=  \frac{1+\beta_t^2}{1-\beta_t^2} \, , \quad n \cdot v_3 = \bar{n} \cdot v_4 = \frac{1-\beta_t\cos\theta_1}{\sqrt{1-\beta_t^2}} \, , \quad \bar{n} \cdot v_3 = n \cdot v_4 = \frac{1+\beta_t\cos\theta_1}{\sqrt{1-\beta_t^2}} \nonumber
  \\
  n \cdot k &= k_0 \, (1-\cos\theta_1) \, , \quad \bar{n} \cdot k = k_0 \, (1+\cos\theta_1) \, , \nonumber
  \\
  v_3 \cdot k &= \frac{k_0}{\sqrt{1-\beta_t^2}} \, (1 - \beta_t \sin\theta_1 \cos\theta_2 \sin\theta - \beta_t \cos\theta_1 \cos\theta) \, , \nonumber
  \\
  v_4 \cdot k &= \frac{k_0}{\sqrt{1-\beta_t^2}} \, (1 + \beta_t \sin\theta_1 \cos\theta_2 \sin\theta + \beta_t \cos\theta_1 \cos\theta) \, .
\end{align}
The integration measure can be written as
\begin{align}
  d^dk \, \delta(k^2) \, \theta(k^0) = \Omega_{d-4} \, \frac{k_0^{1-2\epsilon}}{2} dk_0 \sin^{1-2\epsilon}\theta_1 d\theta_1 \sin^{-2\epsilon}\theta_2 d\theta_2 \, ,
\end{align}
where
\begin{align}
  \Omega_{d-4} = \frac{2\pi^{\frac{1}{2}-\epsilon}}{\Gamma\left(\frac{1}{2}-\epsilon\right)} = \frac{2^{1-2\epsilon} \, \pi^{-\epsilon} \, \Gamma(1-\epsilon)}{\Gamma(1-2\epsilon)} \, .
\end{align}
We now demonstrate the calculation of the integrals, taking $\tilde{I}_{13}$ as an example, which is
\begin{align}
  \tilde{I}_{13} &= -\frac{\mu^{2\epsilon}e^{\epsilon\gamma_E}}{\pi^{1-\epsilon}} \, n \cdot v_3 \, \Omega_{d-4} \int k_0^{1-2\epsilon}dk_0 \sin^{1-2\epsilon}\theta_1 d\theta_1 \sin^{-2\epsilon}\theta_2 d\theta_2 \, \delta(k_0^2\sin^2\theta_1-q_T^2) \nonumber
  \\
  &\hspace{6em} \times \frac{\nu^\alpha}{k_0^{2+\alpha}} \, \frac{1}{(1-\cos\theta_1)^{1+\alpha}} \, \frac{\sqrt{1-\beta_t^2}}{1 - \beta_t \sin\theta_1 \cos\theta_2 \sin\theta - \beta_t \cos\theta_1 \cos\theta} \, .
\end{align}
Performing the integration over $k_0$ using the $\delta$-function, we obtain
\begin{align}
  \tilde{I}_{13} &= \frac{1}{q_T^2} \left( \frac{\mu^2}{q_T^2} \right)^{\epsilon+\alpha/2} \left( \frac{\nu^2}{\mu^2} \right)^{\alpha/2} \tilde{I}'_{13}\ ,
\end{align}
where
\begin{align}
  \tilde{I}'_{13} &= -\frac{2^{-2\epsilon} \, e^{\epsilon\gamma_E} \, \Gamma(1-\epsilon)}{\pi \, \Gamma(1-2\epsilon)} \, (1-\beta_t\cos\theta) \int_0^\pi \sin^{1+\alpha}\theta_1 d\theta_1 \int_0^\pi \sin^{-2\epsilon}\theta_2 d\theta_2 \nonumber
  \\
  &\hspace{6em} \times \frac{1}{(1 - \cos\theta_1)^{1+\alpha}} \, \frac{1}{(1 - \beta_t\sin\theta\sin\theta_1\cos\theta_2 - \beta_t\cos\theta\cos\theta_1)} \nonumber
  \\
  &= -\frac{e^{\epsilon\gamma_E}}{\Gamma(1-\epsilon)} \int_0^\pi d\theta_1 \left( \cot\frac{\theta_1}{2} \right)^{1+\alpha} \frac{1-\beta_t\cos\theta}{1-\beta_t\cos(\theta+\theta_1)} \nonumber
  \\
  &\hspace{6em} \times  {}_2F_1 \Biggl( 1 , \frac{1}{2}-\epsilon , 1-2\epsilon , \frac{2\beta_t\sin\theta\sin\theta_1}{1-\beta_t\cos(\theta+\theta_1)} \Biggr) \, .
\end{align}
We redefine the integration variable as
\begin{align}
  \theta_1 = 2 \arctan\left(\frac{1 - y}{y}\right) , \quad
  \int_0^\pi d\theta_1 = \int_0^1 \frac{2}{1-2y+2y^2} \, dy \, .
\end{align}
The integral then becomes
\begin{align}
  \tilde{I}'_{13} &= - \frac{2e^{\gamma_E \epsilon} }{\Gamma(1-\epsilon)} \, (1-\beta_t\cos\theta) \int_0^1 dy \, y^{1+\alpha} \, (1-y)^{-1-\alpha} \nn
    \\
    &\hspace{3em} \times \frac{{}_2F_{1} \Big( 1, \frac{1}{2}-\epsilon, 1-2\epsilon, \frac{4y(1-y)\beta_t\sin\theta}{1 - 2y(1-y) + (1-2y)\beta_t\cos\theta + 2y(1-y)\beta_t\sin\theta} \Big)}{1 - 2y(1-y) + (1-2y)\beta_t \cos\theta + 2y(1-y) \beta_t \sin\theta} \, .
\end{align}
We use the Mathematica package HypExp \cite{Huber:2005yg} to expand the hypergeometric functions as a series in $\epsilon$. After integrating over $y$, we obtain the explicit expression for $\tilde{I}_{13}$, which is
\begin{align}
  \tilde{I}_{13} = \frac{1}{q_T^2} \left( \frac{\mu^2}{q_T^2} \right)^{\epsilon+\alpha/2} \left( \frac{\nu^2}{\mu^2} \right)^{\alpha/2} \left[ \frac{2}{\alpha} \, \frac{e^{\epsilon\gamma_E}}{\Gamma(1-\epsilon)} - 2 \ln\frac{-t_1}{m_tM} + \epsilon f_{13} \right] .
\end{align}
where
\begin{align}
  f_{13} = \mathrm{Li}_{2}\left( -\frac{\beta_t^2\sin^2\theta}{1-\beta_t^2} \right)  .
\end{align}
With similar calculations, the other soft integrals can be obtained, which are
\begin{align}
  \tilde{I}_{23} &= \frac{1}{q_T^2} \left( \frac{\mu^2}{q_T^2} \right)^{\epsilon+\alpha/2} \left( \frac{\nu^2}{\mu^2} \right)^{\alpha/2} \left[ -\frac{2}{\alpha} \, \frac{e^{\epsilon\gamma_E}}{\Gamma(1-\epsilon)} - 2 \ln\frac{-u_1}{m_tM} + \epsilon f_{23} \right] , \nonumber
  \\
  \tilde{I}_{33} &= \frac{1}{q_T^2} \left( \frac{\mu^2}{q_T^2} \right)^{\epsilon} \left[ -2 + \epsilon f_{33} \right] , \nonumber
  \\
  \tilde{I}_{34} &= \frac{1}{q_T^2} \left( \frac{\mu^2}{q_T^2} \right)^{\epsilon} \left[ \frac{1+\beta_t^2}{\beta_t} \ln \frac{1+\beta_t}{1-\beta_t} + \epsilon f_{34} \right] .
\end{align}
where
\begin{align}
   f_{23} &= \mathrm{Li}_{2} \left( -\frac{\beta_t^2\sin^2\theta}{1-\beta_t^2} \right) , \nonumber
   \\
   f_{33} &= 2 \ln \frac{1-\beta_t^2}{1-\beta_t^2\cos^2\theta} \, , \nonumber
   \\
   f_{34} &= \frac{1+\beta_t^2}{\beta_t} \left[ 4 \ln \frac{1+\beta_t}{1-\beta_t} \, \ln \sec \frac{\theta}{2} - \mathrm{Li}_2 \left( -\frac{1-\beta_t}{1+\beta_t} \tan^2 \frac{\theta}{2} \right) + \mathrm{Li}_2 \left( -\frac{1+\beta_t}{1-\beta_t} \tan^2 \frac{\theta}{2} \right) \right] .
\end{align}
We can then perform the Fourier transform to obtain the results in Eq.~(\ref{eq:soft_Iij}).

\bibliographystyle{apsrev4-1}
\bibliography{ttbar_pt_resum}

\begin{thebibliography}{105}%
\makeatletter
\providecommand \@ifxundefined [1]{%
 \@ifx{#1\undefined}
}%
\providecommand \@ifnum [1]{%
 \ifnum #1\expandafter \@firstoftwo
 \else \expandafter \@secondoftwo
 \fi
}%
\providecommand \@ifx [1]{%
 \ifx #1\expandafter \@firstoftwo
 \else \expandafter \@secondoftwo
 \fi
}%
\providecommand \natexlab [1]{#1}%
\providecommand \enquote  [1]{``#1''}%
\providecommand \bibnamefont  [1]{#1}%
\providecommand \bibfnamefont [1]{#1}%
\providecommand \citenamefont [1]{#1}%
\providecommand \href@noop [0]{\@secondoftwo}%
\providecommand \href [0]{\begingroup \@sanitize@url \@href}%
\providecommand \@href[1]{\@@startlink{#1}\@@href}%
\providecommand \@@href[1]{\endgroup#1\@@endlink}%
\providecommand \@sanitize@url [0]{\catcode `\\12\catcode `\$12\catcode
  `\&12\catcode `\#12\catcode `\^12\catcode `\_12\catcode `\%12\relax}%
\providecommand \@@startlink[1]{}%
\providecommand \@@endlink[0]{}%
\providecommand \url  [0]{\begingroup\@sanitize@url \@url }%
\providecommand \@url [1]{\endgroup\@href {#1}{\urlprefix }}%
\providecommand \urlprefix  [0]{URL }%
\providecommand \Eprint [0]{\href }%
\providecommand \doibase [0]{http://dx.doi.org/}%
\providecommand \selectlanguage [0]{\@gobble}%
\providecommand \bibinfo  [0]{\@secondoftwo}%
\providecommand \bibfield  [0]{\@secondoftwo}%
\providecommand \translation [1]{[#1]}%
\providecommand \BibitemOpen [0]{}%
\providecommand \bibitemStop [0]{}%
\providecommand \bibitemNoStop [0]{.\EOS\space}%
\providecommand \EOS [0]{\spacefactor3000\relax}%
\providecommand \BibitemShut  [1]{\csname bibitem#1\endcsname}%
\let\auto@bib@innerbib\@empty
\bibitem [{\citenamefont {Nason}\ \emph {et~al.}(1988)\citenamefont {Nason},
  \citenamefont {Dawson},\ and\ \citenamefont {Ellis}}]{Nason:1987xz}%
  \BibitemOpen
  \bibfield  {author} {\bibinfo {author} {\bibfnamefont {P.}~\bibnamefont
  {Nason}}, \bibinfo {author} {\bibfnamefont {S.}~\bibnamefont {Dawson}}, \
  and\ \bibinfo {author} {\bibfnamefont {R.~K.}\ \bibnamefont {Ellis}},\ }\href
  {\doibase 10.1016/0550-3213(88)90422-1} {\bibfield  {journal} {\bibinfo
  {journal} {Nucl.Phys.}\ }\textbf {\bibinfo {volume} {B303}},\ \bibinfo
  {pages} {607} (\bibinfo {year} {1988})}\BibitemShut {NoStop}%
\bibitem [{\citenamefont {Beenakker}\ \emph {et~al.}(1989)\citenamefont
  {Beenakker}, \citenamefont {Kuijf}, \citenamefont {van Neerven},\ and\
  \citenamefont {Smith}}]{Beenakker:1988bq}%
  \BibitemOpen
  \bibfield  {author} {\bibinfo {author} {\bibfnamefont {W.}~\bibnamefont
  {Beenakker}}, \bibinfo {author} {\bibfnamefont {H.}~\bibnamefont {Kuijf}},
  \bibinfo {author} {\bibfnamefont {W.}~\bibfnamefont {L.}~\bibnamefont {van Neerven}}, \ and\
  \bibinfo {author} {\bibfnamefont {J.}~\bibnamefont {Smith}},\ }\href
  {\doibase 10.1103/PhysRevD.40.54} {\bibfield  {journal} {\bibinfo  {journal}
  {Phys.Rev.}\ }\textbf {\bibinfo {volume} {D40}},\ \bibinfo {pages} {54}
  (\bibinfo {year} {1989})}\BibitemShut {NoStop}%
\bibitem [{\citenamefont {Beenakker}\ \emph {et~al.}(1991)\citenamefont
  {Beenakker}, \citenamefont {van Neerven}, \citenamefont {Meng}, \citenamefont
  {Schuler},\ and\ \citenamefont {Smith}}]{Beenakker:1990maa}%
  \BibitemOpen
  \bibfield  {author} {\bibinfo {author} {\bibfnamefont {W.}~\bibnamefont
  {Beenakker}}, \bibinfo {author} {\bibfnamefont {W.}~\bibnamefont {van
  Neerven}}, \bibinfo {author} {\bibfnamefont {R.}~\bibnamefont {Meng}},
  \bibinfo {author} {\bibfnamefont {G.}~\bibnamefont {Schuler}}, \ and\
  \bibinfo {author} {\bibfnamefont {J.}~\bibnamefont {Smith}},\ }\href
  {\doibase 10.1016/S0550-3213(05)80032-X} {\bibfield  {journal} {\bibinfo
  {journal} {Nucl.Phys.}\ }\textbf {\bibinfo {volume} {B351}},\ \bibinfo
  {pages} {507} (\bibinfo {year} {1991})}\BibitemShut {NoStop}%
\bibitem [{\citenamefont {Beenakker}\ \emph {et~al.}(1994)\citenamefont
  {Beenakker}, \citenamefont {Denner}, \citenamefont {Hollik}, \citenamefont
  {Mertig}, \citenamefont {Sack} \emph {et~al.}}]{Beenakker:1993yr}%
  \BibitemOpen
  \bibfield  {author} {\bibinfo {author} {\bibfnamefont {W.}~\bibnamefont
  {Beenakker}}, \bibinfo {author} {\bibfnamefont {A.}~\bibnamefont {Denner}},
  \bibinfo {author} {\bibfnamefont {W.}~\bibnamefont {Hollik}}, \bibinfo
  {author} {\bibfnamefont {R.}~\bibnamefont {Mertig}}, \bibinfo {author}
  {\bibfnamefont {T.}~\bibnamefont {Sack}},  \emph {et~al.},\ }\href {\doibase
  10.1016/0550-3213(94)90454-5} {\bibfield  {journal} {\bibinfo  {journal}
  {Nucl.Phys.}\ }\textbf {\bibinfo {volume} {B411}},\ \bibinfo {pages} {343}
  (\bibinfo {year} {1994})}\BibitemShut {NoStop}%
\bibitem [{\citenamefont {Bernreuther}\ \emph {et~al.}(2006)\citenamefont
  {Bernreuther}, \citenamefont {Fuecker},\ and\ \citenamefont
  {Si}}]{Bernreuther:2005is}%
  \BibitemOpen
  \bibfield  {author} {\bibinfo {author} {\bibfnamefont {W.}~\bibnamefont
  {Bernreuther}}, \bibinfo {author} {\bibfnamefont {M.}~\bibnamefont
  {Fuecker}}, \ and\ \bibinfo {author} {\bibfnamefont {Z.}~\bibnamefont {Si}},\
  }\href {\doibase 10.1016/j.physletb.2005.11.056} {\bibfield  {journal}
  {\bibinfo  {journal} {Phys.Lett.}\ }\textbf {\bibinfo {volume} {B633}},\
  \bibinfo {pages} {54} (\bibinfo {year} {2006})},\ \Eprint
  {http://arxiv.org/abs/hep-ph/0508091} {arXiv:hep-ph/0508091 [hep-ph]}
  \BibitemShut {NoStop}%
\bibitem [{\citenamefont {Melnikov}\ and\ \citenamefont
  {Schulze}(2009)}]{Melnikov:2009dn}%
  \BibitemOpen
  \bibfield  {author} {\bibinfo {author} {\bibfnamefont {K.}~\bibnamefont
  {Melnikov}}\ and\ \bibinfo {author} {\bibfnamefont {M.}~\bibnamefont
  {Schulze}},\ }\href {\doibase 10.1088/1126-6708/2009/08/049} {\bibfield
  {journal} {\bibinfo  {journal} {JHEP}\ }\textbf {\bibinfo {volume} {0908}},\
  \bibinfo {pages} {049} (\bibinfo {year} {2009})},\ \Eprint
  {http://arxiv.org/abs/0907.3090} {arXiv:0907.3090 [hep-ph]} \BibitemShut
  {NoStop}%
\bibitem [{\citenamefont {Bernreuther}\ and\ \citenamefont
  {Si}(2010)}]{Bernreuther:2010ny}%
  \BibitemOpen
  \bibfield  {author} {\bibinfo {author} {\bibfnamefont {W.}~\bibnamefont
  {Bernreuther}}\ and\ \bibinfo {author} {\bibfnamefont {Z.-G.}\ \bibnamefont
  {Si}},\ }\href {\doibase 10.1016/j.nuclphysb.2010.05.001} {\bibfield
  {journal} {\bibinfo  {journal} {Nucl.Phys.}\ }\textbf {\bibinfo {volume}
  {B837}},\ \bibinfo {pages} {90} (\bibinfo {year} {2010})},\ \Eprint
  {http://arxiv.org/abs/1003.3926} {arXiv:1003.3926 [hep-ph]} \BibitemShut
  {NoStop}%
\bibitem [{\citenamefont {Campbell}\ and\ \citenamefont
  {Ellis}(2012)}]{Campbell:2012uf}%
  \BibitemOpen
  \bibfield  {author} {\bibinfo {author} {\bibfnamefont {J.~M.}\ \bibnamefont
  {Campbell}}\ and\ \bibinfo {author} {\bibfnamefont {R.~K.}\ \bibnamefont
  {Ellis}},\ }\href@noop {} {\  (\bibinfo {year} {2012})},\ \Eprint
  {http://arxiv.org/abs/1204.1513} {arXiv:1204.1513 [hep-ph]} \BibitemShut
  {NoStop}%
\bibitem [{\citenamefont {Denner}\ \emph {et~al.}(2011)\citenamefont {Denner},
  \citenamefont {Dittmaier}, \citenamefont {Kallweit},\ and\ \citenamefont
  {Pozzorini}}]{Denner:2010jp}%
  \BibitemOpen
  \bibfield  {author} {\bibinfo {author} {\bibfnamefont {A.}~\bibnamefont
  {Denner}}, \bibinfo {author} {\bibfnamefont {S.}~\bibnamefont {Dittmaier}},
  \bibinfo {author} {\bibfnamefont {S.}~\bibnamefont {Kallweit}}, \ and\
  \bibinfo {author} {\bibfnamefont {S.}~\bibnamefont {Pozzorini}},\ }\href
  {\doibase 10.1103/PhysRevLett.106.052001} {\bibfield  {journal} {\bibinfo
  {journal} {Phys.Rev.Lett.}\ }\textbf {\bibinfo {volume} {106}},\ \bibinfo
  {pages} {052001} (\bibinfo {year} {2011})},\ \Eprint
  {http://arxiv.org/abs/1012.3975} {arXiv:1012.3975 [hep-ph]} \BibitemShut
  {NoStop}%
\bibitem [{\citenamefont {Bevilacqua}\ \emph {et~al.}(2011)\citenamefont
  {Bevilacqua}, \citenamefont {Czakon}, \citenamefont {van Hameren},
  \citenamefont {Papadopoulos},\ and\ \citenamefont
  {Worek}}]{Bevilacqua:2010qb}%
  \BibitemOpen
  \bibfield  {author} {\bibinfo {author} {\bibfnamefont {G.}~\bibnamefont
  {Bevilacqua}}, \bibinfo {author} {\bibfnamefont {M.}~\bibnamefont {Czakon}},
  \bibinfo {author} {\bibfnamefont {A.}~\bibnamefont {van Hameren}}, \bibinfo
  {author} {\bibfnamefont {C.~G.}\ \bibnamefont {Papadopoulos}}, \ and\
  \bibinfo {author} {\bibfnamefont {M.}~\bibnamefont {Worek}},\ }\href
  {\doibase 10.1007/JHEP02(2011)083} {\bibfield  {journal} {\bibinfo  {journal}
  {JHEP}\ }\textbf {\bibinfo {volume} {1102}},\ \bibinfo {pages} {083}
  (\bibinfo {year} {2011})},\ \Eprint {http://arxiv.org/abs/1012.4230}
  {arXiv:1012.4230 [hep-ph]} \BibitemShut {NoStop}%
\bibitem [{\citenamefont {Falgari}\ \emph {et~al.}(2013)\citenamefont
  {Falgari}, \citenamefont {Papanastasiou},\ and\ \citenamefont
  {Signer}}]{Falgari:2013gwa}%
  \BibitemOpen
  \bibfield  {author} {\bibinfo {author} {\bibfnamefont {P.}~\bibnamefont
  {Falgari}}, \bibinfo {author} {\bibfnamefont {A.}~\bibnamefont
  {Papanastasiou}}, \ and\ \bibinfo {author} {\bibfnamefont {A.}~\bibnamefont
  {Signer}},\ }\href {\doibase 10.1007/JHEP05(2013)156} {\bibfield  {journal}
  {\bibinfo  {journal} {JHEP}\ }\textbf {\bibinfo {volume} {1305}},\ \bibinfo
  {pages} {156} (\bibinfo {year} {2013})},\ \Eprint
  {http://arxiv.org/abs/1303.5299} {arXiv:1303.5299 [hep-ph]} \BibitemShut
  {NoStop}%
\bibitem [{\citenamefont {Czakon}\ \emph {et~al.}(2009)\citenamefont {Czakon},
  \citenamefont {Mitov},\ and\ \citenamefont {Sterman}}]{Czakon:2009zw}%
  \BibitemOpen
  \bibfield  {author} {\bibinfo {author} {\bibfnamefont {M.}~\bibnamefont
  {Czakon}}, \bibinfo {author} {\bibfnamefont {A.}~\bibnamefont {Mitov}}, \
  and\ \bibinfo {author} {\bibfnamefont {G.~F.}\ \bibnamefont {Sterman}},\
  }\href {\doibase 10.1103/PhysRevD.80.074017} {\bibfield  {journal} {\bibinfo
  {journal} {Phys.Rev.}\ }\textbf {\bibinfo {volume} {D80}},\ \bibinfo {pages}
  {074017} (\bibinfo {year} {2009})},\ \Eprint {http://arxiv.org/abs/0907.1790}
  {arXiv:0907.1790 [hep-ph]} \BibitemShut {NoStop}%
\bibitem [{\citenamefont {Beneke}\ \emph {et~al.}(2010)\citenamefont {Beneke},
  \citenamefont {Falgari},\ and\ \citenamefont {Schwinn}}]{Beneke:2009rj}%
  \BibitemOpen
  \bibfield  {author} {\bibinfo {author} {\bibfnamefont {M.}~\bibnamefont
  {Beneke}}, \bibinfo {author} {\bibfnamefont {P.}~\bibnamefont {Falgari}}, \
  and\ \bibinfo {author} {\bibfnamefont {C.}~\bibnamefont {Schwinn}},\ }\href
  {\doibase 10.1016/j.nuclphysb.2009.11.004} {\bibfield  {journal} {\bibinfo
  {journal} {Nucl.Phys.}\ }\textbf {\bibinfo {volume} {B828}},\ \bibinfo
  {pages} {69} (\bibinfo {year} {2010})},\ \Eprint
  {http://arxiv.org/abs/0907.1443} {arXiv:0907.1443 [hep-ph]} \BibitemShut
  {NoStop}%
\bibitem [{\citenamefont {Ahrens}\ \emph {et~al.}(2010)\citenamefont {Ahrens},
  \citenamefont {Ferroglia}, \citenamefont {Neubert}, \citenamefont {Pecjak},\
  and\ \citenamefont {Yang}}]{Ahrens:2010zv}%
  \BibitemOpen
  \bibfield  {author} {\bibinfo {author} {\bibfnamefont {V.}~\bibnamefont
  {Ahrens}}, \bibinfo {author} {\bibfnamefont {A.}~\bibnamefont {Ferroglia}},
  \bibinfo {author} {\bibfnamefont {M.}~\bibnamefont {Neubert}}, \bibinfo
  {author} {\bibfnamefont {B.~D.}\ \bibnamefont {Pecjak}}, \ and\ \bibinfo
  {author} {\bibfnamefont {L.~L.}\ \bibnamefont {Yang}},\ }\href {\doibase
  10.1007/JHEP09(2010)097} {\bibfield  {journal} {\bibinfo  {journal} {JHEP}\
  }\textbf {\bibinfo {volume} {1009}},\ \bibinfo {pages} {097} (\bibinfo {year}
  {2010})},\ \Eprint {http://arxiv.org/abs/1003.5827} {arXiv:1003.5827
  [hep-ph]} \BibitemShut {NoStop}%
\bibitem [{\citenamefont {Kidonakis}(2010)}]{Kidonakis:2010dk}%
  \BibitemOpen
  \bibfield  {author} {\bibinfo {author} {\bibfnamefont {N.}~\bibnamefont
  {Kidonakis}},\ }\href {\doibase 10.1103/PhysRevD.82.114030} {\bibfield
  {journal} {\bibinfo  {journal} {Phys.Rev.}\ }\textbf {\bibinfo {volume}
  {D82}},\ \bibinfo {pages} {114030} (\bibinfo {year} {2010})},\ \Eprint
  {http://arxiv.org/abs/1009.4935} {arXiv:1009.4935 [hep-ph]} \BibitemShut
  {NoStop}%
\bibitem [{\citenamefont {Cacciari}\ \emph {et~al.}(2012)\citenamefont
  {Cacciari}, \citenamefont {Czakon}, \citenamefont {Mangano}, \citenamefont
  {Mitov},\ and\ \citenamefont {Nason}}]{Cacciari:2011hy}%
  \BibitemOpen
  \bibfield  {author} {\bibinfo {author} {\bibfnamefont {M.}~\bibnamefont
  {Cacciari}}, \bibinfo {author} {\bibfnamefont {M.}~\bibnamefont {Czakon}},
  \bibinfo {author} {\bibfnamefont {M.}~\bibnamefont {Mangano}}, \bibinfo
  {author} {\bibfnamefont {A.}~\bibnamefont {Mitov}}, \ and\ \bibinfo {author}
  {\bibfnamefont {P.}~\bibnamefont {Nason}},\ }\href {\doibase
  10.1016/j.physletb.2012.03.013} {\bibfield  {journal} {\bibinfo  {journal}
  {Phys.Lett.}\ }\textbf {\bibinfo {volume} {B710}},\ \bibinfo {pages} {612}
  (\bibinfo {year} {2012})},\ \Eprint {http://arxiv.org/abs/1111.5869}
  {arXiv:1111.5869 [hep-ph]} \BibitemShut {NoStop}%
\bibitem [{\citenamefont {Baernreuther}\ \emph {et~al.}(2012)\citenamefont
  {Baernreuther}, \citenamefont {Czakon},\ and\ \citenamefont
  {Mitov}}]{Baernreuther:2012ws}%
  \BibitemOpen
  \bibfield  {author} {\bibinfo {author} {\bibfnamefont {P.}~\bibnamefont
  {Barnreuther}}, \bibinfo {author} {\bibfnamefont {M.}~\bibnamefont
  {Czakon}}, \ and\ \bibinfo {author} {\bibfnamefont {A.}~\bibnamefont
  {Mitov}},\ }\href {\doibase 10.1103/PhysRevLett.109.132001} {\bibfield
  {journal} {\bibinfo  {journal} {Phys.Rev.Lett.}\ }\textbf {\bibinfo {volume}
  {109}},\ \bibinfo {pages} {132001} (\bibinfo {year} {2012})},\ \Eprint
  {http://arxiv.org/abs/1204.5201} {arXiv:1204.5201 [hep-ph]} \BibitemShut
  {NoStop}%
\bibitem [{\citenamefont {Czakon}\ and\ \citenamefont
  {Mitov}(2012)}]{Czakon:2012zr}%
  \BibitemOpen
  \bibfield  {author} {\bibinfo {author} {\bibfnamefont {M.}~\bibnamefont
  {Czakon}}\ and\ \bibinfo {author} {\bibfnamefont {A.}~\bibnamefont {Mitov}},\
  }\href {\doibase 10.1007/JHEP12(2012)054} {\bibfield  {journal} {\bibinfo
  {journal} {JHEP}\ }\textbf {\bibinfo {volume} {1212}},\ \bibinfo {pages}
  {054} (\bibinfo {year} {2012})},\ \Eprint {http://arxiv.org/abs/1207.0236}
  {arXiv:1207.0236 [hep-ph]} \BibitemShut {NoStop}%
\bibitem [{\citenamefont {Czakon}\ and\ \citenamefont
  {Mitov}(2013)}]{Czakon:2012pz}%
  \BibitemOpen
  \bibfield  {author} {\bibinfo {author} {\bibfnamefont {M.}~\bibnamefont
  {Czakon}}\ and\ \bibinfo {author} {\bibfnamefont {A.}~\bibnamefont {Mitov}},\
  }\href {\doibase 10.1007/JHEP01(2013)080} {\bibfield  {journal} {\bibinfo
  {journal} {JHEP}\ }\textbf {\bibinfo {volume} {1301}},\ \bibinfo {pages}
  {080} (\bibinfo {year} {2013})},\ \Eprint {http://arxiv.org/abs/1210.6832}
  {arXiv:1210.6832 [hep-ph]} \BibitemShut {NoStop}%
\bibitem [{\citenamefont {Czakon}\ \emph {et~al.}(2013)\citenamefont {Czakon},
  \citenamefont {Fiedler},\ and\ \citenamefont {Mitov}}]{Czakon:2013goa}%
  \BibitemOpen
  \bibfield  {author} {\bibinfo {author} {\bibfnamefont {M.}~\bibnamefont
  {Czakon}}, \bibinfo {author} {\bibfnamefont {P.}~\bibnamefont {Fiedler}}, \
  and\ \bibinfo {author} {\bibfnamefont {A.}~\bibnamefont {Mitov}},\
  }\href@noop {} {\  (\bibinfo {year} {2013})},\ \Eprint
  {http://arxiv.org/abs/1303.6254} {arXiv:1303.6254 [hep-ph]} \BibitemShut
  {NoStop}%
\bibitem [{\citenamefont {Gao}\ \emph {et~al.}(2013)\citenamefont {Gao},
  \citenamefont {Li},\ and\ \citenamefont {Zhu}}]{Gao:2012ja}%
  \BibitemOpen
  \bibfield  {author} {\bibinfo {author} {\bibfnamefont {J.}~\bibnamefont
  {Gao}}, \bibinfo {author} {\bibfnamefont {C.~S.}\ \bibnamefont {Li}}, \ and\
  \bibinfo {author} {\bibfnamefont {H.~X.}\ \bibnamefont {Zhu}},\ }\href
  {\doibase 10.1103/PhysRevLett.110.042001} {\bibfield  {journal} {\bibinfo
  {journal} {Phys.Rev.Lett.}\ }\textbf {\bibinfo {volume} {110}},\ \bibinfo
  {pages} {042001} (\bibinfo {year} {2013})},\ \Eprint
  {http://arxiv.org/abs/1210.2808} {arXiv:1210.2808 [hep-ph]} \BibitemShut
  {NoStop}%
\bibitem [{\citenamefont {Brucherseifer}\ \emph {et~al.}(2013)\citenamefont
  {Brucherseifer}, \citenamefont {Caola},\ and\ \citenamefont
  {Melnikov}}]{Brucherseifer:2013iv}%
  \BibitemOpen
  \bibfield  {author} {\bibinfo {author} {\bibfnamefont {M.}~\bibnamefont
  {Brucherseifer}}, \bibinfo {author} {\bibfnamefont {F.}~\bibnamefont
  {Caola}}, \ and\ \bibinfo {author} {\bibfnamefont {K.}~\bibnamefont
  {Melnikov}},\ }\href {\doibase 10.1007/JHEP04(2013)059} {\bibfield  {journal}
  {\bibinfo  {journal} {JHEP}\ }\textbf {\bibinfo {volume} {1304}},\ \bibinfo
  {pages} {059} (\bibinfo {year} {2013})},\ \Eprint
  {http://arxiv.org/abs/1301.7133} {arXiv:1301.7133 [hep-ph]} \BibitemShut
  {NoStop}%
\bibitem [{\citenamefont {Aad}\ \emph {et~al.}(2013{\natexlab{a}})\citenamefont
  {Aad} \emph {et~al.}}]{Aad:2012hg}%
  \BibitemOpen
  \bibfield  {author} {\bibinfo {author} {\bibfnamefont {G.}~\bibnamefont
  {Aad}} \emph {et~al.} (\bibinfo {collaboration} {ATLAS Collaboration}),\
  }\href {\doibase 10.1140/epjc/s10052-012-2261-1} {\bibfield  {journal}
  {\bibinfo  {journal} {Eur.Phys.J.}\ }\textbf {\bibinfo {volume} {C73}},\
  \bibinfo {pages} {2261} (\bibinfo {year} {2013}{\natexlab{a}})},\ \Eprint
  {http://arxiv.org/abs/1207.5644} {arXiv:1207.5644 [hep-ex]} \BibitemShut
  {NoStop}%
\bibitem [{\citenamefont {Aad}\ \emph {et~al.}(2013{\natexlab{b}})\citenamefont
  {Aad} \emph {et~al.}}]{Aad:2012vip}%
  \BibitemOpen
  \bibfield  {author} {\bibinfo {author} {\bibfnamefont {G.}~\bibnamefont
  {Aad}} \emph {et~al.} (\bibinfo {collaboration} {ATLAS Collaboration}),\
  }\href {\doibase 10.1140/epjc/s10052-013-2328-7} {\bibfield  {journal}
  {\bibinfo  {journal} {Eur.Phys.J.}\ }\textbf {\bibinfo {volume} {C73}},\
  \bibinfo {pages} {2328} (\bibinfo {year} {2013}{\natexlab{b}})},\ \Eprint
  {http://arxiv.org/abs/1211.7205} {arXiv:1211.7205 [hep-ex]} \BibitemShut
  {NoStop}%
\bibitem [{\citenamefont {Chatrchyan}\ \emph {et~al.}(2012)\citenamefont
  {Chatrchyan} \emph {et~al.}}]{Chatrchyan:2012bra}%
  \BibitemOpen
  \bibfield  {author} {\bibinfo {author} {\bibfnamefont {S.}~\bibnamefont
  {Chatrchyan}} \emph {et~al.} (\bibinfo {collaboration} {CMS Collaboration}),\
  }\href {\doibase 10.1007/JHEP11(2012)067} {\bibfield  {journal} {\bibinfo
  {journal} {JHEP}\ }\textbf {\bibinfo {volume} {1211}},\ \bibinfo {pages}
  {067} (\bibinfo {year} {2012})},\ \Eprint {http://arxiv.org/abs/1208.2671}
  {arXiv:1208.2671 [hep-ex]} \BibitemShut {NoStop}%
\bibitem [{\citenamefont {{The CMS
  Collaboration}}(2013{\natexlab{a}})}]{CMS:pas-top-12-027}%
  \BibitemOpen
  \bibfield  {author} {\bibinfo {author} {\bibnamefont {{The CMS
  Collaboration}}},\ }\href@noop {} {\enquote {\bibinfo {title} {{Measurement
  of differential top-quark pair production cross sections in the lepton+jets
  channel in pp collisions at 8 TeV}},}\ } (\bibinfo {year}
  {2013}{\natexlab{a}}),\ \Eprint {http://arxiv.org/abs/CMS PAS TOP-12-027}
  {CMS PAS TOP-12-027} \BibitemShut {NoStop}%
\bibitem [{\citenamefont {{The CMS
  Collaboration}}(2013{\natexlab{b}})}]{CMS:pas-top-12-028}%
  \BibitemOpen
  \bibfield  {author} {\bibinfo {author} {\bibnamefont {{The CMS
  Collaboration}}},\ }\href@noop {} {\enquote {\bibinfo {title} {{Measurement
  of the differential ttbar cross section in the dilepton channel at 8 TeV}},}\
  } (\bibinfo {year} {2013}{\natexlab{b}}),\ \Eprint
  {http://arxiv.org/abs/CMS-PAS-TOP-12-028} {CMS-PAS-TOP-12-028} \BibitemShut
  {NoStop}%
\bibitem [{\citenamefont {Aaltonen}\ \emph {et~al.}(2013)\citenamefont
  {Aaltonen} \emph {et~al.}}]{Aaltonen:2012it}%
  \BibitemOpen
  \bibfield  {author} {\bibinfo {author} {\bibfnamefont {T.}~\bibnamefont
  {Aaltonen}} \emph {et~al.} (\bibinfo {collaboration} {CDF Collaboration}),\
  }\href {\doibase 10.1103/PhysRevD.87.092002} {\bibfield  {journal} {\bibinfo
  {journal} {Phys.Rev.}\ }\textbf {\bibinfo {volume} {D87}},\ \bibinfo {pages}
  {092002} (\bibinfo {year} {2013})},\ \Eprint {http://arxiv.org/abs/1211.1003}
  {arXiv:1211.1003 [hep-ex]} \BibitemShut {NoStop}%
\bibitem [{\citenamefont {Abazov}\ \emph {et~al.}(2011)\citenamefont {Abazov}
  \emph {et~al.}}]{Abazov:2011rq}%
  \BibitemOpen
  \bibfield  {author} {\bibinfo {author} {\bibfnamefont {V.~M.}\ \bibnamefont
  {Abazov}} \emph {et~al.} (\bibinfo {collaboration} {D0 Collaboration}),\
  }\href {\doibase 10.1103/PhysRevD.84.112005} {\bibfield  {journal} {\bibinfo
  {journal} {Phys.Rev.}\ }\textbf {\bibinfo {volume} {D84}},\ \bibinfo {pages}
  {112005} (\bibinfo {year} {2011})},\ \Eprint {http://arxiv.org/abs/1107.4995}
  {arXiv:1107.4995 [hep-ex]} \BibitemShut {NoStop}%
\bibitem [{\citenamefont {Kuhn}\ and\ \citenamefont
  {Rodrigo}(2012)}]{Kuhn:2011ri}%
  \BibitemOpen
  \bibfield  {author} {\bibinfo {author} {\bibfnamefont {J.~H.}\ \bibnamefont
  {Kuhn}}\ and\ \bibinfo {author} {\bibfnamefont {G.}~\bibnamefont {Rodrigo}},\
  }\href {\doibase 10.1007/JHEP01(2012)063} {\bibfield  {journal} {\bibinfo
  {journal} {JHEP}\ }\textbf {\bibinfo {volume} {1201}},\ \bibinfo {pages}
  {063} (\bibinfo {year} {2012})},\ \Eprint {http://arxiv.org/abs/1109.6830}
  {arXiv:1109.6830 [hep-ph]} \BibitemShut {NoStop}%
\bibitem [{\citenamefont {Alvarez}(2012)}]{Alvarez:2012vq}%
  \BibitemOpen
  \bibfield  {author} {\bibinfo {author} {\bibfnamefont {E.}~\bibnamefont
  {Alvarez}},\ }\href {\doibase 10.1103/PhysRevD.85.094026} {\bibfield
  {journal} {\bibinfo  {journal} {Phys.Rev.}\ }\textbf {\bibinfo {volume}
  {D85}},\ \bibinfo {pages} {094026} (\bibinfo {year} {2012})},\ \Eprint
  {http://arxiv.org/abs/1202.6622} {arXiv:1202.6622 [hep-ph]} \BibitemShut
  {NoStop}%
\bibitem [{\citenamefont {Collins}\ and\ \citenamefont
  {Soper}(1981)}]{Collins:1981uk}%
  \BibitemOpen
  \bibfield  {author} {\bibinfo {author} {\bibfnamefont {J.~C.}\ \bibnamefont
  {Collins}}\ and\ \bibinfo {author} {\bibfnamefont {D.~E.}\ \bibnamefont
  {Soper}},\ }\href {\doibase 10.1016/0550-3213(81)90339-4} {\bibfield
  {journal} {\bibinfo  {journal} {Nucl.Phys.}\ }\textbf {\bibinfo {volume}
  {B193}},\ \bibinfo {pages} {381} (\bibinfo {year} {1981})}\BibitemShut
  {NoStop}%
\bibitem [{\citenamefont {Collins}\ and\ \citenamefont
  {Soper}(1982)}]{Collins:1981uw}%
  \BibitemOpen
  \bibfield  {author} {\bibinfo {author} {\bibfnamefont {J.~C.}\ \bibnamefont
  {Collins}}\ and\ \bibinfo {author} {\bibfnamefont {D.~E.}\ \bibnamefont
  {Soper}},\ }\href {\doibase 10.1016/0550-3213(82)90021-9} {\bibfield
  {journal} {\bibinfo  {journal} {Nucl.Phys.}\ }\textbf {\bibinfo {volume}
  {B194}},\ \bibinfo {pages} {445} (\bibinfo {year} {1982})}\BibitemShut
  {NoStop}%
\bibitem [{\citenamefont {Collins}\ \emph {et~al.}(1985)\citenamefont
  {Collins}, \citenamefont {Soper},\ and\ \citenamefont
  {Sterman}}]{Collins:1984kg}%
  \BibitemOpen
  \bibfield  {author} {\bibinfo {author} {\bibfnamefont {J.~C.}\ \bibnamefont
  {Collins}}, \bibinfo {author} {\bibfnamefont {D.~E.}\ \bibnamefont {Soper}},
  \ and\ \bibinfo {author} {\bibfnamefont {G.~F.}\ \bibnamefont {Sterman}},\
  }\href {\doibase 10.1016/0550-3213(85)90479-1} {\bibfield  {journal}
  {\bibinfo  {journal} {Nucl.Phys.}\ }\textbf {\bibinfo {volume} {B250}},\
  \bibinfo {pages} {199} (\bibinfo {year} {1985})}\BibitemShut {NoStop}%
\bibitem [{\citenamefont {Collins}(2011)}]{Collins:2011zzd}%
  \BibitemOpen
  \bibfield  {author} {\bibinfo {author} {\bibfnamefont {J.}~\bibnamefont
  {Collins}},\ }\href@noop {} {\emph {\bibinfo {title} {{Foundations of
  Perturbative QCD}}}}\ (\bibinfo  {publisher} {Cambridge University Press},\
  \bibinfo {year} {2011})\BibitemShut {NoStop}%
\bibitem [{\citenamefont {Balazs}\ \emph {et~al.}(1998)\citenamefont {Balazs},
  \citenamefont {Berger}, \citenamefont {Mrenna},\ and\ \citenamefont
  {Yuan}}]{Balazs:1997hv}%
  \BibitemOpen
  \bibfield  {author} {\bibinfo {author} {\bibfnamefont {C.}~\bibnamefont
  {Balazs}}, \bibinfo {author} {\bibfnamefont {E.~L.}\ \bibnamefont {Berger}},
  \bibinfo {author} {\bibfnamefont {S.}~\bibnamefont {Mrenna}}, \ and\ \bibinfo
  {author} {\bibfnamefont {C.~P.}~\bibnamefont {Yuan}},\ }\href {\doibase
  10.1103/PhysRevD.57.6934} {\bibfield  {journal} {\bibinfo  {journal}
  {Phys.Rev.}\ }\textbf {\bibinfo {volume} {D57}},\ \bibinfo {pages} {6934}
  (\bibinfo {year} {1998})},\ \Eprint {http://arxiv.org/abs/hep-ph/9712471}
  {arXiv:hep-ph/9712471 [hep-ph]} \BibitemShut {NoStop}%
\bibitem [{\citenamefont {Balazs}\ \emph {et~al.}(2000)\citenamefont {Balazs},
  \citenamefont {Nadolsky}, \citenamefont {Schmidt},\ and\ \citenamefont
  {Yuan}}]{Balazs:1999yf}%
  \BibitemOpen
  \bibfield  {author} {\bibinfo {author} {\bibfnamefont {C.}~\bibnamefont
  {Balazs}}, \bibinfo {author} {\bibfnamefont {P.~M.}\ \bibnamefont
  {Nadolsky}}, \bibinfo {author} {\bibfnamefont {C.}~\bibnamefont {Schmidt}}, \
  and\ \bibinfo {author} {\bibfnamefont {C.}~\bibnamefont {Yuan}},\ }\href
  {\doibase 10.1016/S0370-2693(00)00934-5} {\bibfield  {journal} {\bibinfo
  {journal} {Phys.Lett.}\ }\textbf {\bibinfo {volume} {B489}},\ \bibinfo
  {pages} {157} (\bibinfo {year} {2000})},\ \Eprint
  {http://arxiv.org/abs/hep-ph/9905551} {arXiv:hep-ph/9905551 [hep-ph]}
  \BibitemShut {NoStop}%
\bibitem [{\citenamefont {Balazs}\ \emph {et~al.}(2001)\citenamefont {Balazs},
  \citenamefont {Huston},\ and\ \citenamefont {Puljak}}]{Balazs:2000sz}%
  \BibitemOpen
  \bibfield  {author} {\bibinfo {author} {\bibfnamefont {C.}~\bibnamefont
  {Balazs}}, \bibinfo {author} {\bibfnamefont {J.}~\bibnamefont {Huston}}, \
  and\ \bibinfo {author} {\bibfnamefont {I.}~\bibnamefont {Puljak}},\ }\href
  {\doibase 10.1103/PhysRevD.63.014021} {\bibfield  {journal} {\bibinfo
  {journal} {Phys.Rev.}\ }\textbf {\bibinfo {volume} {D63}},\ \bibinfo {pages}
  {014021} (\bibinfo {year} {2000})},\ \Eprint
  {http://arxiv.org/abs/hep-ph/0002032} {arXiv:hep-ph/0002032 [hep-ph]}
  \BibitemShut {NoStop}%
\bibitem [{\citenamefont {Balazs}\ and\ \citenamefont
  {Yuan}(2000)}]{Balazs:2000wv}%
  \BibitemOpen
  \bibfield  {author} {\bibinfo {author} {\bibfnamefont {C.}~\bibnamefont
  {Balazs}}\ and\ \bibinfo {author} {\bibfnamefont {C.}~\bibnamefont {Yuan}},\
  }\href {\doibase 10.1016/S0370-2693(00)00270-7} {\bibfield  {journal}
  {\bibinfo  {journal} {Phys.Lett.}\ }\textbf {\bibinfo {volume} {B478}},\
  \bibinfo {pages} {192} (\bibinfo {year} {2000})},\ \Eprint
  {http://arxiv.org/abs/hep-ph/0001103} {arXiv:hep-ph/0001103 [hep-ph]}
  \BibitemShut {NoStop}%
\bibitem [{\citenamefont {Nadolsky}\ and\ \citenamefont
  {Schmidt}(2003)}]{Nadolsky:2002gj}%
  \BibitemOpen
  \bibfield  {author} {\bibinfo {author} {\bibfnamefont {P.~M.}\ \bibnamefont
  {Nadolsky}}\ and\ \bibinfo {author} {\bibfnamefont {C.}~\bibnamefont
  {Schmidt}},\ }\href {\doibase 10.1016/S0370-2693(03)00218-1} {\bibfield
  {journal} {\bibinfo  {journal} {Phys.Lett.}\ }\textbf {\bibinfo {volume}
  {B558}},\ \bibinfo {pages} {63} (\bibinfo {year} {2003})},\ \Eprint
  {http://arxiv.org/abs/hep-ph/0211398} {arXiv:hep-ph/0211398 [hep-ph]}
  \BibitemShut {NoStop}%
\bibitem [{\citenamefont {Bozzi}\ \emph {et~al.}(2003)\citenamefont {Bozzi},
  \citenamefont {Catani}, \citenamefont {de~Florian},\ and\ \citenamefont
  {Grazzini}}]{Bozzi:2003jy}%
  \BibitemOpen
  \bibfield  {author} {\bibinfo {author} {\bibfnamefont {G.}~\bibnamefont
  {Bozzi}}, \bibinfo {author} {\bibfnamefont {S.}~\bibnamefont {Catani}},
  \bibinfo {author} {\bibfnamefont {D.}~\bibnamefont {de~Florian}}, \ and\
  \bibinfo {author} {\bibfnamefont {M.}~\bibnamefont {Grazzini}},\ }\href
  {\doibase 10.1016/S0370-2693(03)00656-7} {\bibfield  {journal} {\bibinfo
  {journal} {Phys.Lett.}\ }\textbf {\bibinfo {volume} {B564}},\ \bibinfo
  {pages} {65} (\bibinfo {year} {2003})},\ \Eprint
  {http://arxiv.org/abs/hep-ph/0302104} {arXiv:hep-ph/0302104 [hep-ph]}
  \BibitemShut {NoStop}%
\bibitem [{\citenamefont {Bozzi}\ \emph {et~al.}(2006)\citenamefont {Bozzi},
  \citenamefont {Catani}, \citenamefont {de~Florian},\ and\ \citenamefont
  {Grazzini}}]{Bozzi:2005wk}%
  \BibitemOpen
  \bibfield  {author} {\bibinfo {author} {\bibfnamefont {G.}~\bibnamefont
  {Bozzi}}, \bibinfo {author} {\bibfnamefont {S.}~\bibnamefont {Catani}},
  \bibinfo {author} {\bibfnamefont {D.}~\bibnamefont {de~Florian}}, \ and\
  \bibinfo {author} {\bibfnamefont {M.}~\bibnamefont {Grazzini}},\ }\href
  {\doibase 10.1016/j.nuclphysb.2005.12.022} {\bibfield  {journal} {\bibinfo
  {journal} {Nucl.Phys.}\ }\textbf {\bibinfo {volume} {B737}},\ \bibinfo
  {pages} {73} (\bibinfo {year} {2006})},\ \Eprint
  {http://arxiv.org/abs/hep-ph/0508068} {arXiv:hep-ph/0508068 [hep-ph]}
  \BibitemShut {NoStop}%
\bibitem [{\citenamefont {Balazs}\ \emph {et~al.}(2006)\citenamefont {Balazs},
  \citenamefont {Berger}, \citenamefont {Nadolsky},\ and\ \citenamefont
  {Yuan}}]{Balazs:2006cc}%
  \BibitemOpen
  \bibfield  {author} {\bibinfo {author} {\bibfnamefont {C.}~\bibnamefont
  {Balazs}}, \bibinfo {author} {\bibfnamefont {E.~L.}\ \bibnamefont {Berger}},
  \bibinfo {author} {\bibfnamefont {P.~M.}\ \bibnamefont {Nadolsky}}, \ and\
  \bibinfo {author} {\bibfnamefont {C.-P.}\ \bibnamefont {Yuan}},\ }\href
  {\doibase 10.1016/j.physletb.2006.04.017} {\bibfield  {journal} {\bibinfo
  {journal} {Phys.Lett.}\ }\textbf {\bibinfo {volume} {B637}},\ \bibinfo
  {pages} {235} (\bibinfo {year} {2006})},\ \Eprint
  {http://arxiv.org/abs/hep-ph/0603037} {arXiv:hep-ph/0603037 [hep-ph]}
  \BibitemShut {NoStop}%
\bibitem [{\citenamefont {Cao}\ \emph {et~al.}(2009)\citenamefont {Cao},
  \citenamefont {Chen}, \citenamefont {Schmidt},\ and\ \citenamefont
  {Yuan}}]{Cao:2009md}%
  \BibitemOpen
  \bibfield  {author} {\bibinfo {author} {\bibfnamefont {Q.-H.}\ \bibnamefont
  {Cao}}, \bibinfo {author} {\bibfnamefont {C.-R.}\ \bibnamefont {Chen}},
  \bibinfo {author} {\bibfnamefont {C.}~\bibnamefont {Schmidt}}, \ and\
  \bibinfo {author} {\bibfnamefont {C.-P.}\ \bibnamefont {Yuan}},\ }\href@noop
  {} {\  (\bibinfo {year} {2009})},\ \Eprint {http://arxiv.org/abs/0909.2305}
  {arXiv:0909.2305 [hep-ph]} \BibitemShut {NoStop}%
\bibitem [{\citenamefont {de~Florian}\ \emph {et~al.}(2011)\citenamefont
  {de~Florian}, \citenamefont {Ferrera}, \citenamefont {Grazzini},\ and\
  \citenamefont {Tommasini}}]{deFlorian:2011xf}%
  \BibitemOpen
  \bibfield  {author} {\bibinfo {author} {\bibfnamefont {D.}~\bibnamefont
  {de~Florian}}, \bibinfo {author} {\bibfnamefont {G.}~\bibnamefont {Ferrera}},
  \bibinfo {author} {\bibfnamefont {M.}~\bibnamefont {Grazzini}}, \ and\
  \bibinfo {author} {\bibfnamefont {D.}~\bibnamefont {Tommasini}},\ }\href
  {\doibase 10.1007/JHEP11(2011)064} {\bibfield  {journal} {\bibinfo  {journal}
  {JHEP}\ }\textbf {\bibinfo {volume} {1111}},\ \bibinfo {pages} {064}
  (\bibinfo {year} {2011})},\ \Eprint {http://arxiv.org/abs/1109.2109}
  {arXiv:1109.2109 [hep-ph]} \BibitemShut {NoStop}%
\bibitem [{\citenamefont {Wang}\ \emph {et~al.}(2012)\citenamefont {Wang},
  \citenamefont {Li}, \citenamefont {Li}, \citenamefont {Yuan},\ and\
  \citenamefont {Li}}]{Wang:2012xs}%
  \BibitemOpen
  \bibfield  {author} {\bibinfo {author} {\bibfnamefont {J.}~\bibnamefont
  {Wang}}, \bibinfo {author} {\bibfnamefont {C.~S.}\ \bibnamefont {Li}},\bibinfo {author}
  {\bibfnamefont {H.~T.}\ \bibnamefont {Li}},
  \bibinfo {author} {\bibfnamefont {Z.}~\bibnamefont {Li}},  \ and\  \bibinfo {author}
  {\bibfnamefont {C.~P.}~\bibnamefont {Yuan}},\ }\href {\doibase
  10.1103/PhysRevD.86.094026} {\bibfield  {journal} {\bibinfo  {journal}
  {Phys.Rev.}\ }\textbf {\bibinfo {volume} {D86}},\ \bibinfo {pages} {094026}
  (\bibinfo {year} {2012})},\ \Eprint {http://arxiv.org/abs/1205.4311}
  {arXiv:1205.4311 [hep-ph]} \BibitemShut {NoStop}%
\bibitem [{\citenamefont {Landry}\ \emph {et~al.}(2001)\citenamefont {Landry},
  \citenamefont {Brock}, \citenamefont {Ladinsky},\ and\ \citenamefont
  {Yuan}}]{Landry:1999an}%
  \BibitemOpen
  \bibfield  {author} {\bibinfo {author} {\bibfnamefont {F.}~\bibnamefont
  {Landry}}, \bibinfo {author} {\bibfnamefont {R.}~\bibnamefont {Brock}},
  \bibinfo {author} {\bibfnamefont {G.}~\bibnamefont {Ladinsky}}, \ and\
  \bibinfo {author} {\bibfnamefont {C.~P.}~\bibnamefont {Yuan}},\ }\href {\doibase
  10.1103/PhysRevD.63.013004} {\bibfield  {journal} {\bibinfo  {journal}
  {Phys.Rev.}\ }\textbf {\bibinfo {volume} {D63}},\ \bibinfo {pages} {013004}
  (\bibinfo {year} {2000})},\ \Eprint {http://arxiv.org/abs/hep-ph/9905391}
  {arXiv:hep-ph/9905391 [hep-ph]} \BibitemShut {NoStop}%
\bibitem [{\citenamefont {Landry}\ \emph {et~al.}(2003)\citenamefont {Landry},
  \citenamefont {Brock}, \citenamefont {Nadolsky},\ and\ \citenamefont
  {Yuan}}]{Landry:2002ix}%
  \BibitemOpen
  \bibfield  {author} {\bibinfo {author} {\bibfnamefont {F.}~\bibnamefont
  {Landry}}, \bibinfo {author} {\bibfnamefont {R.}~\bibnamefont {Brock}},
  \bibinfo {author} {\bibfnamefont {P.~M.}\ \bibnamefont {Nadolsky}}, \ and\
  \bibinfo {author} {\bibfnamefont {C.~P.}~\bibnamefont {Yuan}},\ }\href {\doibase
  10.1103/PhysRevD.67.073016} {\bibfield  {journal} {\bibinfo  {journal}
  {Phys.Rev.}\ }\textbf {\bibinfo {volume} {D67}},\ \bibinfo {pages} {073016}
  (\bibinfo {year} {2003})},\ \Eprint {http://arxiv.org/abs/hep-ph/0212159}
  {arXiv:hep-ph/0212159 [hep-ph]} \BibitemShut {NoStop}%
\bibitem [{\citenamefont {Collins}\ and\ \citenamefont
  {Qiu}(2007)}]{Collins:2007nk}%
  \BibitemOpen
  \bibfield  {author} {\bibinfo {author} {\bibfnamefont {J.}~\bibnamefont
  {Collins}}\ and\ \bibinfo {author} {\bibfnamefont {J.-W.}\ \bibnamefont
  {Qiu}},\ }\href {\doibase 10.1103/PhysRevD.75.114014} {\bibfield  {journal}
  {\bibinfo  {journal} {Phys.Rev.}\ }\textbf {\bibinfo {volume} {D75}},\
  \bibinfo {pages} {114014} (\bibinfo {year} {2007})},\ \Eprint
  {http://arxiv.org/abs/0705.2141} {arXiv:0705.2141 [hep-ph]} \BibitemShut
  {NoStop}%
\bibitem [{\citenamefont {Rogers}\ and\ \citenamefont
  {Mulders}(2010)}]{Rogers:2010dm}%
  \BibitemOpen
  \bibfield  {author} {\bibinfo {author} {\bibfnamefont {T.~C.}\ \bibnamefont
  {Rogers}}\ and\ \bibinfo {author} {\bibfnamefont {P.~J.}\ \bibnamefont
  {Mulders}},\ }\href {\doibase 10.1103/PhysRevD.81.094006} {\bibfield
  {journal} {\bibinfo  {journal} {Phys.Rev.}\ }\textbf {\bibinfo {volume}
  {D81}},\ \bibinfo {pages} {094006} (\bibinfo {year} {2010})},\ \Eprint
  {http://arxiv.org/abs/1001.2977} {arXiv:1001.2977 [hep-ph]} \BibitemShut
  {NoStop}%
\bibitem [{\citenamefont {Berger}\ and\ \citenamefont
  {Meng}(1994)}]{Berger:1993yp}%
  \BibitemOpen
  \bibfield  {author} {\bibinfo {author} {\bibfnamefont {E.~L.}\ \bibnamefont
  {Berger}}\ and\ \bibinfo {author} {\bibfnamefont {R.-b.}\ \bibnamefont
  {Meng}},\ }\href {\doibase 10.1103/PhysRevD.49.3248} {\bibfield  {journal}
  {\bibinfo  {journal} {Phys.Rev.}\ }\textbf {\bibinfo {volume} {D49}},\
  \bibinfo {pages} {3248} (\bibinfo {year} {1994})},\ \Eprint
  {http://arxiv.org/abs/hep-ph/9310341} {arXiv:hep-ph/9310341 [hep-ph]}
  \BibitemShut {NoStop}%
\bibitem [{\citenamefont {Mrenna}\ and\ \citenamefont
  {Yuan}(1997)}]{Mrenna:1996cz}%
  \BibitemOpen
  \bibfield  {author} {\bibinfo {author} {\bibfnamefont {S.}~\bibnamefont
  {Mrenna}}\ and\ \bibinfo {author} {\bibfnamefont {C.~P.}~\bibnamefont {Yuan}},\
  }\href {\doibase 10.1103/PhysRevD.55.120} {\bibfield  {journal} {\bibinfo
  {journal} {Phys.Rev.}\ }\textbf {\bibinfo {volume} {D55}},\ \bibinfo {pages}
  {120} (\bibinfo {year} {1997})},\ \Eprint
  {http://arxiv.org/abs/hep-ph/9606363} {arXiv:hep-ph/9606363 [hep-ph]}
  \BibitemShut {NoStop}%
\bibitem [{\citenamefont {Bauer}\ \emph {et~al.}(2001)\citenamefont {Bauer},
  \citenamefont {Fleming}, \citenamefont {Pirjol},\ and\ \citenamefont
  {Stewart}}]{Bauer:2000yr}%
  \BibitemOpen
  \bibfield  {author} {\bibinfo {author} {\bibfnamefont {C.~W.}\ \bibnamefont
  {Bauer}}, \bibinfo {author} {\bibfnamefont {S.}~\bibnamefont {Fleming}},
  \bibinfo {author} {\bibfnamefont {D.}~\bibnamefont {Pirjol}}, \ and\ \bibinfo
  {author} {\bibfnamefont {I.~W.}\ \bibnamefont {Stewart}},\ }\href {\doibase
  10.1103/PhysRevD.63.114020} {\bibfield  {journal} {\bibinfo  {journal}
  {Phys.Rev.}\ }\textbf {\bibinfo {volume} {D63}},\ \bibinfo {pages} {114020}
  (\bibinfo {year} {2001})},\ \Eprint {http://arxiv.org/abs/hep-ph/0011336}
  {arXiv:hep-ph/0011336 [hep-ph]} \BibitemShut {NoStop}%
\bibitem [{\citenamefont {Bauer}\ \emph {et~al.}(2002)\citenamefont {Bauer},
  \citenamefont {Pirjol},\ and\ \citenamefont {Stewart}}]{Bauer:2001yt}%
  \BibitemOpen
  \bibfield  {author} {\bibinfo {author} {\bibfnamefont {C.~W.}\ \bibnamefont
  {Bauer}}, \bibinfo {author} {\bibfnamefont {D.}~\bibnamefont {Pirjol}}, \
  and\ \bibinfo {author} {\bibfnamefont {I.~W.}\ \bibnamefont {Stewart}},\
  }\href {\doibase 10.1103/PhysRevD.65.054022} {\bibfield  {journal} {\bibinfo
  {journal} {Phys.Rev.}\ }\textbf {\bibinfo {volume} {D65}},\ \bibinfo {pages}
  {054022} (\bibinfo {year} {2002})},\ \Eprint
  {http://arxiv.org/abs/hep-ph/0109045} {arXiv:hep-ph/0109045 [hep-ph]}
  \BibitemShut {NoStop}%
\bibitem [{\citenamefont {Beneke}\ \emph {et~al.}(2002)\citenamefont {Beneke},
  \citenamefont {Chapovsky}, \citenamefont {Diehl},\ and\ \citenamefont
  {Feldmann}}]{Beneke:2002ph}%
  \BibitemOpen
  \bibfield  {author} {\bibinfo {author} {\bibfnamefont {M.}~\bibnamefont
  {Beneke}}, \bibinfo {author} {\bibfnamefont {A.}~\bibnamefont {Chapovsky}},
  \bibinfo {author} {\bibfnamefont {M.}~\bibnamefont {Diehl}}, \ and\ \bibinfo
  {author} {\bibfnamefont {T.}~\bibnamefont {Feldmann}},\ }\href {\doibase
  10.1016/S0550-3213(02)00687-9} {\bibfield  {journal} {\bibinfo  {journal}
  {Nucl.Phys.}\ }\textbf {\bibinfo {volume} {B643}},\ \bibinfo {pages} {431}
  (\bibinfo {year} {2002})},\ \Eprint {http://arxiv.org/abs/hep-ph/0206152}
  {arXiv:hep-ph/0206152 [hep-ph]} \BibitemShut {NoStop}%
\bibitem [{\citenamefont {Gao}\ \emph {et~al.}(2005)\citenamefont {Gao},
  \citenamefont {Li},\ and\ \citenamefont {Liu}}]{Gao:2005iu}%
  \BibitemOpen
  \bibfield  {author} {\bibinfo {author} {\bibfnamefont {Y.}~\bibnamefont
  {Gao}}, \bibinfo {author} {\bibfnamefont {C.~S.}\ \bibnamefont {Li}}, \ and\
  \bibinfo {author} {\bibfnamefont {J.~J.}\ \bibnamefont {Liu}},\ }\href
  {\doibase 10.1103/PhysRevD.72.114020} {\bibfield  {journal} {\bibinfo
  {journal} {Phys.Rev.}\ }\textbf {\bibinfo {volume} {D72}},\ \bibinfo {pages}
  {114020} (\bibinfo {year} {2005})},\ \Eprint
  {http://arxiv.org/abs/hep-ph/0501229} {arXiv:hep-ph/0501229 [hep-ph]}
  \BibitemShut {NoStop}%
\bibitem [{\citenamefont {Idilbi}\ \emph {et~al.}(2005)\citenamefont {Idilbi},
  \citenamefont {Ji},\ and\ \citenamefont {Yuan}}]{Idilbi:2005er}%
  \BibitemOpen
  \bibfield  {author} {\bibinfo {author} {\bibfnamefont {A.}~\bibnamefont
  {Idilbi}}, \bibinfo {author} {\bibfnamefont {X.-d.}\ \bibnamefont {Ji}}, \
  and\ \bibinfo {author} {\bibfnamefont {F.}~\bibnamefont {Yuan}},\ }\href
  {\doibase 10.1016/j.physletb.2005.08.038} {\bibfield  {journal} {\bibinfo
  {journal} {Phys.Lett.}\ }\textbf {\bibinfo {volume} {B625}},\ \bibinfo
  {pages} {253} (\bibinfo {year} {2005})},\ \Eprint
  {http://arxiv.org/abs/hep-ph/0507196} {arXiv:hep-ph/0507196 [hep-ph]}
  \BibitemShut {NoStop}%
\bibitem [{\citenamefont {Becher}\ and\ \citenamefont
  {Neubert}(2011)}]{Becher:2010tm}%
  \BibitemOpen
  \bibfield  {author} {\bibinfo {author} {\bibfnamefont {T.}~\bibnamefont
  {Becher}}\ and\ \bibinfo {author} {\bibfnamefont {M.}~\bibnamefont
  {Neubert}},\ }\href {\doibase 10.1140/epjc/s10052-011-1665-7} {\bibfield
  {journal} {\bibinfo  {journal} {Eur.Phys.J.}\ }\textbf {\bibinfo {volume}
  {C71}},\ \bibinfo {pages} {1665} (\bibinfo {year} {2011})},\ \Eprint
  {http://arxiv.org/abs/1007.4005} {arXiv:1007.4005 [hep-ph]} \BibitemShut
  {NoStop}%
\bibitem [{\citenamefont {Becher}\ \emph {et~al.}(2012)\citenamefont {Becher},
  \citenamefont {Neubert},\ and\ \citenamefont {Wilhelm}}]{Becher:2011xn}%
  \BibitemOpen
  \bibfield  {author} {\bibinfo {author} {\bibfnamefont {T.}~\bibnamefont
  {Becher}}, \bibinfo {author} {\bibfnamefont {M.}~\bibnamefont {Neubert}}, \
  and\ \bibinfo {author} {\bibfnamefont {D.}~\bibnamefont {Wilhelm}},\ }\href
  {\doibase 10.1007/JHEP02(2012)124} {\bibfield  {journal} {\bibinfo  {journal}
  {JHEP}\ }\textbf {\bibinfo {volume} {1202}},\ \bibinfo {pages} {124}
  (\bibinfo {year} {2012})},\ \Eprint {http://arxiv.org/abs/1109.6027}
  {arXiv:1109.6027 [hep-ph]} \BibitemShut {NoStop}%
\bibitem [{\citenamefont {Echevarria}\ \emph
  {et~al.}(2012{\natexlab{a}})\citenamefont {Echevarria}, \citenamefont
  {Idilbi},\ and\ \citenamefont {Scimemi}}]{GarciaEchevarria:2011rb}%
  \BibitemOpen
  \bibfield  {author} {\bibinfo {author} {\bibfnamefont {M.~G.}\ \bibnamefont
  {Echevarria}}, \bibinfo {author} {\bibfnamefont {A.}~\bibnamefont {Idilbi}},
  \ and\ \bibinfo {author} {\bibfnamefont {I.}~\bibnamefont {Scimemi}},\ }\href
  {\doibase 10.1007/JHEP07(2012)002} {\bibfield  {journal} {\bibinfo  {journal}
  {JHEP}\ }\textbf {\bibinfo {volume} {1207}},\ \bibinfo {pages} {002}
  (\bibinfo {year} {2012}{\natexlab{a}})},\ \Eprint
  {http://arxiv.org/abs/1111.4996} {arXiv:1111.4996 [hep-ph]} \BibitemShut
  {NoStop}%
\bibitem [{\citenamefont {Chiu}\ \emph {et~al.}(2012)\citenamefont {Chiu},
  \citenamefont {Jain}, \citenamefont {Neill},\ and\ \citenamefont
  {Rothstein}}]{Chiu:2012ir}%
  \BibitemOpen
  \bibfield  {author} {\bibinfo {author} {\bibfnamefont {J.-Y.}\ \bibnamefont
  {Chiu}}, \bibinfo {author} {\bibfnamefont {A.}~\bibnamefont {Jain}}, \bibinfo
  {author} {\bibfnamefont {D.}~\bibnamefont {Neill}}, \ and\ \bibinfo {author}
  {\bibfnamefont {I.~Z.}\ \bibnamefont {Rothstein}},\ }\href {\doibase
  10.1007/JHEP05(2012)084} {\bibfield  {journal} {\bibinfo  {journal} {JHEP}\
  }\textbf {\bibinfo {volume} {1205}},\ \bibinfo {pages} {084} (\bibinfo {year}
  {2012})},\ \Eprint {http://arxiv.org/abs/1202.0814} {arXiv:1202.0814
  [hep-ph]} \BibitemShut {NoStop}%
\bibitem [{\citenamefont {Becher}\ \emph {et~al.}(2013)\citenamefont {Becher},
  \citenamefont {Neubert},\ and\ \citenamefont {Wilhelm}}]{Becher:2012yn}%
  \BibitemOpen
  \bibfield  {author} {\bibinfo {author} {\bibfnamefont {T.}~\bibnamefont
  {Becher}}, \bibinfo {author} {\bibfnamefont {M.}~\bibnamefont {Neubert}}, \
  and\ \bibinfo {author} {\bibfnamefont {D.}~\bibnamefont {Wilhelm}},\ }\href
  {\doibase 10.1007/JHEP05(2013)110} {\bibfield  {journal} {\bibinfo  {journal}
  {JHEP}\ }\textbf {\bibinfo {volume} {1305}},\ \bibinfo {pages} {110}
  (\bibinfo {year} {2013})},\ \Eprint {http://arxiv.org/abs/1212.2621}
  {arXiv:1212.2621 [hep-ph]} \BibitemShut {NoStop}%
\bibitem [{\citenamefont {Zhu}\ \emph {et~al.}(2013)\citenamefont {Zhu},
  \citenamefont {Li}, \citenamefont {Li}, \citenamefont {Shao},\ and\
  \citenamefont {Yang}}]{Zhu:2012ts}%
  \BibitemOpen
  \bibfield  {author} {\bibinfo {author} {\bibfnamefont {H.~X.}\ \bibnamefont
  {Zhu}}, \bibinfo {author} {\bibfnamefont {C.~S.}\ \bibnamefont {Li}},
  \bibinfo {author} {\bibfnamefont {H.~T.}\ \bibnamefont {Li}}, \bibinfo
  {author} {\bibfnamefont {D.~Y.}\ \bibnamefont {Shao}}, \ and\ \bibinfo
  {author} {\bibfnamefont {L.~L.}\ \bibnamefont {Yang}},\ }\href {\doibase
  10.1103/PhysRevLett.110.082001} {\bibfield  {journal} {\bibinfo  {journal}
  {Phys.Rev.Lett.}\ }\textbf {\bibinfo {volume} {110}},\ \bibinfo {pages}
  {082001} (\bibinfo {year} {2013})},\ \Eprint {http://arxiv.org/abs/1208.5774}
  {arXiv:1208.5774 [hep-ph]} \BibitemShut {NoStop}%
\bibitem [{\citenamefont {Catani}\ and\ \citenamefont
  {Grazzini}(2007)}]{Catani:2007vq}%
  \BibitemOpen
  \bibfield  {author} {\bibinfo {author} {\bibfnamefont {S.}~\bibnamefont
  {Catani}}\ and\ \bibinfo {author} {\bibfnamefont {M.}~\bibnamefont
  {Grazzini}},\ }\href {\doibase 10.1103/PhysRevLett.98.222002} {\bibfield
  {journal} {\bibinfo  {journal} {Phys.Rev.Lett.}\ }\textbf {\bibinfo {volume}
  {98}},\ \bibinfo {pages} {222002} (\bibinfo {year} {2007})},\ \Eprint
  {http://arxiv.org/abs/hep-ph/0703012} {arXiv:hep-ph/0703012 [hep-ph]}
  \BibitemShut {NoStop}%
\bibitem [{\citenamefont {Mitov}\ and\ \citenamefont
  {Sterman}(2012)}]{Mitov:2012gt}%
  \BibitemOpen
  \bibfield  {author} {\bibinfo {author} {\bibfnamefont {A.}~\bibnamefont
  {Mitov}}\ and\ \bibinfo {author} {\bibfnamefont {G.}~\bibnamefont
  {Sterman}},\ }\href {\doibase 10.1103/PhysRevD.86.114038} {\bibfield
  {journal} {\bibinfo  {journal} {Phys.Rev.}\ }\textbf {\bibinfo {volume}
  {D86}},\ \bibinfo {pages} {114038} (\bibinfo {year} {2012})},\ \Eprint
  {http://arxiv.org/abs/1209.5798} {arXiv:1209.5798 [hep-ph]} \BibitemShut
  {NoStop}%
\bibitem [{\citenamefont {Eichten}\ and\ \citenamefont
  {Hill}(1990)}]{Eichten:1989zv}%
  \BibitemOpen
  \bibfield  {author} {\bibinfo {author} {\bibfnamefont {E.}~\bibnamefont
  {Eichten}}\ and\ \bibinfo {author} {\bibfnamefont {B.~R.}\ \bibnamefont
  {Hill}},\ }\href {\doibase 10.1016/0370-2693(90)92049-O} {\bibfield
  {journal} {\bibinfo  {journal} {Phys.Lett.}\ }\textbf {\bibinfo {volume}
  {B234}},\ \bibinfo {pages} {511} (\bibinfo {year} {1990})}\BibitemShut
  {NoStop}%
\bibitem [{\citenamefont {Georgi}(1990)}]{Georgi:1990um}%
  \BibitemOpen
  \bibfield  {author} {\bibinfo {author} {\bibfnamefont {H.}~\bibnamefont
  {Georgi}},\ }\href {\doibase 10.1016/0370-2693(90)91128-X} {\bibfield
  {journal} {\bibinfo  {journal} {Phys.Lett.}\ }\textbf {\bibinfo {volume}
  {B240}},\ \bibinfo {pages} {447} (\bibinfo {year} {1990})}\BibitemShut
  {NoStop}%
\bibitem [{\citenamefont {Grinstein}\ \emph {et~al.}(1990)\citenamefont
  {Grinstein}, \citenamefont {Springer},\ and\ \citenamefont
  {Wise}}]{Grinstein:1990tj}%
  \BibitemOpen
  \bibfield  {author} {\bibinfo {author} {\bibfnamefont {B.}~\bibnamefont
  {Grinstein}}, \bibinfo {author} {\bibfnamefont {R.~P.}\ \bibnamefont
  {Springer}}, \ and\ \bibinfo {author} {\bibfnamefont {M.~B.}\ \bibnamefont
  {Wise}},\ }\href {\doibase 10.1016/0550-3213(90)90350-M} {\bibfield
  {journal} {\bibinfo  {journal} {Nucl.Phys.}\ }\textbf {\bibinfo {volume}
  {B339}},\ \bibinfo {pages} {269} (\bibinfo {year} {1990})}\BibitemShut
  {NoStop}%
\bibitem [{\citenamefont {Mannel}\ \emph {et~al.}(1992)\citenamefont {Mannel},
  \citenamefont {Roberts},\ and\ \citenamefont {Ryzak}}]{Mannel:1991mc}%
  \BibitemOpen
  \bibfield  {author} {\bibinfo {author} {\bibfnamefont {T.}~\bibnamefont
  {Mannel}}, \bibinfo {author} {\bibfnamefont {W.}~\bibnamefont {Roberts}}, \
  and\ \bibinfo {author} {\bibfnamefont {Z.}~\bibnamefont {Ryzak}},\ }\href
  {\doibase 10.1016/0550-3213(92)90204-O} {\bibfield  {journal} {\bibinfo
  {journal} {Nucl.Phys.}\ }\textbf {\bibinfo {volume} {B368}},\ \bibinfo
  {pages} {204} (\bibinfo {year} {1992})}\BibitemShut {NoStop}%
\bibitem [{\citenamefont {Neubert}(1994)}]{Neubert:1993mb}%
  \BibitemOpen
  \bibfield  {author} {\bibinfo {author} {\bibfnamefont {M.}~\bibnamefont
  {Neubert}},\ }\href {\doibase 10.1016/0370-1573(94)90091-4} {\bibfield
  {journal} {\bibinfo  {journal} {Phys.Rept.}\ }\textbf {\bibinfo {volume}
  {245}},\ \bibinfo {pages} {259} (\bibinfo {year} {1994})},\ \Eprint
  {http://arxiv.org/abs/hep-ph/9306320} {arXiv:hep-ph/9306320 [hep-ph]}
  \BibitemShut {NoStop}%
\bibitem [{\citenamefont {Manohar}\ and\ \citenamefont
  {Wise}(2000)}]{Manohar:2000dt}%
  \BibitemOpen
  \bibfield  {author} {\bibinfo {author} {\bibfnamefont {A.~V.}\ \bibnamefont
  {Manohar}}\ and\ \bibinfo {author} {\bibfnamefont {M.~B.}\ \bibnamefont
  {Wise}},\ }\href@noop {} {\bibfield  {journal} {\bibinfo  {journal}
  {Camb.Monogr.Part.Phys.Nucl.Phys.Cosmol.}\ }\textbf {\bibinfo {volume}
  {10}},\ \bibinfo {pages} {1} (\bibinfo {year} {2000})}\BibitemShut {NoStop}%
\bibitem [{\citenamefont {Yang}\ \emph {et~al.}(2006)\citenamefont {Yang},
  \citenamefont {Li}, \citenamefont {Gao},\ and\ \citenamefont
  {Liu}}]{Yang:2006gs}%
  \BibitemOpen
  \bibfield  {author} {\bibinfo {author} {\bibfnamefont {L.~L.}\ \bibnamefont
  {Yang}}, \bibinfo {author} {\bibfnamefont {C.~S.}\ \bibnamefont {Li}},
  \bibinfo {author} {\bibfnamefont {Y.}~\bibnamefont {Gao}}, \ and\ \bibinfo
  {author} {\bibfnamefont {J.~J.}\ \bibnamefont {Liu}},\ }\href {\doibase
  10.1103/PhysRevD.73.074017} {\bibfield  {journal} {\bibinfo  {journal}
  {Phys.Rev.}\ }\textbf {\bibinfo {volume} {D73}},\ \bibinfo {pages} {074017}
  (\bibinfo {year} {2006})},\ \Eprint {http://arxiv.org/abs/hep-ph/0601180}
  {arXiv:hep-ph/0601180 [hep-ph]} \BibitemShut {NoStop}%
\bibitem [{\citenamefont {Ahrens}\ \emph
  {et~al.}(2011{\natexlab{a}})\citenamefont {Ahrens}, \citenamefont
  {Ferroglia}, \citenamefont {Neubert}, \citenamefont {Pecjak},\ and\
  \citenamefont {Yang}}]{Ahrens:2011mw}%
  \BibitemOpen
  \bibfield  {author} {\bibinfo {author} {\bibfnamefont {V.}~\bibnamefont
  {Ahrens}}, \bibinfo {author} {\bibfnamefont {A.}~\bibnamefont {Ferroglia}},
  \bibinfo {author} {\bibfnamefont {M.}~\bibnamefont {Neubert}}, \bibinfo
  {author} {\bibfnamefont {B.~D.}\ \bibnamefont {Pecjak}}, \ and\ \bibinfo
  {author} {\bibfnamefont {L.-L.}\ \bibnamefont {Yang}},\ }\href {\doibase
  10.1007/JHEP09(2011)070} {\bibfield  {journal} {\bibinfo  {journal} {JHEP}\
  }\textbf {\bibinfo {volume} {1109}},\ \bibinfo {pages} {070} (\bibinfo {year}
  {2011}{\natexlab{a}})},\ \Eprint {http://arxiv.org/abs/1103.0550}
  {arXiv:1103.0550 [hep-ph]} \BibitemShut {NoStop}%
\bibitem [{\citenamefont {Ahrens}\ \emph
  {et~al.}(2011{\natexlab{b}})\citenamefont {Ahrens}, \citenamefont
  {Ferroglia}, \citenamefont {Neubert}, \citenamefont {Pecjak},\ and\
  \citenamefont {Yang}}]{Ahrens:2011px}%
  \BibitemOpen
  \bibfield  {author} {\bibinfo {author} {\bibfnamefont {V.}~\bibnamefont
  {Ahrens}}, \bibinfo {author} {\bibfnamefont {A.}~\bibnamefont {Ferroglia}},
  \bibinfo {author} {\bibfnamefont {M.}~\bibnamefont {Neubert}}, \bibinfo
  {author} {\bibfnamefont {B.~D.}\ \bibnamefont {Pecjak}}, \ and\ \bibinfo
  {author} {\bibfnamefont {L.~L.}\ \bibnamefont {Yang}},\ }\href {\doibase
  10.1016/j.physletb.2011.07.058} {\bibfield  {journal} {\bibinfo  {journal}
  {Phys.Lett.}\ }\textbf {\bibinfo {volume} {B703}},\ \bibinfo {pages} {135}
  (\bibinfo {year} {2011}{\natexlab{b}})},\ \Eprint
  {http://arxiv.org/abs/1105.5824} {arXiv:1105.5824 [hep-ph]} \BibitemShut
  {NoStop}%
\bibitem [{\citenamefont {Ahrens}\ \emph
  {et~al.}(2011{\natexlab{c}})\citenamefont {Ahrens}, \citenamefont
  {Ferroglia}, \citenamefont {Neubert}, \citenamefont {Pecjak},\ and\
  \citenamefont {Yang}}]{Ahrens:2011uf}%
  \BibitemOpen
  \bibfield  {author} {\bibinfo {author} {\bibfnamefont {V.}~\bibnamefont
  {Ahrens}}, \bibinfo {author} {\bibfnamefont {A.}~\bibnamefont {Ferroglia}},
  \bibinfo {author} {\bibfnamefont {M.}~\bibnamefont {Neubert}}, \bibinfo
  {author} {\bibfnamefont {B.~D.}\ \bibnamefont {Pecjak}}, \ and\ \bibinfo
  {author} {\bibfnamefont {L.~L.}\ \bibnamefont {Yang}},\ }\href {\doibase
  10.1103/PhysRevD.84.074004} {\bibfield  {journal} {\bibinfo  {journal}
  {Phys.Rev.}\ }\textbf {\bibinfo {volume} {D84}},\ \bibinfo {pages} {074004}
  (\bibinfo {year} {2011}{\natexlab{c}})},\ \Eprint
  {http://arxiv.org/abs/1106.6051} {arXiv:1106.6051 [hep-ph]} \BibitemShut
  {NoStop}%
\bibitem [{\citenamefont {Ferroglia}\ \emph
  {et~al.}(2012{\natexlab{a}})\citenamefont {Ferroglia}, \citenamefont
  {Pecjak},\ and\ \citenamefont {Yang}}]{Ferroglia:2012ku}%
  \BibitemOpen
  \bibfield  {author} {\bibinfo {author} {\bibfnamefont {A.}~\bibnamefont
  {Ferroglia}}, \bibinfo {author} {\bibfnamefont {B.~D.}\ \bibnamefont
  {Pecjak}}, \ and\ \bibinfo {author} {\bibfnamefont {L.~L.}\ \bibnamefont
  {Yang}},\ }\href {\doibase 10.1103/PhysRevD.86.034010} {\bibfield  {journal}
  {\bibinfo  {journal} {Phys.Rev.}\ }\textbf {\bibinfo {volume} {D86}},\
  \bibinfo {pages} {034010} (\bibinfo {year} {2012}{\natexlab{a}})},\ \Eprint
  {http://arxiv.org/abs/1205.3662} {arXiv:1205.3662 [hep-ph]} \BibitemShut
  {NoStop}%
\bibitem [{\citenamefont {Ferroglia}\ \emph
  {et~al.}(2012{\natexlab{b}})\citenamefont {Ferroglia}, \citenamefont
  {Pecjak}, \citenamefont {Yang}, \citenamefont {Pecjak},\ and\ \citenamefont
  {Yang}}]{Ferroglia:2012uy}%
  \BibitemOpen
  \bibfield  {author} {\bibinfo {author} {\bibfnamefont {A.}~\bibnamefont
  {Ferroglia}}, \bibinfo {author} {\bibfnamefont {B.~D.}\ \bibnamefont
  {Pecjak}}, \bibinfo {author} {\bibfnamefont {L.~L.}\ \bibnamefont {Yang}},
  \bibinfo {author} {\bibfnamefont {B.~D.}\ \bibnamefont {Pecjak}}, \ and\
  \bibinfo {author} {\bibfnamefont {L.~L.}\ \bibnamefont {Yang}},\ }\href
  {\doibase 10.1007/JHEP10(2012)180} {\bibfield  {journal} {\bibinfo  {journal}
  {JHEP}\ }\textbf {\bibinfo {volume} {1210}},\ \bibinfo {pages} {180}
  (\bibinfo {year} {2012}{\natexlab{b}})},\ \Eprint
  {http://arxiv.org/abs/1207.4798} {arXiv:1207.4798 [hep-ph]} \BibitemShut
  {NoStop}%
\bibitem [{\citenamefont {Catani}\ and\ \citenamefont
  {Seymour}(1996)}]{Catani:1996jh}%
  \BibitemOpen
  \bibfield  {author} {\bibinfo {author} {\bibfnamefont {S.}~\bibnamefont
  {Catani}}\ and\ \bibinfo {author} {\bibfnamefont {M.}~\bibnamefont
  {Seymour}},\ }\href {\doibase 10.1016/0370-2693(96)00425-X} {\bibfield
  {journal} {\bibinfo  {journal} {Phys.Lett.}\ }\textbf {\bibinfo {volume}
  {B378}},\ \bibinfo {pages} {287} (\bibinfo {year} {1996})},\ \Eprint
  {http://arxiv.org/abs/hep-ph/9602277} {arXiv:hep-ph/9602277 [hep-ph]}
  \BibitemShut {NoStop}%
\bibitem [{\citenamefont {Catani}\ and\ \citenamefont
  {Seymour}(1997)}]{Catani:1996vz}%
  \BibitemOpen
  \bibfield  {author} {\bibinfo {author} {\bibfnamefont {S.}~\bibnamefont
  {Catani}}\ and\ \bibinfo {author} {\bibfnamefont {M.}~\bibnamefont
  {Seymour}},\ }\href {\doibase 10.1016/S0550-3213(96)00589-5} {\bibfield
  {journal} {\bibinfo  {journal} {Nucl.Phys.}\ }\textbf {\bibinfo {volume}
  {B485}},\ \bibinfo {pages} {291} (\bibinfo {year} {1997})},\ \Eprint
  {http://arxiv.org/abs/hep-ph/9605323} {arXiv:hep-ph/9605323 [hep-ph]}
  \BibitemShut {NoStop}%
\bibitem [{\citenamefont {Becher}\ and\ \citenamefont
  {Bell}(2012)}]{Becher:2011dz}%
  \BibitemOpen
  \bibfield  {author} {\bibinfo {author} {\bibfnamefont {T.}~\bibnamefont
  {Becher}}\ and\ \bibinfo {author} {\bibfnamefont {G.}~\bibnamefont {Bell}},\
  }\href {\doibase 10.1016/j.physletb.2012.05.016} {\bibfield  {journal}
  {\bibinfo  {journal} {Phys.Lett.}\ }\textbf {\bibinfo {volume} {B713}},\
  \bibinfo {pages} {41} (\bibinfo {year} {2012})},\ \Eprint
  {http://arxiv.org/abs/1112.3907} {arXiv:1112.3907 [hep-ph]} \BibitemShut
  {NoStop}%
\bibitem [{\citenamefont {Echevarria}\ \emph
  {et~al.}(2012{\natexlab{b}})\citenamefont {Echevarria}, \citenamefont
  {Idilbi},\ and\ \citenamefont {Scimemi}}]{Echevarria:2012qe}%
  \BibitemOpen
  \bibfield  {author} {\bibinfo {author} {\bibfnamefont {M.~G.}\ \bibnamefont
  {Echevarria}}, \bibinfo {author} {\bibfnamefont {A.}~\bibnamefont {Idilbi}},
  \ and\ \bibinfo {author} {\bibfnamefont {I.}~\bibnamefont {Scimemi}},\ }\href
  {\doibase 10.1142/S2010194512009130} {\bibfield  {journal} {\bibinfo
  {journal} {Int.J.Mod.Phys.Conf.Ser.}\ }\textbf {\bibinfo {volume} {20}},\
  \bibinfo {pages} {92} (\bibinfo {year} {2012}{\natexlab{b}})},\ \Eprint
  {http://arxiv.org/abs/1209.3892} {arXiv:1209.3892 [hep-ph]} \BibitemShut
  {NoStop}%
\bibitem [{\citenamefont {Catani}\ and\ \citenamefont
  {Grazzini}(2011)}]{Catani:2010pd}%
  \BibitemOpen
  \bibfield  {author} {\bibinfo {author} {\bibfnamefont {S.}~\bibnamefont
  {Catani}}\ and\ \bibinfo {author} {\bibfnamefont {M.}~\bibnamefont
  {Grazzini}},\ }\href {\doibase 10.1016/j.nuclphysb.2010.12.007} {\bibfield
  {journal} {\bibinfo  {journal} {Nucl.Phys.}\ }\textbf {\bibinfo {volume}
  {B845}},\ \bibinfo {pages} {297} (\bibinfo {year} {2011})},\ \Eprint
  {http://arxiv.org/abs/1011.3918} {arXiv:1011.3918 [hep-ph]} \BibitemShut
  {NoStop}%
\bibitem [{\citenamefont {Becher}\ and\ \citenamefont
  {Neubert}(2012)}]{Becher:2012qa}%
  \BibitemOpen
  \bibfield  {author} {\bibinfo {author} {\bibfnamefont {T.}~\bibnamefont
  {Becher}}\ and\ \bibinfo {author} {\bibfnamefont {M.}~\bibnamefont
  {Neubert}},\ }\href {\doibase 10.1007/JHEP07(2012)108} {\bibfield  {journal}
  {\bibinfo  {journal} {JHEP}\ }\textbf {\bibinfo {volume} {1207}},\ \bibinfo
  {pages} {108} (\bibinfo {year} {2012})},\ \Eprint
  {http://arxiv.org/abs/1205.3806} {arXiv:1205.3806 [hep-ph]} \BibitemShut
  {NoStop}%
\bibitem [{\citenamefont {Ferroglia}\ \emph {et~al.}(2009)\citenamefont
  {Ferroglia}, \citenamefont {Neubert}, \citenamefont {Pecjak},\ and\
  \citenamefont {Yang}}]{Ferroglia:2009ii}%
  \BibitemOpen
  \bibfield  {author} {\bibinfo {author} {\bibfnamefont {A.}~\bibnamefont
  {Ferroglia}}, \bibinfo {author} {\bibfnamefont {M.}~\bibnamefont {Neubert}},
  \bibinfo {author} {\bibfnamefont {B.~D.}\ \bibnamefont {Pecjak}}, \ and\
  \bibinfo {author} {\bibfnamefont {L.~L.}\ \bibnamefont {Yang}},\ }\href
  {\doibase 10.1088/1126-6708/2009/11/062} {\bibfield  {journal} {\bibinfo
  {journal} {JHEP}\ }\textbf {\bibinfo {volume} {0911}},\ \bibinfo {pages}
  {062} (\bibinfo {year} {2009})},\ \Eprint {http://arxiv.org/abs/0908.3676}
  {arXiv:0908.3676 [hep-ph]} \BibitemShut {NoStop}%
\bibitem [{\citenamefont {Buras}\ \emph {et~al.}(1992)\citenamefont {Buras},
  \citenamefont {Jamin}, \citenamefont {Lautenbacher},\ and\ \citenamefont
  {Weisz}}]{Buras:1991jm}%
  \BibitemOpen
  \bibfield  {author} {\bibinfo {author} {\bibfnamefont {A.~J.}\ \bibnamefont
  {Buras}}, \bibinfo {author} {\bibfnamefont {M.}~\bibnamefont {Jamin}},
  \bibinfo {author} {\bibfnamefont {M.}~\bibnamefont {Lautenbacher}}, \ and\
  \bibinfo {author} {\bibfnamefont {P.~H.}\ \bibnamefont {Weisz}},\ }\href
  {\doibase 10.1016/0550-3213(92)90345-C} {\bibfield  {journal} {\bibinfo
  {journal} {Nucl.Phys.}\ }\textbf {\bibinfo {volume} {B370}},\ \bibinfo
  {pages} {69} (\bibinfo {year} {1992})}\BibitemShut {NoStop}%
\bibitem [{\citenamefont {Buchalla}\ \emph {et~al.}(1996)\citenamefont
  {Buchalla}, \citenamefont {Buras},\ and\ \citenamefont
  {Lautenbacher}}]{Buchalla:1995vs}%
  \BibitemOpen
  \bibfield  {author} {\bibinfo {author} {\bibfnamefont {G.}~\bibnamefont
  {Buchalla}}, \bibinfo {author} {\bibfnamefont {A.~J.}\ \bibnamefont {Buras}},
  \ and\ \bibinfo {author} {\bibfnamefont {M.~E.}\ \bibnamefont
  {Lautenbacher}},\ }\href {\doibase 10.1103/RevModPhys.68.1125} {\bibfield
  {journal} {\bibinfo  {journal} {Rev.Mod.Phys.}\ }\textbf {\bibinfo {volume}
  {68}},\ \bibinfo {pages} {1125} (\bibinfo {year} {1996})},\ \Eprint
  {http://arxiv.org/abs/hep-ph/9512380} {arXiv:hep-ph/9512380 [hep-ph]}
  \BibitemShut {NoStop}%
\bibitem [{\citenamefont {Campbell}\ and\ \citenamefont
  {Ellis}(2000)}]{Campbell:2000bg}%
  \BibitemOpen
  \bibfield  {author} {\bibinfo {author} {\bibfnamefont {J.~M.}\ \bibnamefont
  {Campbell}}\ and\ \bibinfo {author} {\bibfnamefont {R.~K.}\ \bibnamefont
  {Ellis}},\ }\href {\doibase 10.1103/PhysRevD.62.114012} {\bibfield  {journal}
  {\bibinfo  {journal} {Phys.Rev.}\ }\textbf {\bibinfo {volume} {D62}},\
  \bibinfo {pages} {114012} (\bibinfo {year} {2000})},\ \Eprint
  {http://arxiv.org/abs/hep-ph/0006304} {arXiv:hep-ph/0006304 [hep-ph]}
  \BibitemShut {NoStop}%
\bibitem [{\citenamefont {Schwartz}(2008)}]{Schwartz:2007ib}%
  \BibitemOpen
  \bibfield  {author} {\bibinfo {author} {\bibfnamefont {M.~D.}\ \bibnamefont
  {Schwartz}},\ }\href {\doibase 10.1103/PhysRevD.77.014026} {\bibfield
  {journal} {\bibinfo  {journal} {Phys.Rev.}\ }\textbf {\bibinfo {volume}
  {D77}},\ \bibinfo {pages} {014026} (\bibinfo {year} {2008})},\ \Eprint
  {http://arxiv.org/abs/0709.2709} {arXiv:0709.2709 [hep-ph]} \BibitemShut
  {NoStop}%
\bibitem [{\citenamefont {Martin}\ \emph {et~al.}(2009)\citenamefont {Martin},
  \citenamefont {Stirling}, \citenamefont {Thorne},\ and\ \citenamefont
  {Watt}}]{Martin:2009iq}%
  \BibitemOpen
  \bibfield  {author} {\bibinfo {author} {\bibfnamefont {A.}~\bibnamefont
  {Martin}}, \bibinfo {author} {\bibfnamefont {W.}~\bibnamefont {Stirling}},
  \bibinfo {author} {\bibfnamefont {R.}~\bibnamefont {Thorne}}, \ and\ \bibinfo
  {author} {\bibfnamefont {G.}~\bibnamefont {Watt}},\ }\href {\doibase
  10.1140/epjc/s10052-009-1072-5} {\bibfield  {journal} {\bibinfo  {journal}
  {Eur.Phys.J.}\ }\textbf {\bibinfo {volume} {C63}},\ \bibinfo {pages} {189}
  (\bibinfo {year} {2009})},\ \Eprint {http://arxiv.org/abs/0901.0002}
  {arXiv:0901.0002 [hep-ph]} \BibitemShut {NoStop}%
\bibitem [{\citenamefont {Gehrmann}\ \emph {et~al.}(2012)\citenamefont
  {Gehrmann}, \citenamefont {Lubbert},\ and\ \citenamefont
  {Yang}}]{Gehrmann:2012ze}%
  \BibitemOpen
  \bibfield  {author} {\bibinfo {author} {\bibfnamefont {T.}~\bibnamefont
  {Gehrmann}}, \bibinfo {author} {\bibfnamefont {T.}~\bibnamefont {Lubbert}}, \
  and\ \bibinfo {author} {\bibfnamefont {L.~L.}\ \bibnamefont {Yang}},\ }\href
  {\doibase 10.1103/PhysRevLett.109.242003} {\bibfield  {journal} {\bibinfo
  {journal} {Phys.Rev.Lett.}\ }\textbf {\bibinfo {volume} {109}},\ \bibinfo
  {pages} {242003} (\bibinfo {year} {2012})},\ \Eprint
  {http://arxiv.org/abs/1209.0682} {arXiv:1209.0682 [hep-ph]} \BibitemShut
  {NoStop}%
\bibitem [{\citenamefont {Nason}(2004)}]{Nason:2004rx}%
  \BibitemOpen
  \bibfield  {author} {\bibinfo {author} {\bibfnamefont {P.}~\bibnamefont
  {Nason}},\ }\href {\doibase 10.1088/1126-6708/2004/11/040} {\bibfield
  {journal} {\bibinfo  {journal} {JHEP}\ }\textbf {\bibinfo {volume} {0411}},\
  \bibinfo {pages} {040} (\bibinfo {year} {2004})},\ \Eprint
  {http://arxiv.org/abs/hep-ph/0409146} {arXiv:hep-ph/0409146 [hep-ph]}
  \BibitemShut {NoStop}%
\bibitem [{\citenamefont {Frixione}\ \emph {et~al.}(2007)\citenamefont
  {Frixione}, \citenamefont {Nason},\ and\ \citenamefont
  {Oleari}}]{Frixione:2007vw}%
  \BibitemOpen
  \bibfield  {author} {\bibinfo {author} {\bibfnamefont {S.}~\bibnamefont
  {Frixione}}, \bibinfo {author} {\bibfnamefont {P.}~\bibnamefont {Nason}}, \
  and\ \bibinfo {author} {\bibfnamefont {C.}~\bibnamefont {Oleari}},\ }\href
  {\doibase 10.1088/1126-6708/2007/11/070} {\bibfield  {journal} {\bibinfo
  {journal} {JHEP}\ }\textbf {\bibinfo {volume} {0711}},\ \bibinfo {pages}
  {070} (\bibinfo {year} {2007})},\ \Eprint {http://arxiv.org/abs/0709.2092}
  {arXiv:0709.2092 [hep-ph]} \BibitemShut {NoStop}%
\bibitem [{\citenamefont {Alioli}\ \emph {et~al.}(2010)\citenamefont {Alioli},
  \citenamefont {Nason}, \citenamefont {Oleari},\ and\ \citenamefont
  {Re}}]{Alioli:2010xd}%
  \BibitemOpen
  \bibfield  {author} {\bibinfo {author} {\bibfnamefont {S.}~\bibnamefont
  {Alioli}}, \bibinfo {author} {\bibfnamefont {P.}~\bibnamefont {Nason}},
  \bibinfo {author} {\bibfnamefont {C.}~\bibnamefont {Oleari}}, \ and\ \bibinfo
  {author} {\bibfnamefont {E.}~\bibnamefont {Re}},\ }\href {\doibase
  10.1007/JHEP06(2010)043} {\bibfield  {journal} {\bibinfo  {journal} {JHEP}\
  }\textbf {\bibinfo {volume} {1006}},\ \bibinfo {pages} {043} (\bibinfo {year}
  {2010})},\ \Eprint {http://arxiv.org/abs/1002.2581} {arXiv:1002.2581
  [hep-ph]} \BibitemShut {NoStop}%
\bibitem [{\citenamefont {Alioli}\ \emph {et~al.}(2012)\citenamefont {Alioli},
  \citenamefont {Moch},\ and\ \citenamefont {Uwer}}]{Alioli:2011as}%
  \BibitemOpen
  \bibfield  {author} {\bibinfo {author} {\bibfnamefont {S.}~\bibnamefont
  {Alioli}}, \bibinfo {author} {\bibfnamefont {S.-O.}\ \bibnamefont {Moch}}, \
  and\ \bibinfo {author} {\bibfnamefont {P.}~\bibnamefont {Uwer}},\ }\href
  {\doibase 10.1007/JHEP01(2012)137} {\bibfield  {journal} {\bibinfo  {journal}
  {JHEP}\ }\textbf {\bibinfo {volume} {1201}},\ \bibinfo {pages} {137}
  (\bibinfo {year} {2012})},\ \Eprint {http://arxiv.org/abs/1110.5251}
  {arXiv:1110.5251 [hep-ph]} \BibitemShut {NoStop}%
\bibitem [{\citenamefont {Dittmaier}\ \emph {et~al.}(2007)\citenamefont
  {Dittmaier}, \citenamefont {Uwer},\ and\ \citenamefont
  {Weinzierl}}]{Dittmaier:2007wz}%
  \BibitemOpen
  \bibfield  {author} {\bibinfo {author} {\bibfnamefont {S.}~\bibnamefont
  {Dittmaier}}, \bibinfo {author} {\bibfnamefont {P.}~\bibnamefont {Uwer}}, \
  and\ \bibinfo {author} {\bibfnamefont {S.}~\bibnamefont {Weinzierl}},\ }\href
  {\doibase 10.1103/PhysRevLett.98.262002} {\bibfield  {journal} {\bibinfo
  {journal} {Phys.Rev.Lett.}\ }\textbf {\bibinfo {volume} {98}},\ \bibinfo
  {pages} {262002} (\bibinfo {year} {2007})},\ \Eprint
  {http://arxiv.org/abs/hep-ph/0703120} {arXiv:hep-ph/0703120 [HEP-PH]}
  \BibitemShut {NoStop}%
\bibitem [{\citenamefont {Sjostrand}\ \emph {et~al.}(2008)\citenamefont
  {Sjostrand}, \citenamefont {Mrenna},\ and\ \citenamefont
  {Skands}}]{Sjostrand:2007gs}%
  \BibitemOpen
  \bibfield  {author} {\bibinfo {author} {\bibfnamefont {T.}~\bibnamefont
  {Sjostrand}}, \bibinfo {author} {\bibfnamefont {S.}~\bibnamefont {Mrenna}}, \
  and\ \bibinfo {author} {\bibfnamefont {P.~Z.}\ \bibnamefont {Skands}},\
  }\href {\doibase 10.1016/j.cpc.2008.01.036} {\bibfield  {journal} {\bibinfo
  {journal} {Comput.Phys.Commun.}\ }\textbf {\bibinfo {volume} {178}},\
  \bibinfo {pages} {852} (\bibinfo {year} {2008})},\ \Eprint
  {http://arxiv.org/abs/0710.3820} {arXiv:0710.3820 [hep-ph]} \BibitemShut
  {NoStop}%
\bibitem [{\citenamefont {Frixione}\ \emph
  {et~al.}(2007{\natexlab{b}})\citenamefont {Frixione}, \citenamefont {Nason},\
  and\ \citenamefont {Ridolfi}}]{Frixione:2007nw}%
  \BibitemOpen
  \bibfield  {author} {\bibinfo {author} {\bibfnamefont {S.}~\bibnamefont
  {Frixione}}, \bibinfo {author} {\bibfnamefont {P.}~\bibnamefont {Nason}}, \
  and\ \bibinfo {author} {\bibfnamefont {G.}~\bibnamefont {Ridolfi}},\ }\href
  {\doibase 10.1088/1126-6708/2007/09/126} {\bibfield  {journal} {\bibinfo
  {journal} {JHEP}\ }\textbf {\bibinfo {volume} {0709}},\ \bibinfo {pages}
  {126} (\bibinfo {year} {2007}{\natexlab{b}})},\ \Eprint
  {http://arxiv.org/abs/0707.3088} {arXiv:0707.3088 [hep-ph]} \BibitemShut
  {NoStop}%
\bibitem [{\citenamefont {Dittmaier}\ \emph {et~al.}(2009)\citenamefont
  {Dittmaier}, \citenamefont {Uwer},\ and\ \citenamefont
  {Weinzierl}}]{Dittmaier:2008uj}%
  \BibitemOpen
  \bibfield  {author} {\bibinfo {author} {\bibfnamefont {S.}~\bibnamefont
  {Dittmaier}}, \bibinfo {author} {\bibfnamefont {P.}~\bibnamefont {Uwer}}, \
  and\ \bibinfo {author} {\bibfnamefont {S.}~\bibnamefont {Weinzierl}},\ }\href
  {\doibase 10.1140/epjc/s10052-008-0816-y} {\bibfield  {journal} {\bibinfo
  {journal} {Eur.Phys.J.}\ }\textbf {\bibinfo {volume} {C59}},\ \bibinfo
  {pages} {625} (\bibinfo {year} {2009})},\ \Eprint
  {http://arxiv.org/abs/0810.0452} {arXiv:0810.0452 [hep-ph]} \BibitemShut
  {NoStop}%
\bibitem [{\citenamefont {Melnikov}\ and\ \citenamefont
  {Schulze}(2010)}]{Melnikov:2010iu}%
  \BibitemOpen
  \bibfield  {author} {\bibinfo {author} {\bibfnamefont {K.}~\bibnamefont
  {Melnikov}}\ and\ \bibinfo {author} {\bibfnamefont {M.}~\bibnamefont
  {Schulze}},\ }\href {\doibase 10.1016/j.nuclphysb.2010.07.003} {\bibfield
  {journal} {\bibinfo  {journal} {Nucl.Phys.}\ }\textbf {\bibinfo {volume}
  {B840}},\ \bibinfo {pages} {129} (\bibinfo {year} {2010})},\ \Eprint
  {http://arxiv.org/abs/1004.3284} {arXiv:1004.3284 [hep-ph]} \BibitemShut
  {NoStop}%
\bibitem [{\citenamefont {Melnikov}\ \emph {et~al.}(2012)\citenamefont
  {Melnikov}, \citenamefont {Scharf},\ and\ \citenamefont
  {Schulze}}]{Melnikov:2011qx}%
  \BibitemOpen
  \bibfield  {author} {\bibinfo {author} {\bibfnamefont {K.}~\bibnamefont
  {Melnikov}}, \bibinfo {author} {\bibfnamefont {A.}~\bibnamefont {Scharf}}, \
  and\ \bibinfo {author} {\bibfnamefont {M.}~\bibnamefont {Schulze}},\ }\href
  {\doibase 10.1103/PhysRevD.85.054002} {\bibfield  {journal} {\bibinfo
  {journal} {Phys.Rev.}\ }\textbf {\bibinfo {volume} {D85}},\ \bibinfo {pages}
  {054002} (\bibinfo {year} {2012})},\ \Eprint {http://arxiv.org/abs/1111.4991}
  {arXiv:1111.4991 [hep-ph]} \BibitemShut {NoStop}%
\bibitem [{\citenamefont {Aaltonen}\ \emph {et~al.}(2011)\citenamefont
  {Aaltonen} \emph {et~al.}}]{Aaltonen:2011kc}%
  \BibitemOpen
  \bibfield  {author} {\bibinfo {author} {\bibfnamefont {T.}~\bibnamefont
  {Aaltonen}} \emph {et~al.} (\bibinfo {collaboration} {CDF Collaboration}),\
  }\href {\doibase 10.1103/PhysRevD.83.112003} {\bibfield  {journal} {\bibinfo
  {journal} {Phys.Rev.}\ }\textbf {\bibinfo {volume} {D83}},\ \bibinfo {pages}
  {112003} (\bibinfo {year} {2011})},\ \Eprint {http://arxiv.org/abs/1101.0034}
  {arXiv:1101.0034 [hep-ex]} \BibitemShut {NoStop}%
\bibitem [{\citenamefont {Korchemskaya}\ and\ \citenamefont
  {Korchemsky}(1992)}]{Korchemskaya:1992je}%
  \BibitemOpen
  \bibfield  {author} {\bibinfo {author} {\bibfnamefont {I.}~\bibnamefont
  {Korchemskaya}}\ and\ \bibinfo {author} {\bibfnamefont {G.}~\bibnamefont
  {Korchemsky}},\ }\href {\doibase 10.1016/0370-2693(92)91895-G} {\bibfield
  {journal} {\bibinfo  {journal} {Phys.Lett.}\ }\textbf {\bibinfo {volume}
  {B287}},\ \bibinfo {pages} {169} (\bibinfo {year} {1992})}\BibitemShut
  {NoStop}%
\bibitem [{\citenamefont {Moch}\ \emph {et~al.}(2004)\citenamefont {Moch},
  \citenamefont {Vermaseren},\ and\ \citenamefont {Vogt}}]{Moch:2004pa}%
  \BibitemOpen
  \bibfield  {author} {\bibinfo {author} {\bibfnamefont {S.}~\bibnamefont
  {Moch}}, \bibinfo {author} {\bibfnamefont {J.}~\bibnamefont {Vermaseren}}, \
  and\ \bibinfo {author} {\bibfnamefont {A.}~\bibnamefont {Vogt}},\ }\href
  {\doibase 10.1016/j.nuclphysb.2004.03.030} {\bibfield  {journal} {\bibinfo
  {journal} {Nucl.Phys.}\ }\textbf {\bibinfo {volume} {B688}},\ \bibinfo
  {pages} {101} (\bibinfo {year} {2004})},\ \Eprint
  {http://arxiv.org/abs/hep-ph/0403192} {arXiv:hep-ph/0403192 [hep-ph]}
  \BibitemShut {NoStop}%
\bibitem [{\citenamefont {Becher}\ and\ \citenamefont
  {Neubert}(2009{\natexlab{a}})}]{Becher:2009qa}%
  \BibitemOpen
  \bibfield  {author} {\bibinfo {author} {\bibfnamefont {T.}~\bibnamefont
  {Becher}}\ and\ \bibinfo {author} {\bibfnamefont {M.}~\bibnamefont
  {Neubert}},\ }\href {\doibase 10.1088/1126-6708/2009/06/081} {\bibfield
  {journal} {\bibinfo  {journal} {JHEP}\ }\textbf {\bibinfo {volume} {0906}},\
  \bibinfo {pages} {081} (\bibinfo {year} {2009}{\natexlab{a}})},\ \Eprint
  {http://arxiv.org/abs/0903.1126} {arXiv:0903.1126 [hep-ph]} \BibitemShut
  {NoStop}%
\bibitem [{\citenamefont {Becher}\ and\ \citenamefont
  {Neubert}(2009{\natexlab{b}})}]{Becher:2009kw}%
  \BibitemOpen
  \bibfield  {author} {\bibinfo {author} {\bibfnamefont {T.}~\bibnamefont
  {Becher}}\ and\ \bibinfo {author} {\bibfnamefont {M.}~\bibnamefont
  {Neubert}},\ }\href {\doibase 10.1103/PhysRevD.79.125004,
  10.1103/PhysRevD.80.109901} {\bibfield  {journal} {\bibinfo  {journal}
  {Phys.Rev.}\ }\textbf {\bibinfo {volume} {D79}},\ \bibinfo {pages} {125004}
  (\bibinfo {year} {2009}{\natexlab{b}})},\ \Eprint
  {http://arxiv.org/abs/0904.1021} {arXiv:0904.1021 [hep-ph]} \BibitemShut
  {NoStop}%
\bibitem [{\citenamefont {Huber}\ and\ \citenamefont
  {Maitre}(2006)}]{Huber:2005yg}%
  \BibitemOpen
  \bibfield  {author} {\bibinfo {author} {\bibfnamefont {T.}~\bibnamefont
  {Huber}}\ and\ \bibinfo {author} {\bibfnamefont {D.}~\bibnamefont {Maitre}},\
  }\href {\doibase 10.1016/j.cpc.2006.01.007} {\bibfield  {journal} {\bibinfo
  {journal} {Comput.Phys.Commun.}\ }\textbf {\bibinfo {volume} {175}},\
  \bibinfo {pages} {122} (\bibinfo {year} {2006})},\ \Eprint
  {http://arxiv.org/abs/hep-ph/0507094} {arXiv:hep-ph/0507094 [hep-ph]}
  \BibitemShut {NoStop}%
\end{thebibliography}%

\end{document}